\documentclass[acmsmall]{acmart}
\usepackage[utf8]{inputenc}
\usepackage{graphicx}
\usepackage{soul} % allows highlighting
\usepackage{verbatim}
\usepackage{xcolor,colortbl}
\usepackage{pgfplotstable,filecontents}
\pgfplotsset{compat=1.9}% suppress warning
\usepackage[inline]{enumitem}
\usepackage{url}
\usepackage{ulem}
\usepackage{lscape}
\hypersetup{colorlinks=true,urlcolor=blue}
\usepackage{wrapfig}
\usepackage{longtable,tabu}

\usepackage{multirow}
\usepackage{multicol}
\usepackage{booktabs}% For formal tables
\usepackage{threeparttable}

\usepackage{caption}
\usepackage{subcaption}
\usepackage[colorinlistoftodos,prependcaption, textsize=small]{todonotes}

\usepackage{microtype}
\usepackage{rotating}

\usepackage{lipsum}

\usepackage{amsmath}

\newcommand{\smalltilde}{\raise.17ex\hbox{$\scriptstyle\mathtt{\sim}$}}

\usepackage{fancyhdr}

\newcommand*{\questionS}[2]{
	\noindent
	\begin{minipage}{\columnwidth}
		\textbf{Q#1.} #2\\
	\end{minipage}
}

\newcommand*{\quest}[3]{
	\noindent
	\begin{minipage}{\columnwidth}
		\textbf{Q#1.} #2\\
		\begin{itemize*}[before=\itshape,label={$\circ$}]
			#3
		\end{itemize*}
		\vspace{\baselineskip}	
	\end{minipage}
}

\newcommand*{\questionC}[3]{
	\noindent
	\begin{minipage}{\columnwidth}
		\textbf{Section #1.} #2\\
		\begin{itemize*}[before=\itshape,label={$\circ$}]
			#3
		\end{itemize*}
		\vspace{\baselineskip}	
	\end{minipage}
}

\newcommand*{\questionB}[3]{
	\noindent
	\begin{minipage}{\columnwidth}
		\textbf{Subcategory #1.} #2\\
		\begin{itemize*}[before=\itshape,label={$\circ$}]
			#3
		\end{itemize*}
		\vspace{\baselineskip}	
	\end{minipage}
}

\usepackage{arydshln}
\makeatletter
\def\adl@drawiv#1#2#3{%
        \hskip.5\tabcolsep
        \xleaders#3{#2.5\@tempdimb #1{1}#2.5\@tempdimb}%
                #2\z@ plus1fil minus1fil\relax
        \hskip.5\tabcolsep}
\newcommand{\cdashlinelr}[1]{%
  \noalign{\vskip\aboverulesep
           \global\let\@dashdrawstore\adl@draw
           \global\let\adl@draw\adl@drawiv}
  \cdashline{#1}
  \noalign{\global\let\adl@draw\@dashdrawstore
           \vskip\belowrulesep}}
\makeatother

\begin{document}

\title{Testing SOAR Tools in Use}
\thanks{Data \& code from this work available at \url{https://github.com/bridgesra/soar_experiment_data_code}. \\
\footnotesize{This manuscript has been co-authored by UT-Battelle LLC under contract DE-AC05-00OR22725 with the US Department of Energy (DOE). The US government retains and the publisher, by accepting the article for publication, acknowledges that the US government retains a nonexclusive, paid-up, irrevocable, worldwide license to publish or reproduce the published form of this manuscript, or allow others to do so, for US government purposes. DOE will provide public access to these results of federally sponsored research in accordance with the DOE Public Access Plan (\url{http://energy.gov/downloads/doe-public-access-plan}).}}

% \author{Long list}

\author{Robert~A.~Bridges}
% \authornote{Corresponding author}
\orcid{0001-7962-6329}
\email{bridgesra@ornl.gov} 

\author{Ashley E. Rice}
\email{riceae@ornl.gov} 

\author{Sean~Oesch}
\email{oeschts@ornl.gov} 

\author{Jeffrey~A.~Nichols}
\email{nicholsja2@ornl.gov}

\author{Cory~Watson}
\email{watsoncl1@ornl.gov}

\author{Kevin~Spakes}
\email{spakeskd@ornl.gov}

\author{Savannah Norem}
\email{savannah.norem@gmail.com} 

\author{Mike~Huettel}
\email{huettelmr@ornl.gov}

\author{Brian~Jewell}
\email{jewellbc@ornl.gov} 

\author{Brian~Weber}%\authornotemark[1]
\email{weberb@ornl.gov}

\author{Connor~Gannon}%\authornotemark[1]
\email{gannoncm@ornl.gov}

\author{Olivia Bizovi}%\authornotemark[1]
\email{bizovio@ornl.gov}

\author{Samuel C Hollifield}%\authornotemark[1]
\email{hollifieldsc@ornl.gov}

\affiliation{%
    \institution{Oak Ridge National Laboratory}
    \state{TN}
    \country{USA}
    }

\author{Samantha Erwin}
\email{samantha.erwin@pnnl.gov} 
\affiliation{%
    \institution{Pacific Northwest National Laboratory}
    \state{WA}
    \country{USA}
}

% \author{Jonathan M. Hodapp}
% \email{ jonathanhodapp@gmail.com} 
% \affiliation{%
%     \institution{Booz Allen Hamilton}
%     \state{CA}
%     \country{USA}
% }

% Short authors for page headers: 
\renewcommand{\shortauthors}{Bridges, et al.}

% - Robert A. Bridges
% - Jeff {cyber} Nichols
% - Sean Oesch
% - Samantha Erwin 
% - Savannah Norem 
% - Ashley Rice
% - Cory Watson 
% - Kevin Spakes
% - Brian Weber
% - Brian Jewell 
% - Mike Huettel 
% - Olivia Bizovi 
% - Samuel Hollifield
% - Connor Gannon 

\begin{abstract}
Investigations within Security Operation Centers (SOCs) are tedious as they rely on manual efforts to query diverse data sources, overlay related logs, correlate the data into information, and then document results in a ticketing system. 
Security Orchestration, Automation, and Response (SOAR) tools are a relatively new technology that promise, with appropriate configuration, to collect, filter,  and display needed diverse information;  automate many of the common tasks that unnecessarily require SOC analysts' time; facilitate SOC collaboration; and, in doing so, improve both efficiency and consistency of SOCs. 
There has been no prior research to test SOAR tools in practice; hence, understanding and evaluation of their effect is nascent and needed.   
In this paper, we design and administer the first hands-on user study of SOAR tools, involving 24 participants and six commercial SOAR tools. 
Our contributions include the experimental design, itemizing six characteristics of SOAR tools, and a methodology for testing them.    
We describe configuration of a cyber range test environment, including network, user, and threat emulation; a full SOC tool suite; and creation of artifacts allowing multiple representative investigation scenarios to permit testing.
We present the first research results on SOAR tools. 
Concisely, our findings are that: per-SOC SOAR configuration is extremely important; SOAR tools increase efficiency and reduce context switching, although with potentially decreased ticketing accuracy/completeness; user preference is slightly negatively correlated with their performance with the tool; internet dependence varies widely among SOAR tools; and balance of automation with assisting decision making is preferred by senior participants.
We deliver a public user- and tool-anonymized and -obfuscated version of the data.
%from the study to assist future research. 
\end{abstract}

%install/configuration, usability, efficiency gains, context switching reduction, investigation quality, and resilience to degraded internet.
\maketitle

% \todo[inline]{running list of todo's:
% \begin{itemize}
%     \item Jeff please proofread Section 3, 4
%     \item results need to be organized. would like to have italicized concise observations at the end of discussions. 
%     \item map/how to use paper section ``For the reader only interested in the experiment and results, see section2 for and section 5 for results)''
%     ``For the reader interested in the network, scenario, SOC workstation emulation see section ''; good words are escaping me for that
%     \item need to reorder the appendix
%     \item collect concise observations into discussion section
%     \item revisit intro and abstract to tie it all up
% \end{itemize}} 

\section{Introduction}
Security Operation Centers (SOCs) are the portion of an enterprise's Information Technology (IT) team responsible for protecting the organization from cyber threats. 
Modern SOCs leverage a wide variety of sensors that generate an enormous quantity of audit logs (e.g., operating system logs from workstations and servers, network flows from switches); intelligence feeds (e.g., vulnerability scanning reports, malware information); and alerts (e.g., from endpoint Anti Virus (AV) and Network Intrusion Detection Systems (NIDS)). Most SOCs are equipped with a Security Information and Event Management (SIEM) system, which aggregates these diverse data feeds into a centralized platform and  provides configurable dashboards and query interfaces to monitor data and sift, sort, and find artifacts from the network. 
However, even with the centralization that the SIEM can provide, the burden of sorting through vast amounts of information across many sources, identifying which logs require further attention, understanding which alerts are related, and executing a solution still falls on operators~\cite{islam_multi-vocal_nodate, fink2006bridging} and requires substantial manual effort \cite{goodall2004, botta2007towards, kokulu2019matched, fink2006bridging, bridges2018information}. 

Security orchestration, automation, and response (SOAR) tools represent a relatively new technology that attempts to automate many of the common manual tasks with the goal of improving both efficiency and consistency, and as a result security.
According to Gartner's report on SOAR \cite{gartner2019}, a SOAR tool enables ``organizations to take inputs from a variety of sources and apply workflows aligned to processes and procedures ... improving efficiency and consistency of people and processes.''
The main difference between a SOAR tool and a SIEM tool is that SOAR tools have configurable workflows or playbooks that guide analysts and automate many investigation and incident response actions, whereas a SIEM tool's primary function is widespread data collection and query. There is overlap, e.g., both require and facilitate diverse data ingestion and provide  customizable data visualizations, in particular  real-time dashboards. 
SOAR tools aim to make security operations 
(1) more efficient by automating common tasks, prioritizing, correlating, and augmenting incoming alerts (e.g., with threat intelligence), and facilitating communication and collaboration between and within SOCs, and 
(2) more consistent by providing a unified, usable interface where analysts interact with augmented intelligence and alerts and by using playbooks to standardize procedures.

The promised benefits of SOAR tools are extensive, particularly for expediting and adding consistency to the work of low-tier analysts who repeatedly perform investigation tasks. 
Because of their integration capabilities, SOAR tools can relieve analysts from manually navigating multiple heterogeneous data sources to piece together alert context by automatically gathering and displaying all---and ideally only---the data needed for investigation into a usable interface. 
Expediting and enhancing quality of low-tier operators is particularly important for large (i.e., dozens to hundreds of analysts) organizations that are expected to perform and document investigations according to strict processes and with similar quality, as well as for organizations that
experience high turnover of junior analysts (e.g., military). 
Furthermore, incident handling often requires several analysts, all of whom may not necessarily be located at the same SOC. 
SOAR tools address this specific need by offering collaboration capabilities that allow analysts, regardless of geographic location, to coordinate synchronously or asynchronously on a single incident. 

While SOAR tools offer potentially significant efficiency gains, they also entail nontrivial costs. 
SOAR tools are expensive products, both financially and in terms of the effort required to ensure a stable installation with proper configuration; 
further, as a centerpiece technology for a SOC, SOAR adoption entails a systemic shift for SOC processes including training, daily operations, and technical debt. 
In addition, the extent to which SOAR tool adoption increases dependence on reliable internet is an important questions for some SOCs. %open question- one of which we address in this study.  

We conjecture that ease of integration and customization is important for adoption, as is the usability of the configuration of the SOAR tool in the eyes of the user. Consequently, it is natural for SOCs to want to understand the benefits of SOAR tools in practice, and our work attempts to provide insight.  
For example, many research questions arise: 
\begin{itemize}
\item How easily do SOAR tools integrate and permit customization to SOC-specific procedures? 
\item Do SOAR tools perform in internet-degraded environments? 
\item Do SOC operators believe current SOAR tools are usable, and would they prefer using them? 
\item How much do SOAR tools enhance efficiency, especially for repeated investigations suitable for lower-tier analysts? 
In particular, can a SOAR tool be used to effectively ``automate-out'' the role of a Tier 1 (junior) analyst, at least for prespecified, well-defined investigation procedures? 
\item Do SOAR tools reduce context switching for analysts? 
\item When using SOAR tools, does investigation quality improve or diminish? 
\item By using SOAR tools, can junior analysts perform investigations with the quality of a more senior analyst?
\item Do SOAR tools improve collaboration?

\end{itemize}
% In this work, we present a novel expei framework constructed from a combination of mathematical, computational, and experimental methods to emperically study  these  questions about SOAR tools.

Despite their apparent usefulness, SOAR tools in practice have hardly been studied; see Related Works, Section \ref{sec:related-works}. 
In part, SOAR tools are simply too new to admit ample research; in fact, the term ``SOAR'' was coined by Gartner only in 2017, and SOAR deployments are still in their infancy. 
Because SOAR tools must integrate with a variety of SOC tools and are designed to assist humans though interactive use, studying SOAR tools imposes significant barriers. 
For example, an interested research team can seek a cooperative SOC with a SOAR deployment  to administer qualitative research of that SOAR tool in situ or, as this paper describes, can build a test environment that emulates a network, users, threats, and the full SOC tool suite allowing deployment and use of the SOAR tool in a controlled environment. 
Both are difficult,  because of SOC operators' time constraints, SOCs' privacy requirements, and a wide variety of logistic issues imposed by such studies. 
As such, SOAR tools remain full of promise but with with no empirical verification of their benefits and pitfalls. 

% \subsection{AI ATAC Challenge 3: Efficiency and Effectiveness Afforded by SOAR Capabilities}
% Our framework can be used by both industry researchers and academia to meaningfully compare SOAR solutions. 
%We developed this framework to support the 
This work details the first scientific evaluation and user study of six commercial SOAR tools as the final phase of the third Artificial Intelligence Applications to Autonomous Cybersecurity  (AI  ATAC 3)  Challenge \cite{aiatac3}.  
The AI ATAC 3 Challenge was  funded by the US Navy, 
%  a multistep project consisting of interdisciplinary efforts from a large team of researchers to evalute SOAR tools. AI ATAC 3 
and it awarded \$750K to the winning SOAR tool. 
AI ATAC 3 was administered to entice the best SOAR tool creators to submit their technology to the competition with the goal of evaluating their benefits and informing potential acquisition decisions for adoption into Navy SOCs. 
In the first phase, Navy security professionals rated and ranked the eligible submissions based on video overviews of the SOAR tools. 
All but the lowest-rated tools were advanced to the hands-on testing (i.e., final) round. 
This down-selection subexperiment  is described in detail in our previous work  \cite{norem2021mathematical} 
and an overview resides in Appendix \ref{sec:appendix-phase1}. 
This paper describes the final, second phase, a hands-on user study and its results. 
Only SOAR tools submitted to the challenge and passing the first round downselection were included in the study described in this paper. 
As required by the AI ATAC Challenges, SOAR tool contenders involved in this evaluation must be held anonymous.

    \subsection{Contributions} 
% To this end, we 
Our approach to the AI ATAC 3 Challenge designed first-of-their-kind methods for evaluating a SOAR tool by administering a user study of real SOC operators working investigations with and without six commercial SOAR tools, in a controlled test environment. 
In doing so, we seek answers to the questions itemized above, share results, and discuss key takeaways. 
We make the following contributions: 

\textit{SOAR Testing Methodology -} We explain  the design and instantiation of our cyber range (i.e., a data center dedicated to network, user, threat emulation for cyber experimentation) to create an environment in which the SOAR tools can be deployed, configured, and used by analysts. 
We itemize particular challenges and solutions for this experimental setup to assist future researchers. 
%Installation of the SOAR tools was a joint endeavor by our internal team members with representatives from the vendor company. 
We provide our experimental design, including SOAR evaluation criteria\textemdash installation and configuration, resilience to degraded internet, usability, efficiency gains, reduction in context switching, and investigation quality\textemdash and their corresponding testing methodologies. 
Lessons learned are given in Section \ref{sec:discussion}.

\textit{SOAR Takeaways -} Overall, the experiment reported here involved 24 participants (three from our lab and the remaining 21 from four US Navy SOCs)   testing six commercial SOAR tools with over 200 total hours spent working investigations while under study. 
The experiment produced 75 raw quantitative measurements per user-tool test and 68 user-tool tests.  
We gather findings from our data including coded results (i.e., categorized themes from analyzing free-response data from participants), ratings from the research team evaluating SOAR tool installations and resilience to degraded internet situations, and statistical results from quantitative measurements during the user study investigating the questions outlined above. 
In the spirit of reproducibility and to assist further investigations, we have made our experimental data public in a user- and tool-anonymized and -obfuscated table (description in Section \ref{sec:public-data}, download link on title page).%and  \hl{include footnote with github page}. 

% \todo{i also think this feels redundant after the last intro paragraph, consider combining or better distinguishing}
% In our work, we propose an \hl{a first-of-its kind} experimental framework for evaluating SOAR tools and use it to compare six leading COTS SOAR solutions, reporting both our results and experience utilizing the framework. 
% \hl{should have some takeaways that are gains for researchers (prioritized direction? generally lessons learned about what SOAR tools can/can't do? .. } 
% This work fills a gap in the literature \hl{fills gap or seminal work?} and provides a tested methodology for SOAR tool comparison. 

    \subsection{How to Navigate this Paper}
The final subsection of this introduction (Section \ref{SOC background}) provides a brief overview of SOCs' data, sensors, job functions, and problems as needed context for understanding the goals and promise of SOAR tools.
For readers with an understanding of SOCs, this background can safely be skipped. 
Section \ref{sec:related-works} positions this paper in a discussion of previous papers contributing to SOC user studies and SOAR tools. 
Figure \ref{fig:infographic} provides a workflow infographic of the whole experiment to be used as a reference throughout the paper. 
In Section \ref{sec:design}, we articulate a definition of SOAR tools, then in Section \ref{sec:target-capabilites}
 list the characteristics and capabilities of SOAR tools we will test and our methods. 
A concise overview of the testing workflow concludes the section. 
The preprocessing and analysis methods for the quantitative data collected in our user study merited its own discussion in Section \ref{sec:data-analysis-methods}. 
For those interested in the design and instantiation of the experimental environment, Section \ref{sec:test-network} describes these aspects in detail, and descriptions of the investigation scenarios used for the test concludes the section. 
Section \ref{sec:data-results} presents the results of our investigation.
We provide a discussion of limitations to this study in Section \ref{sec:discussion}.
Those interested only in our takeaways may choose to focus on the italicized ``\textit{Observation}'' statements throughout Section \ref{sec:data-results} and the summarized takeaways in Section \ref{sec:discussion}. 
    \subsection{Background: An Overview of Security Operation Centers} 
\label{SOC background}
In order to understand the need for and potential of SOAR tools, one must understand the data, tasks, workflows, and difficulties of a modern SOC; 
hence, we provide a brief overview based on our experience in studying or working with many SOCs \cite{bridges2018information} and from the academic literature of SOC studies \cite{goodall2004, werlinger2009integrated, werlinger2009security, werlinger2010preparation, botta2007towards, sundaramurthy2016turning, islam_multi-vocal_nodate, adetoye2023building, nyre-yu_identifying_nodate}.

\subsubsection{Overarching Goal}
SOCS or Security Operation Centers are IT personnel tasked with maintaining  an enterprise network's systems and data \textit{confidentiality} (only those with authorized access can access the systems and data), \textit{integrity} (the data and the systems' functioning are not corrupted / do not lack fidelity), and to a lesser extent the \textit{availability} (those with access can use the data and systems), as \textit{availability} can sometimes fall into the purview of the Network Operations Center (NOC), a collaborating but separate team in many organizations. 
Each organization has different requirements, systems, data and levels of control over the networked computing assets (e.g., universities do not own or control the software present on student's devices, which are allowed on the network, whereas government facilities' may both own and control much of the software present on its employees computers). Hence, the subsystems, data, workflows, and difficulties will vary. In this SOC overview discussion, as with in our experimental design, we provide a representative sample of each.

\subsubsection{Sensors \& a Sea of Data} 
SOCs deploy a set of sensors within their network to both report audit logs and also automatically block or take actions. 
To aid this discussion, we provide a representative network diagram in Appendix \ref{sec:appendix-network} itemizing alert and logging tools in a network topology. 
As a first line of defense, SOCs will stereotypically deploy and configure at the network level, a firewall, (e.g., pfSense, blocking all but white-listed traffic), rule-based Network Intrusion Detection/Prevention System (NIDS) (e.g., Suricata or SNORT, blocking traffic that violates a pre-specified rule) and for all hosts the SOC controls, endpoint detection and response (e.g., a host-based toolsuite providing firewall, anti-virus/malware detection and prevention, rogue device detection, policy configuration compliance, other telemetry data). 
Services needed for the enterprise network will be in part configured and deployed by the SOC, e.g, email servers (which have automatic blocking, spam detection configurations) or configuration management servers (e.g. Salt Stack). 
Further, most SOCs leverage subscriptions to vulnerability management (e.g. Nessus, which scan network devices and provide reports on known vulnerabilities). 
These are the basic alert, logging, and prevention tools, and many additional are used based on the SOC's needs and resources, e.g., anomaly detectors at the network level, network flow logging, audit log collection for high-value servers/hosts, data level protection to name a few. 
Further, tools to aid investigations post initial detection are common for SOCs with sufficient monetary and human resources, e.g., malware sandboxes for dynamic analysis such as ThreatGrid, or hard drive and memory snapshot and analysis tools. 

The SOC's job is to root out and remediate any breaches to confidentiality, integrity, and availability, and this is made possible by the array of sensors and protection systems. Yet, all the alert and logging systems  produce a \textit{high-volume} (up to 100GB/day in one SOC we visited) of diverse data, spread across many different servers and in various formats. 
Further, an individual log of a network event (e.g., a connection of an internal IP to an external IP, a user login, a file download, ...), without context is difficult to assess. 
The perennial problem for SOC operators gain understanding of what is and is not OK on their network, but this requires use of their sea of data that must be queried, filtered, correlated, and contextualized (e.g., overlaying network flows, firewall logs, NIDS alerts, vulnerabilities, authentications, and running processes for an IP or host of interest) in order to is convert the data into information. 

In our experience, nearly all SOCs have a Security Incident and Event Management System or SIEM (e.g. Elasticsearch), which allows continual ingestion and indexing of the diverse alert and logging data, as well as rapid query and configurable, real-time-updating dashboards. The limitation is on how much data the SIEM can accommodate  and how many logging systems the SOC can purchase and maintain, both based on financial and human resources. 
While the use of a well-configured SIEM usually solves the problem of consolidating and providing visualizations of the diverse data, the shear volume and lack of context are persistent problems.

\subsubsection{Day-to-Day SOC Tasks}
Now armed with  sensors and widespread data collection, query and visualization capabilities, SOCs generally have junior (or ``Tier 1'') analysts perform triage---watching alerts or suspicious logs, and perform a roughly scripted set of investigation activities according to a Standard Operating Procedure (SOP) to determine if the event can be safely ignored or merits further action, e.g., elevation of the event to an ``incident'' that is usually handled by a intermediate-level (``Tier 2'') or senior (``Tier 3'') or perhaps is sent to a specialist on the sensor that requires tweaking  as the event is a false positive. 
Like many teams, SOCs generally employ a ticketing system (e.g,. JIRA) for documenting progress and assigning tasks to their workers. 

Notably, our experience is that for collaboration, especially handoff of an investigation and reporting, some SOCs leverage manually curated reports that must be in a precise format to enable machine-readability. 
Manually typing many details into a format suitable for machine readability is both tedious and prone to downstream errors. 

We refer readers to consider the investigation scenarios described in Section \ref{sec:scenarios} that were designed, based on conversations with many SOC members, to be representative workflows that occur very often in SOCs. 
As one can see, the workflow often begins with a tip from an alert, log, or ticket assigned from another operator, and requires (at least without a SOAR tool) manually gathering and understanding many other correlated data sources  to contextualize, make a decision, and take action. 

\subsubsection{Perennial Problems}
Problems of ``alert fatigue'' stemming from the incredible volume of data and often mundane tasks are common findings in the literature, leading to low quality investigations and/or documentation \cite{sundaramurthy2016turning}.
Further, problems with correlation and contextualization are common as context switching (looking at a different type of data, UI, system, etc.) provides dissonance---the operator must hold their investigation hypotheses and any newly-gained information in their head while simultaneously searching through heterogeneous but possibly related data in possibly disparate systems \cite{goodall2004}. 
Zooming out from the individual to the SOC level, problems of consistency in process, investigation quality, and documentation emerge as the number of SOC personnel grows. 
Recent studies are showing that retention of SOC staff is an issue \cite{adetoye2023building}. 
For military SOCs in particular, high turnover of SOC personnel due to rotation or end of duty is problematic for both consistency and resources needed for constant training of new personnel. 
Investigations that require handoffs (e.g., upon elevation to a more experienced operator or perhaps due to a shift ending/starting) or  collaboration of multiple operators pose difficulties that are exacerbated as consistency of processes and documentation falters. 
SOCs with geographically disparate units (e.g. in military or large organizations) also require technology for handoff and collaboration.  
Finally, because of the uniqueness of each network, sensors require continual  tweaking of configurations to reduce false positives and retain relevance.

\section{Related Work}
\label{sec:related-works}
In this section, we present a concise survey of related research, organized from studies seeking to understand and identify issues in SOCs (which provide context and motivation for SOAR tools), through developments of SOAR concepts and components to use cases for SOAR tools.
Our literature survey reveals 
no academic literature focused on evaluating SOAR tools (save our preliminary work of 
Norem et al.  \cite{norem2021mathematical} discussed in detail in Appendix \ref{sec:phase1}); further, there is no academic literature that proposes and tests experimental frameworks for SOAR tool comparison. 
Rather, related works include  user studies focusing on SOCs (without SOAR tools) that motivate the need for SOAR tools, or develop concepts of, components of, or use cases for SOAR tools as discussed in this section.
Our work, therefore, fills a gap in existing literature by presenting the first experimental framework to evaluate and compare SOAR solutions. 

%To ensure a comprehensive analysis of prior work in this area, a \url{scholar.google.com} alert on \texttt{``SOAR'' AND ``Security''} was employed, allowing us to retrieve matching articles that would then be manually filtered to identify relevant works.
%that SOAR tools are an exciting technology holding much promise for the cyber industry but that there is 
% a lack of research evaluating the actual effectiveness of SOAR solutions.  
% Previous work has been done involving defining the scope of SOAR tools, 
% detailing their benefits, and studying their applications across a wide array of different areas. 
% In fact, there is  

\subsection{SOC Studies \& the Need for Orchestration, Automation}
Existing user studies of SOCs have the illuminated the fact that 
(1) the sheer amount of data at the SOC operators' fingertips is enormous and increasing, yet converting disparate data sources into actionable information requires tacit network-specific knowledge, (2) manual effort is unable to keep pace with SOC needs, and  (3) mental correlation of diverse data source remains a challenge. Further, investigations that require multiple analysts need tools that facilitate collaborations \cite{goodall2004, werlinger2009integrated, werlinger2009security, werlinger2010preparation, botta2007towards, bridges2018information, sundaramurthy2016turning}.
Most directly relevant to SOAR tools is the work of Nyre-Yu  \cite{nyre2021identifying}, who performs a SOC user study followed by a market analysis of security tools, concluding that SOAR tools aim to address needs of incident response teams.

Goodall et al. \cite{goodall2004} note that SOCs generally organize operators into Tiers 1--3, from junior to experienced members. 
The user studies of SOCs drive home the fact that network-specific experience (e.g., as common with Tier 3 analysts)  provides invaluable knowledge when performing investigations \cite{goodall2004, werlinger2010preparation, bridges2018information}. 
Operators must be consistently diligent in monitoring incoming information and must have strong pattern recognition skills to be able to construct a comprehensive overview of a security threat from many different sources \cite{botta2007towards}. 
As such, operator fatigue is common, and recent user studies of SOCs are finding that fatigue is exacerbated by incommensurate demands (e.g., satisfying goals from management that do not align with SOC needs \cite{sundaramurthy2016turning}). 
Further, there is evidence from Werlinger et al.'s 2009 study \cite{werlinger2009security} that SOC tools are not conducive to collaboration.

Common problems within SOCs largely revolve around consistency among operators, along with operator fatigue associated with navigating excessive amounts of data from disparate sources and completing many repetitive tasks during their shift. 
These among other issues are leading to high turnover rates in SOCs according to Adetoye et al. \cite{adetoye2023building}, who mention SOAR technologies as a potential step to address the problem. 
Many research works seek to define and develop methods for orchestration and automation \cite{young_automated_2019,brewer_could_2019,tankard_pandemic_2020, fransen2019security}. 
Orchestration refers to the integration of separate tools into a central platform, and automation refers to the tool's ability to complete certain tasks without necessary aid from an operator. 
In general, orchestration is often considered the driver of automation, and the two terms are frequently used interchangeably \cite{Microsoft, Markets, islam_multi-vocal_nodate}. 
Notably, none focus on evaluating SOAR tools, nor their automation. 
Automation is instrumental in protecting operators from becoming overwhelmed by great swaths of data when implemented alongside a filtering mechanism such that the alerts that reach an operator are almost guaranteed to be legitimate threats that need attention \cite{young_automated_2019, fransen2019security}. 
However, it is worth noting that, while automation of specific tasks can significantly reduce operator burden, it should be implemented with caution. 
The cybersecurity domain requires some level of human oversight due to the inherent uncertainty. As such, automation should be employed as an aid to human effort, rather than as a replacement \cite{schneier, young_automated_2019}. 
Nonetheless, these definitions of orchestration and automation were critical in constructing the submission criteria for this competition.
% the AI ATAC 3 Challenge. 

\subsection{Research Developing SOAR Concepts \& Components}
Many articles discuss orchestration or automation for security \cite{islam_multi-vocal_nodate, nyre-yu_identifying_nodate, nyre-yu_informing_2019, njogu2013comprehensive, sadamatsu2016practice, luo2016orchestration, koyama2015security}, and some even argue that SOAR tools are critical for the modern SOC (e.g.,  to reduce alert fatigue, unify data feeds into a common User Interface (UI), aid in SOC staff retention, and automate common tasks)~\cite{young_automated_2019,brewer_could_2019,tankard_pandemic_2020, fransen2019security, Microsoft, Markets, nyre-yu_identifying_nodate, adetoye2023building}.  
Islam et al. attempt to define SOAR in their survey paper~\cite{islam_multi-vocal_nodate}, in which they conclude that unification, orchestration, and automation are central to SOAR platforms. 
They also provide a conceptual diagram of a SOAR system in a related paper~\cite{islam_architecture-centric_2020}, which argues that such a diagram helps SOAR platform architects build better platforms and that an architectural approach allows SOC staff to better compare potential SOAR solutions for their environment. 
Literature related to SOC effectiveness and the future of SOCs~\cite{erdur_thesis_nodate, kaur_introduction_2021, singh_application_nodate, perera_next_2021} provides helpful context for our study, though it did not directly inform it.
Work proposing SOAR solutions or machine learning methods for detecting malicious and anomalous activity in the SOC is out of the scope of the present work.

Mavroeidis et al. \cite{mavroeidis2020nonproprietary} develop a framework for sharing playbooks. 

Leite et al. \cite{leite2023actionable} pioneer a method for contextualizing security events and identifying threat patterns by correlating threat intelligence (techniques, tactics, and procedures) to network logging data in an automated fashion. 

Configuring a SOAR tool is important for adoption and involves manually identifying the proper Application Programming Interface (API) for many different tools for integration with SOAR tools. 
Sworna et al. \cite{sworna} introduce APIRO (API Recommendation for security Orchestration, automation, and response), a system that aids SOAR configuration through the use of a convolutional neural network.
% APIRO learns  data from the to-be-integrated tools and makes recommendations of the API to use for each. However, even so, an operator must be present to make the final decision on the API, and thus there is still a strong dependence on highly-skilled individuals to ensure robust and reliable deployment, operation, and maintenance of SOAR technology. 

Johnson et al. \cite{johnson2023soar4der} leverage a distributed energy testbed implemented with Zeek (producing traffic logs), Filebeat (producing host logs), and SNORT (a rule-based NIDS) to exhibit a SOAR tool with many playbooks automating detection, response, possibly prevention of a handful of attacks. Notably, the attacks were implemented in the testbed. This work proves the concept that SOAR tools can successful advance such an IOT system's security in terms of speed and automation in responding to attacks, but it does not include any study of how it is used by SOC operators (as is the focus of our study).

\subsection{Research on SOAR Use Cases}
Even though SOAR tools themselves are still a new concept, specialized use cases are already being proposed and developed in the research literature.
Cheng-Hsiang Yeh \cite{cheng2022adopting} presents a case study of a SOC's effort to automate external access via dynamic enterprise firewall re-configuration using SIEM and SOAR, which were already in use at the SOC, compared to manual efforts. 
The contribution focuses on automating the enterprise security architecture with a single SOAR tool.

Bilali et al. \cite{bilali2023iris} provide a brief description of a system designed to  help orchestrate and facilitate collaboration on cyber incidents.

Bearicade is a SOAR system integrated with artificial intelligence technology that is specifically designed to meet the security needs of high-performance computing  systems \cite{miles-board}. 

Another specialized case for SOAR is use in organizations that may lack a large budget for cybersecurity solutions but would still benefit from the offerings of a SOAR tool. 
As such, Gibadullin et al. \cite{gibadullin} propose a solution in which an automated incident management system was constructed using only open-source software. 

Vast et al. \cite{vast_artificial_2021} propose an AI-informed SOAR tool that implements deep learning to assign a risk score to a detected threat, along with natural language processing--driven translators to translate the threat intelligence necessary for sufficient alert context, if it is in a language different from computer settings. 
% The applications of SOAR tools are far-reaching, across multiple industries within varying disciplines. 
% For example, in the realm of autonomous computing, they may be integral for automating vulnerability patches when exposed to a zero-day attack. 
% These automated solutions may be particularly useful for 

Applications of SOAR are being developed to aid autonomous computing, specifically by managing fragmentation issues, in which the attacker identifies weak areas in an organization's cyber kill chain and exploits these vulnerabilities to gain access to and/or control of a system \cite{donevski_survey_2018}. 
SOAR tools can also offer a cybersecurity solution for energy microgrids---cyber-physical systems designed to deliver power---by
%from renewable sources to urban or rural areas, and as such, are susceptible to cyber attacks that can impact their ability to do so. 
% In this setting, SOAR tools can 
providing a centralized dashboard with information from the various data sources in the energy microgrid with which it is integrated. 
Early detection of security anomalies in microgrid devices, followed by at least a partially automated response, can help mitigate the effects of an attack, if not prevent it entirely \cite{dutta_cybersecurity_2020}.

\begin{figure}
    \centering
    \includegraphics[width=\textwidth]{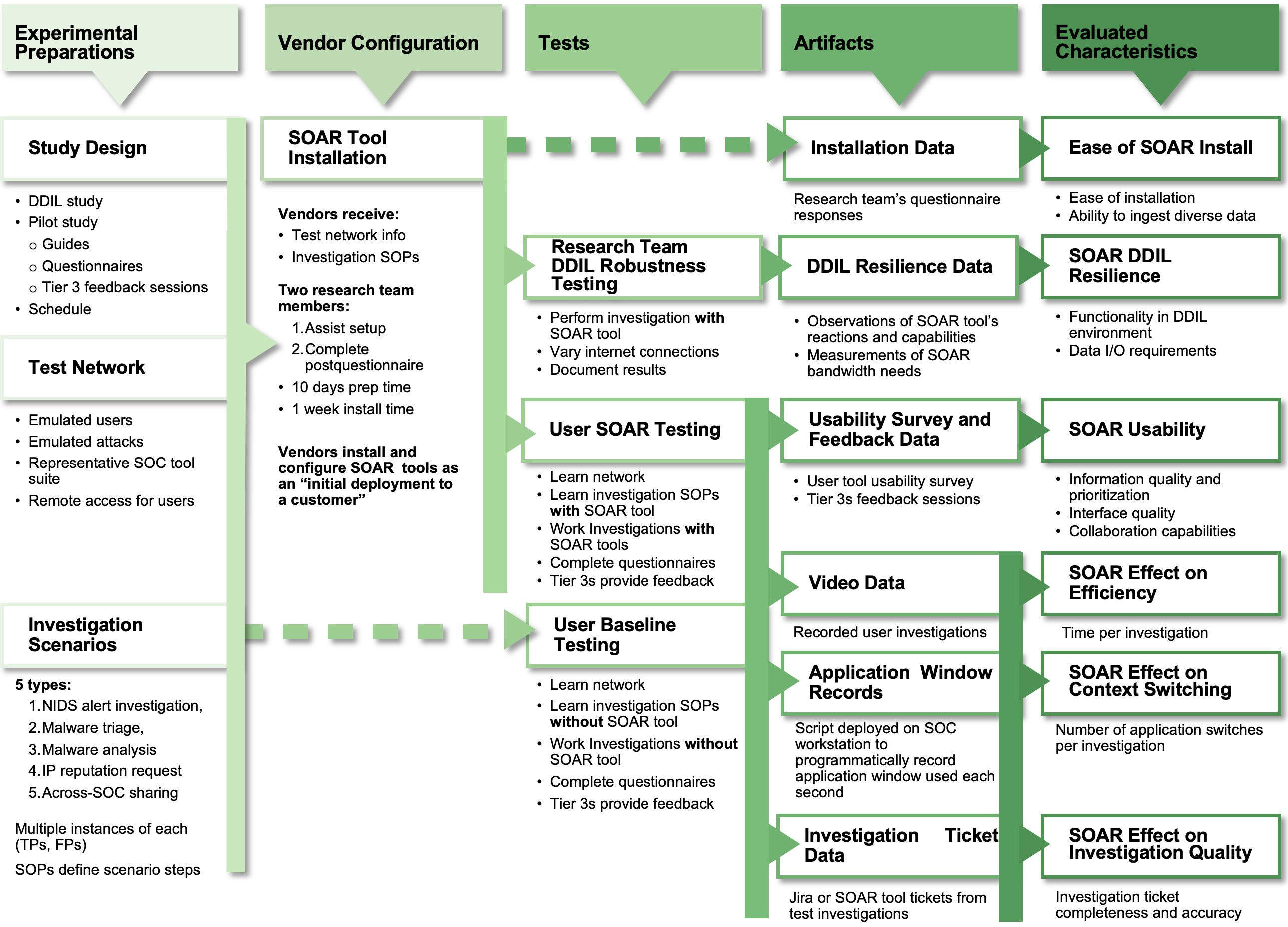}
    \caption{SOAR experiment workflow depicting (left to right): study design, test network, and investigation scenarios;
    %that were created to facilitate testing; 
    SOAR vendor installations and  tests administered, including  DDIL (denied, degraded, intermittent, or limited bandwidth) resilience testing by the research team and user study tests involving naval  SOC operators  who worked investigations in the test network with and without SOAR tools; 
    artifacts and data collected from install and tests; and
    SOAR capabilities under study.}
    \label{fig:infographic}
\end{figure}

\section{Experimental Design}
\label{sec:design}
As mentioned in the introduction, this user study was the second phase of the AI ATAC 3 Competition, in which the US Navy solicited SOAR tool submissions to be evaluated. 
All submissions must remain anonymous as a constraint of the competition. 
Those submissions that were deemed a SOAR tool, as defined below in \ref{sec:defining-aspect}, were considered eligible and underwent a first phase downselection process based on Navy SOC operators rating and ranking video overviews of the submissions---see Appendix \ref{sec:appendix-phase1} and our previous work \cite{norem2021mathematical}. 
After phase one, six SOAR tools were selected for evaluation in the user study described in this paper.
This section describes our design of the user study administered to these six SOAR tools
and  should provide a reference for understanding the experiment and results. 
It provides the SOAR tool definition in \ref{sec:defining-aspect},  methodological details in \ref{sec:target-capabilites} (to be referenced when observing the corresponding result Subsections \ref{sec:data-results}), participant information in  \ref{sec:recruitment}, and puts the pieces together into a workflow, depicted in Figure \ref{fig:infographic} and discussed in the final Subsection \ref{sec:workflow}. 
Notably, data analytics methodologies merit a section of their own to follow (Section \ref{sec:data-analysis-methods}).

    \subsection{Defining Aspects of SOAR Tools}
\label{sec:defining-aspect}
Our first task was to articulate the defining aspects of a SOAR tool to  delineate what is eligible for consideration and identify the capabilities of these tools that could be studied. 
Starting with the discussion of SOAR tools above, and after consulting the related work on SOAR (see Section~\ref{sec:related-works}), we developed the below-listed defining aspects of a SOAR tool, which comprised the necessary and sufficient components for eligibility as a SOAR tool in our study:
\begin{itemize}
\item \textbf{Ingest data:}
Ingest logs and alerts from a wide variety of security tools in a SOC;
ingest or provide threat intelligence information from both internal and external sources.
\item \textbf{Correlate and prioritize data:}
Combine, coordinate, and enrich logging, alert, and threat data in the tool’s UI;
identify common attack patterns and highlight them in the tool’s UI.
\item \textbf{Automate processes:}
Automate and ultimately simplify the alert triage and incident response processes via preset and configurable workflows or playbooks;
automate and expedite documentation of triage, incident response, and forensics (e.g., via ticketing); 
reduce context switching by automatically obtaining and presenting all the necessary information in a single interface. 
\item\textbf{Facilitate collaboration:} 
Facilitate investigation collaborations by different, potentially geographically disparate, operators; 
support both synchronous (i.e., working together) and asynchronous (i.e., sharing or handing off cases) teamwork. 
\end{itemize} 

What makes SOAR tools security-specific is that they natively accept security-related concepts in their data such as IPs, files, hosts, are geared towards organizing information around security-related concepts (e.g., timelines of information for a given host/IP, or prioritization based on risk from a threat intelligence feed) and are designed to assist security operators detect, triage, and remediate threats efficiently.

    \subsection{What Capabilities to Test / Data to Collect?}
\label{sec:target-capabilites} 
% Each of these components is designed to address specific and known problems within SOCs. 
Recent research works on SOCs theorize that SOAR tools or more generally orchestration and automation capabilities are needed to address specific and known problems within SOCs; especially, see Section 5 of Islam et al. \cite{islam_multi-vocal_nodate} and Nyre-Yu \cite{nyre2021identifying} who states ``Future work in this area could focus on ... an empirical study of how SOAR platforms minimize expertise gaps while also improving analyst effectiveness.''
More specifically, the ability to ingest data from a wide array of sources into a single platform and assist with correlation of data is critical to alleviate the currently tedious and distracting context switch between tools to obtain and correlate data, as the current practice leads to an incomplete picture of the security status of their network and slow incident response times \cite{islam_multi-vocal_nodate, zonouz2015, goodall2004}. 
Furthermore, as SOC analysts are often inundated with massive amounts of data
%much of it being false positive alerts
that require an investigation by a human operator, the ability to prioritize data such that their focus can be more directed towards true threats is of high importance \cite{islam_multi-vocal_nodate, forte2017}. Similarly, automation of repetitive and manual processes allows analysts to focus their efforts on true threats and also decreases response time along with operator error \cite{islam_multi-vocal_nodate, rapid7, forescout}. Finally, timely sharing of threat intelligence and collaboration with other operators on decision-making can be critical to successful remediation, and SOAR tools make this possible \cite{islam_multi-vocal_nodate, rapid7, greenfield}.

The capabilities we targeted were driven by these needs, the goals of a SOAR tool\textemdash to enhance efficiency and quality of investigations by providing a more usable interface that integrates multiple data systems and promotes collaboration---and by the research questions itemized in our introduction. 
Here we describe each target capability and the methodology used to produce data for testing them. 
To provide a mental map, the capabilities below can be partitioned into two groups---those tested only by the research team with no user study (Ease of Install/Configuration and Resilience to Degraded Internet) versus those that are tested by the participants in the user study (Usability, Investigation Time, Context Switching, and Investigation Quality). 
These last four capabilities produce quantitative data that requires preprocessing and analysis, which is described in Section \ref{sec:data-analysis-methods}.

\subsubsection{Ease of Install, Integration/Ingestion Abilities, and Configurability} 
\label{sec:install-test}
% Aside from purchasing fees, proper use requires ample 
SOAR tool installation requires integration with a SOC-specific tapestry of tools and data feeds, followed by configuration of dashboards and automation playbooks unique to the particular SOC's operating procedures. 
As evidenced from a recent study of a SOC that adopted a tool by Sundaramurthy et al. \cite{sundaramurthy2016turning}, improper or outdated configuration of promising SOC tools can quickly lead the organization to abandon the tool's use; hence, installation and configurability of the SOAR tool are worthwhile characteristics to study. 

Each vendor was assigned a 13-business-day period for familiarization, configuration, installation, and training development. 
The vendors were hands-on in the environment for only the final seven days of this period.  
At the beginning of the 13-day period, vendors received an information packet complete with descriptions of investigation scenarios and SOPs with which their tool would need to assist during evaluation. 
The packet also included information about the platforms in our test environment with which their tool would need to integrate.
The vendors appointed an installation team on their end who worked together for the first week on preparation of their tool. 
This involved designing and configuring playbooks or automation workflows custom to the SOPs and SOC tool suite. 
The following week, the vendor's installation team met with our assigned research team, who assisted them with configuring their tool in our environment and performing final checks. 
% At the end of the installation week, the vendor submitted training videos demonstrating each of the scenarios with their tools that would be used to familiarize the operators prior to performing the evaluation. 
% The training scenarios were similar, but not identical, to the evaluation scenarios.

This design served two purposes: (1) standardizing and limiting the length of time a vendor had to accomplish everything, thus informing us about ease of configuration; and (2) enabling us to focus our limited resources on each vendor in linear fashion, so they could each be satisfied they had received full attention and had been able to present their tools to the analysts as thoroughly as possible. 
Because the vendors did their own configuration, installation, and training, they were entirely responsible for the quality of the presentation of their tool. 
We included questions in the user surveys about the presentation and training quality. 
The research team who assisted each vendor during their installation week completed a survey (Appendix~\ref{installsurvey}) on the ease of installation and integration in our environment.
This information was used to evaluate the ease of installation for the tool.

From the vendors’ point of view, installation is a combination of technical and design challenges. Each SOAR tool has a different setup for display of data, playbook interaction, and automated-plus-manual ticket completion. The vendors needed to design an interactive environment in their SOAR tool platform to facilitate and semi-automate the SOPs for the investigations. The technical challenges are getting these functionalities to work correctly. Technical challenges involve getting unidirectional and sometimes bidirectional functionality with each system in the SOC toolsuite required for the investigation. Some examples include: automated query by the SOAR tool of the SIEM; submission of a file to ThreatGrid for analysis then receipt of results; submission / query of  threat intelligence websites such as NVD, CVE, or VirusTotal and receipt of results. Each of the integrations, when used in an investigation, would then possible need to trigger a next-step automation based on results and provide appropriate displays.

% To test this, we provided to vendors the environment's components to each vendor beforehand so that they could prepare their installations with extra plug-ins.      
% SOAR vendors were also given the Standard Operating Procedures (SOPs) defining the investigations used in our test and were asked to install their tool in our test environment. 
% 
% To evaluate ease of install, ability to ingest and integrate with diverse data sources/systems, and configurability, we assigned a two-person research team to interface with each SOAR vendor during installation in our test environment.
% The two install team members shepherded the vendor through the entire process, so they were qualified to make judgements about the tools.  
% Research team members completed a survey evaluating the quality of the installation and integration processes and highlighting any issues. 
% Appendix \ref{sec:vendor-install} provides the details for the vendor installation process and the evaluation survey. 

For analytical purposes, after all tools were configured a separate team member read the survey documents filled out by two install team members, discussed the tool installation and configuration process with the install team, and distilling results into three categories: data ingestion, ticketing/Jira integration, and playbook setups. 
Questions relating to data ingestion addressed how easily the tool could integrate with multiple data sources. 
Similarly, ease of integration with our ticketing system, Jira, was also evaluated. 
Questions related to playbook setup addressed playbook flexibility, vendor support, documentation, and ease and speed of deployment.  
In each of these categories, the raw data for each tool installation was distilled into a score of -1 (poor), 0 (neutral), or 1 (good)  performance, relative to the other tools.  
If there were plug-ins provided that interacted correctly with the data sources we provided, and this configuration was relatively fast compared to the other tools, then they were rated highly (1) in the data ingestion category. 
Conversely, those that had insufficient integration (e.g., inability to correctly interact with other needed systems) even after the week of iteration with the install team received a (-1). 
Similar relative rankings based on time, effort, and correct functionality with our ticketing tool Jira (for instance, some tools did not correctly populate custom fields of Jira) and with designing and implementing playbooks that could correctly enable the SOPs resulted in the -1,0,1 trichotomy ratings for the ticketing integration and playbook setup scores. 
% Whereas, tool installations that could not produce correct functionality---population of Jira custom fields, for instance, within the allotted week of iteration with the two-person research team---rated poorly (-1). 
% Each vendor had differences in how they expressed the programming for playbooks. 
% The install team was asked to judge playbook flexibility based on how easy or hard it would be to implement changes.

Overall configuration was defined as the average of the scores in each category (Table \ref{tab:config}).  
Results are reported in \ref{sec:install-results}. 
This targets the first research question about SOAR tools in the introduction.

\subsubsection{Resilience to a DDIL (Denied, Disrupted, Intermittent, or Limited bandwidth) environment.}
Sub-experiments itemized below refer to different investigation scenarios that are described in Section \ref{sec:scenarios}. See results in Section \ref{sec:ddil-results}
\label{sec:ddil-test}
\begin{itemize}
    \item \textbf{Denied Internet:} With the SOAR tool inside a fully functioning network, the internet connection was severed by removing the connection on the WAN side of  the network's border firewall. An Malware Investigation Scenario  was then executed within the network to start the SOAR tool's playbook for that scenario.  The tools were evaluated on how gracefully they handled lack of access to outside data sources and tools.
    \item \textbf{Disrupted / Intermittent Internet:} Network access to the SOAR tool was disabled while malware was injected into the environment, initiating a Malware Triage Scenario ,  and then network access was restored. The evaluation looked at how well the SOAR tool could recover events that occur during and after impairment. 
    \item \textbf{Limited Bandwidth, Steady-State:} We simply took measurements to quantify the  ``steady-state'' bandwidth used by the SOAR tool during normal operations.  
    We report the tools' \texttt{mean} $\pm$ \texttt{standard deviation}   load on a network in megabytes (MB); i.e., the amount of bandwidth it used to monitor events.
    \item \textbf{Limited Bandwidth, Working Test:} Similarly, we measured the amount of data consumed while working an NIDS Scenario with the SOAR tool. Then, we quantified ``working'' bandwidth as the minimum bandwidth required to completely work a given scenario. 
\end{itemize} 
These experiments and results quantify second research question about SOAR tools in the introduction.

\subsubsection{Usability, Including Information Quality, Interface Quality, Collaboration Capabilities, and Other Features} 
\label{sec:usability-test}
During the user study, Likert scale surveys, pertaining to the overall usability of the tools, with questions specifically targeting information quality, interface quality, and collaboration capacity, among other features of the tools, were administered. 
The questionnaire appears in Appendix \ref{sec:appendix-likert}. 
For each tool an analyst tested, we also conducted a semistructured interview after testing with the participant to better understand their experience with and perspectives on the tool. 
See  Appendix~\ref{semistructured} for the interview guide.

Senior or ``Tier 3'' participants were given access to all tools and encouraged to conduct a more open-ended  exploration to gain detailed understanding of the tools' functionalities. 
These senior analysts completed the questionnaires and participated in a roundtable discussion led by the research team to explain their experiences and preferences with regard to each tool. 

We use this data in two methods of analysis: 
\begin{itemize} 
\item \textbf{Usability Qualitative Analysis:} 
A research team member coded the various free response data to better understand the key themes that emerged with positive, negative, or neutral sentiment. 
Section \ref{sec:qualitiative-results} presents the coded results. 

\item \textbf{Usability Quantitative Analysis:} The textual, free-response data recorded from the each user on each tool was mapped to a 1--5 scale with a sentiment analysis algorithm. This now-numeric data was included alongside the other quantitative data (introduced below).
Details on the sentiment analysis algorithm, and more generally pre- and post-processing all quantitative data collected data into results is the topic of Section \ref{sec:data-and-preprocessing}. 
\end{itemize}
These results target the third and last research questions in the introduction.

\subsubsection{Investigation Completion Time} 
Time duration of an investigation is a quantifiable metric related directly to efficiency. 
As such, time to complete an investigation was recorded programmatically for each scenario using a background script  started once an investigation was initiated and terminated once the investigation was complete. 
This script recorded the time in seconds of each investigation and provided a helpful metric for evaluating the ability of the tool to improve response time when compared to a baseline environment (with no SOAR tool), as well as how the SOAR tool compared to others being evaluated. 
This targets the fourth research question in the introduction. 

\subsubsection{Window Switching Counts}
By monitoring the number of times an operator has to switch between windows during an investigation, one can obtain a quantity, namely a lower bound, on the number of context switches the operator underwent in performing an investigation. 
For our experiment, the number of window switches in each investigation was recorded using the same background script that measured time spent on an investigation.
% A benefit of SOAR tools is that they promote a centralized platform for an operator to obtain all the necessary information to conduct an investigation instead of having to navigate several different screens to obtain the same information from separate interfaces. 
This targets the fifth research question in the introduction.  

\subsubsection{Ticket Completeness \& Accuracy}
\label{sec:ticket-test}
Ticket completeness and accuracy provide a window into the quality of the investigation. 
Because users followed an SOP document for each investigation, we were able to create a rubric for the ``gold standard'' ticket of that investigation based on our SOP. % that were provided to the vendors during at the beginning of their installation period. 
The rubric included fields that were based on vendor-submitted SOPs, wherein they detailed any additional information for their tool that should be included in the ticket. All tickets generated by each tool were aggregated and assessed for accuracy and completeness according to the rubric for that scenario. 
%A score was assigned to each ticket based on the fraction of correctly completed fields. The scores were normalized by the average score across scenarios to account for varying levels of difficulty of scenarios that each tool completed. 

    \subsection{Recruitment, Ethics, \& Participant Demographics}
\label{sec:recruitment}
As we conducted this study to evaluate SOAR tools for our Navy sponsor, we targeted as many Naval SOC operators to participate as possible. 
We invited analysts currently employed in five of the sponsor's SOCs to participate, and 21 Navy analysts from four Naval SOCs participated. 
In total we had 24 participants (Table \ref{tab:demographics}) in the task-based evaluations, with 21 Navy analysts and three ORNL personnel who also assisted on the research team. 
All three ORNL participants either had experience as a SOC analyst or experience studying SOC analysts. 

We administered a demographic  survey to participants, which appears in the Appendix \ref{demographicsurvey}. 
\begin{wraptable}[7]{r}{0.65\textwidth}
\vspace{-.25cm}
\caption{Demographics of participants: job role \& SOAR familiarity}
\label{tab:demographics}
\begin{tabular} {ll}
    \begin{tabular}{lr}
    \toprule
    Job role                   &  Count \\
    \midrule
    Network operator &     3 \\
    Security operator &    11 \\
    Other &     10 \\ %% 9 found in demographics surveys, evidently one participant didn't respond. My code reports we have missing demographics from users [2, 10, 20]
    \bottomrule 
    \end{tabular}
&
    \begin{tabular}{lr}
    \toprule
    SOAR familiarity &  Count \\
    \midrule
           Expert user &      1 \\
            Never used &      9 \\
        Somewhat familiar &     12 \\
    \bottomrule
    \end{tabular}
\end{tabular}
\end{wraptable} 
Two participants did not fill out the demographic survey. 
The remaining participants had a mean $\pm$ one standard deviation of $5.14 \pm 5.82$ years of experience, with a minimum of five months and a maximum of 21 years of experience. 

%\subsubsection{IRB Approval}
This user study received Institutional Review Board (IRB) approval\footnote{Oak Ridge Sitewide IRB00000547; submission ID  ORAU000785; title \textit{Evaluations of Tools that Enhance Security Operations Center Effectiveness} }
and was determined to be ``Exempt human subjects research''. 
Written consent was not required by participants but did require informed consent indicated by their level of participation.
I.e., we provided participants with an information sheet detailing the nature of the research and their rights prior to their agreement to join the study. 
Participants could opt out of any/all parts of the experiment.

% \begin{wrapfigure}{r}{0.4\textwidth}
%     \centering
%     \includegraphics[width=.48\textwidth]{images/years-in-job.png}
%     \caption{}
%     \label{fig:jobyears}
% \end{wrapfigure}

 % IRB stuff and data on demographics of participant
    \subsection{Study Workflow} 
\label{sec:workflow} 
To accommodate a user study of the SOAR capabilities/characteristics above, we designed our test framework to encompass the entire process of acquiring and utilizing a SOAR tool, from vendor installation into our test environment, to SOC participants working in realistic scenarios with the tool.
SOAR vendors were asked to treat installation of their tool into the test environment as a fresh install. They were given all information needed for integrating with SOC tools and data feeds, as well as the SOPs describing investigation steps so they could create automated playbooks. The vendors worked with the research team to install the tool, and the research team documented the process as described in Section \ref{sec:install-test}. 
Once the SOAR tool was established in the test environment, it was tested by our research team for DDIL resilience as described in Section \ref{sec:ddil-test}. 
Separately, a user study was administered with  SOC operators and a few team members, which corresponds with evaluations described in Sections \ref{sec:usability-test}--\ref{sec:ticket-test}. 

The workflow for a participant was to remotely log into a SOC workstation in our environment and share screen with two research team members who administered the test. 
At the initial log-in, a hands-on tutorial was performed by the participant to learn the tools in the environment (e.g., how to search in the SIEM, use the malware sandbox tool)  and to learn how to do a handful of investigation  scenarios. 
Section \ref{sec:scenarios} provides descriptions of the investigation scenario types.  
Similarly, during a test with a SOAR tool, a vendor-provided tutorial on how to use the SOAR tool was completed before testing. 
SOPs, which served as step-by-step guides for each type of investigation scenario, were provided by the research team and followed by the participants. 
In addition, the SOPs served as a ``blueprint'' for the SOAR vendors to design playbooks and automated workflows. 
Each participant completed multiple scenarios for each tool they tested. These scenarios consisted of two NIDS scenarios, two Malware Triage scenarios, a Malware Investigation scenario, an IP Report scenario, and an Across-SOC Sharing scenario. 
During the investigations, participants shared and ran a script that recorded their active window each second (allowing completion time and window-swapping counts to be computed). 
All investigations require documentation in a ticket or report, allowing a data source to be collected for quality and accuracy of the conclusions. 
At stages throughout the test, we administered usability questionnaires (Appendix~\ref{usabilityandranking}) and conducted a semistructured interview (Appendix~\ref{semistructured}) to better understand the participant's experience of the tool, in addition to recording their usage during testing. 

To finish the workflow the collected data is analyzed into results. 
Data collected from the experiment is described and motivated in Section \ref{sec:target-capabilites}  along with methodology for obtaining results from those data types not requiring extensive processing and analysis descriptions. 
Methodology for quantitative data requiring substantial preprocessing and/or statistical analysis is in Section \ref{sec:data-analysis-methods}.

% We also measured task completion time, context switching required, and ticket completeness for each of the four scenarios as quantitative measures of tool efficiency and investigation completeness. These four scenarios were carefully constructed to evaluate the following properties: automation and playbooks, ticketing, prioritization of alerts, facilitation of communication, usability and unity of the user interface, and alert enrichment and contextualization. 

% It took approximately 1-6 hours (skewed towards 6)  per participant, per tool to complete both training for that tool and  all four investigation scenarios using the tool. 
% To accompany the scenario testing, we included a Standard Operating Procedure (SOP) document provided by the vendor. 

\section{Quantitative Data Preprocessing \& Analysis Methodologies}
\label{sec:data-and-preprocessing}
\label{sec:data-analysis-methods}

% For qualitative data collected (i.e., verbal/textual feedback from the participants to open response questions) analysis can also be dichotomized into qualitative analysis (coding the data for themes) versus quantitative analysis (performing statistics on numerical representations derived from sentiment analysis). 
% \todo{do we want to make a tree or colored table to make this visualizable?) }

To enable statistical analysis, the quantitative data was parsed from various raw formats into three tables (not presented but discussed): 
\begin{itemize} 
\item The \textit{user table} provides columns for each participant's unique ID number and their responses to the demographic survey. 
Section \ref{sec:appendix-demographic-survey} of the appendix lists demographics questionnaire items, and Section \ref{sec:recruitment} gives an overview of the participants' demographics. 
\item Similarly, the \textit{tool table} provides fields for the tool's index number and their scores for the four configuration categories: data integration, ticketing/Jira integration, playbook setup, and overall configuration (an average of previous three). 
Section \ref{sec:appendix-install} of the appendix lists installation questionnaire items completed by our research team upon SOAR install teams, and summary results regarding installation and configuration are given in Section \ref{sec:install-results}. 
\item Finally, the \textit{results table} houses the quantitative testing data (with collection described in Sections \ref{sec:usability-test}-\ref{sec:ticket-test}).  
Each row is indexed by a user--tool test. It has   75 columns for the measurement data (collected during that user's test of that tool). 
\end{itemize}
This section details the \textit{results table} data preprocessing and analysis methodologies. 

\subsection{Measurement Categories and Preprocessing}
\label{measurement-categories}
The 75 measurements used in this evaluation are broken naturally into five categories and, in turn, multiple subcategories; preprocessing varies per category. The five categories are: 
\begin{itemize}
    \item \textbf{Likert ([1--5]) responses data} to the 32 ``usability'' survey questions was normalized to prevent user bias by subtracting the  user's mean (i.e., the mean of all Likert responses for all tools by that user), as some users are overall more critical/charitable in their responses. All data was scaled and translated into the interval [0,1]. As shown in Section \ref{sec:appendix-likert} of the appendix, these questions naturally fall into seven subcategories. We considered averages of responses to questions pertaining to the following subcategories: training quality, ability to automate recurring (``1,000 times per day'') tasks, collaboration capabilities, ability to handle advanced persistent threats (APTs),    usability, information quality, and interface quality. 
    \item \textbf{Free response sentiment values} were gathered from the semistructured interviews and comprise 20 columns (20 different response values) in total. We applied sentiment analysis using TweetEval\footnote{Code for this sentiment analyzer is available: \url{https://huggingface.co/cardiffnlp/twitter-roberta-base-sentiment}.},  a roBERTa-based sentiment analyzer \cite{barbieri2020tweeteval}, mapping the raw text to a value in the interval [-1, 1], indicating negative (-1) to neutral (0) to positive (1) sentiment. As before, we normalized the data by subtracting the users' mean to account for user bias and then mapped the values linearly to [0,1]. As shown in Section \ref{sec:appendix-sentiment} of the appendix, these questions naturally fall into five subcategories. We will consider averages of responses to questions pertaining to the following subcategories:  ability to automate recurring tasks, collaboration capabilities, information quality, attack scenarios, information quality, and a general category comprised of questions about the tool. 
    \item \textbf{Ticket completion ratios} were collected by scoring the tickets produced by investigations for accuracy and completeness based on a rubric we created from the SOP for each attack scenario. In total, the ticketing data accounts for seven columns. Because scenarios vary widely in difficulty, ticket completion scores are not comparable across scenarios. Consequently,  we normalize per scenario by subtracting the scenario's mean ticket completeness across all users and then linearly map back to [0,1]. (One may wish to normalize by each user's mean per scenario, but, as many users only worked tests with one to a few tools, this was not viable.) 
    We considered subcategories averages (for each user and tool, we average the responses for each subcategory) of each scenario type. 
    \item \textbf{Window swapping counts} were parsed from a custom Python script that programmatically recorded the user's active window during each investigation. In total, eight columns were attributed to window-swapping data. Similar to the ticket completion scores, the window swap counts varied based on scenario, and we normalized by subtracting the  per-scenario means. The normalized counts were scaled and shifted to [0,1] and then inverted via $x\mapsto (1-x)$, maintaining the convention that larger scores indicate better performance. 
    \item \textbf{Investigation completion time}, as with window swapping counts, was computed from the Python script, normalized per scenario,  mapped linearly to [0,1], and inverted via $x\mapsto (1-x)$. The same eight columns were reported for the investigation time data. We considered subcategories averages for results of each scenario type. 
\end{itemize}
\begin{wrapfigure}[29]{r}{0.5\textwidth}
    \vspace{-.1cm}
    \includegraphics[width = \linewidth]{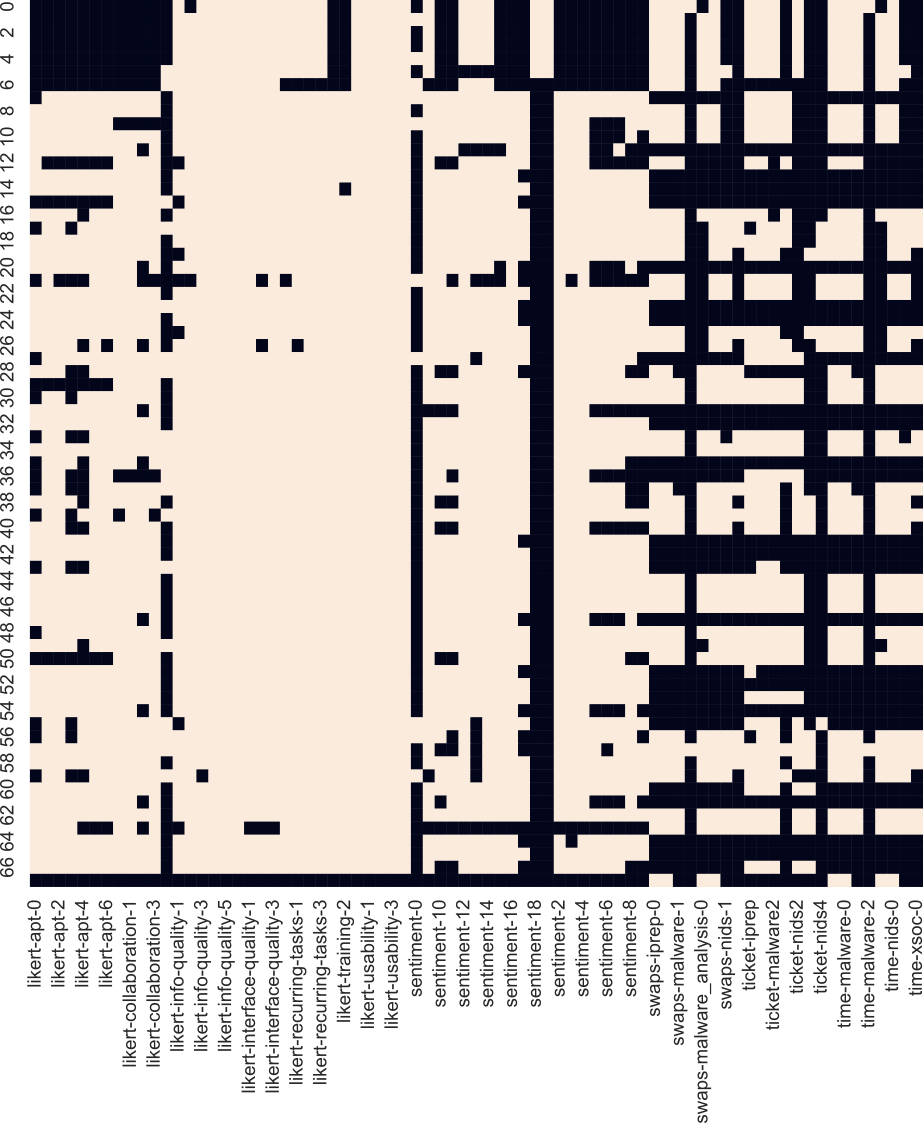}
    \caption{Binary heatmap of normalized data table (68 rows for tests, each indexed by user $\times$ tool, by 75 measurements, not all measurements applicable to each test) shows distribution of missing data. Top six rows are baseline tests for which many questions are not applicable; missing data on right half of matrix is attributed to different scenario versions being worked per test and a few operators working investigations with tools for which monitoring scripts for data collection were not possible.} 
    \label{fig:where-is-data}
\end{wrapfigure}

Once processed, we had a results table with 68 rows (user--tool tests), and 75 measurements columns. 
All  measurement values reside in [0,1], with smaller/larger values indicating worse/better performance in that category. 
Each column attains the minimum (0) and maximum (1). 
There are 3,517 filled of 5,100 total cells in this table (69\%). 
Missing data can be explained by the fact that 
many Likert and free response category questions were not applicable to baseline (i.e., no SOAR tool) tests; each test involved different, specific scenarios leading to different filled/missing columns of ticket completion, window swap counts, and time duration; three Tier 3 (i.e., senior) analysts worked investigations with every tool, but (because of logistical constraints) duration and windows-swapping data was not collected for these tests; and
some participants neglected to answer some questions. 
See Figure \ref{fig:where-is-data}. 

% Broadly, our data falls into five overarching categories: Likert survey responses, free response survey responses, ticket completion fraction, window swap counts, and investigation time. %), and each, in turn, natural subcategories. 
% The latter three categories have subcategories per scenario type, and see Appendix Sec. \ref{sec:appendix-likert} \& \ref{sec:appendix-sentiment} for Likert and Free Response subcategories. 
% Consequently, we often coarsen our view from the normalized data to the per-test (per-user, tool) averages per category ($68\times 5$) or subcategory ($68\times 25$), which naturally eliminates many missing cells. 

\subsection{Public Data} 
\label{sec:public-data}
To make this dataset  public, we combined the normalized results table with the user demographic table (see Section \ref{sec:appendix-demographic-survey} of the appendix for descriptions of the questions that led to the data in this table), which adds four columns providing info on the user in the test to the 75 columns of user-tool tests, allowing investigations of correlations across user demographics and testing results. 
To protect the users' and vendors' privacy, we excluded user and tool identifiers and coarsened the added demographic columns to binary values and then applied a bit of noise to all of the data. This  user- and tool-anonymized and -obfuscated table has been made public  (link on first page in the title notes),
with greater explanation, and we have provided code to reproduce some of the correlation results, showing that data utility was not lost in the obfuscation process.

\subsection{Processing into Subcategories}
For each row in the normalized results data table, which corresponds to a 75-length vector of user--tool test results, we shortened the vector by averaging those columns from the same subcategory. 
For example, multiple questions about how well the tool assists recurring SOC tasks are given to each user in each tool test, allowing us in this analysis to average these few questions' response scores to get that users' ``likert-recurring-task-ave'' score. 
Similarly, as two different NIDS and Malware Triage scenarios were completed in each user-tool test, we averaged the corresponding measurements per scenario (giving a NIDS scenario average and a Malware triage scenario average for each measurement type). 
This method allowed for aggregating multiple measurements in a single category, reduced the number of features, and improved the ratio of missing data (e.g., if a user declined to answer one question, we could still compute an average from their other responses in that question's subcategory).

\subsection{Handling Missing Data with Multiple Imputation} Direct computation of correlation of two vectors with missing data (i.e., by simply ignoring components with missing values) overweights the contributions of the present components. 
Consequently, we employed Multiple Imputation (MI), a standard technique for statistical analysis with missing data \cite{schafer1999multiple, austin2021missing}, and compared the MI results to the naive statistics computed with missing values when drawing conclusions. 
MI estimates  the desired statistic (in our case, correlation) by averaging  samples from a Monte Carlo simulation of the missing data. 
Given two vectors $x, y$ (in our case, columns of a results table) with missing data we impute (i.e., simulate) the missing components by sampling each $m=1,000$ times. 
Furnished with now-complete, partially imputed versions of our data $\{(x^i, y^i\}_{i = 1}^m$,  the average correlation $(1/m)\sum_{i = 1}^m\text{corr}(x^i, y^i)$ was used, and we leveraged the imputed sample statistics to quantify confidence of findings with hypothesis tests and confidence intervals. 
We used $m=1,000$ by investigating convergence empirically---plotting and seeking negligible  average $\ell_1$ change of the average correlation upon adding more samples. See Appendix \ref{sec:appendix-mi} for details.

This method requires a distribution from which to sample the imputed values. 
% For each missing cell of a data table, we require a probability distribution on $[0,1]$ from which we can sample imputed values. 
We analyzed the effects and considered results from two imputation sampling approaches: random and kernel density estimate (KDE) sampling  (ultimately we used the latter).  
Recall the fact that all measurements were normalized and thus lie in $[0,1]$. 
The first approach involved sampling uniformly at random  ($p(x) = 1 $ for all $x\in[0,1]$), which quantifies results from the standpoint that all possible values of a missing data point are equally likely. 
The second sampling method for a missing target value leveraged all data for that user--tool test (row)---the user demographics data, tool configuration data, and all result columns except the target column---to estimate a probability distribution that is more precise.
A decision tree was trained to predict the target column from all other columns (using the known data), and KDEs were used to produce a probability distribution from the training samples residing in each leaf of the tree. 
\begin{wrapfigure}[13]{r}{.35\textwidth}
    \vspace{-.2cm}
    \centering
    \includegraphics[width = \linewidth]{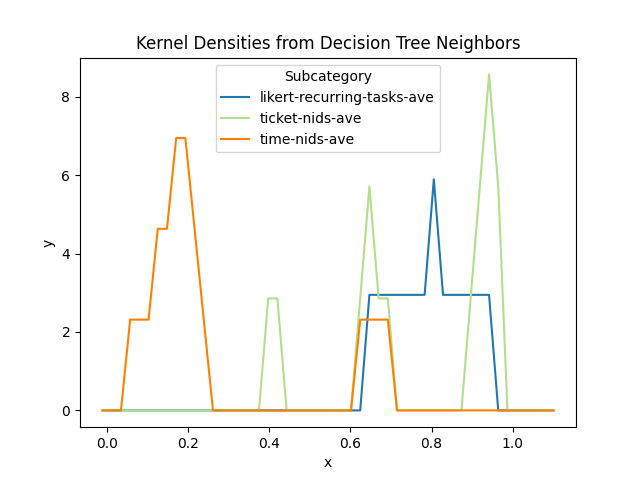}
    \caption{Kernel densities estimated  from leaf nodes of a decision tree regressor, depicted for three subcategories. Sampling from these estimations is used for imputation of missing data.}
    \label{fig:kdes}
\end{wrapfigure}
Imputed values were then sampled from the KDE corresponding to its leaf node, that is, from the KDE informed only by  the known data that had been learned to best predict the value of the target column for those with similar other columns. 
See Figure \ref{fig:kdes}.

Full details of the imputation experiments and approach are provided in Appendix \ref{sec:appendix-mi}, in particular comparison of the subcategory averages under both the uniform random (not used in this paper's results) and the KDE sampling (used in this paper's results) techniques.

\subsection{Analysis Methodologies} 
Each subsubsection here describes a data analysis approach and corresponds to a results subsubsection. 

\subsubsection{Analysis Method: Averages per Category and Tool} 
Our first quantitative results are simple averages for each tool across categories and subcategories. 
Results lie in Section \ref{sec:averages}.

\subsubsection{Analysis Method: Correlations of Subcategories (Across Tools)}
\label{sec:correlation-approach} 
Rather than looking at data per tool as in Figure \ref{fig:bigfig},  we next explored correlations of subcategories across all tools. 
This allows us to identify relationships and trends in the results of the experiment across all SOAR tools. 
%, and we coarsen to the ``subcategorical'' level by averaging similar measurements in each user-tool test.
Results lie in Section \ref{correlation-results}.

\subsubsection{Analysis Method: Correlations of User \& Configuration Data with Quantitative Results}
\label{sec:corr-user-data-results-methodology}
In Section \ref{sec:correlation-test-user-data-results} we present results of  correlations among user demographics data and tool configuration data, with the quantitative test results data. This allows investigation of how well/poorly operators of different backgrounds and experience levels, performed with SOAR tools. 
Also, it allows investigation of how the installation of the SOAR tool trends with quantitative measures of performance. 
These target the first and fourth research questions in the introduction. 

\subsubsection{Analysis Method: Regression Analysis} 
We use a threshold on the install and configuration results to dichotomize all tools as ``well-configured'' SOAR tools, ``poorly configured'' SOAR tools.%, and our no-SOAR-tool baseline environment. 
For each we regress: (1) the window swap counts (a quantification representing context switching) onto our investigation time metric (a quantification of efficiency) to see if an investigation with a fixed number of window swaps can be performed faster/slower with/without SOAR tools  and (2) the ticket completion ratio (a metric measuring investigation quality)  onto operator experience to investigate the question ``Can junior analysts perform investigations with the quality of a senior analyst when using a SOAR tool?''. 
See Section \ref{sec:results-improved-efficiency} for results. 

 %overview of all measurement data, preprocessing, normalization, missing data

\begin{figure}
    \vspace{-.2cm}
    \centering
    \includegraphics[scale = 0.55]{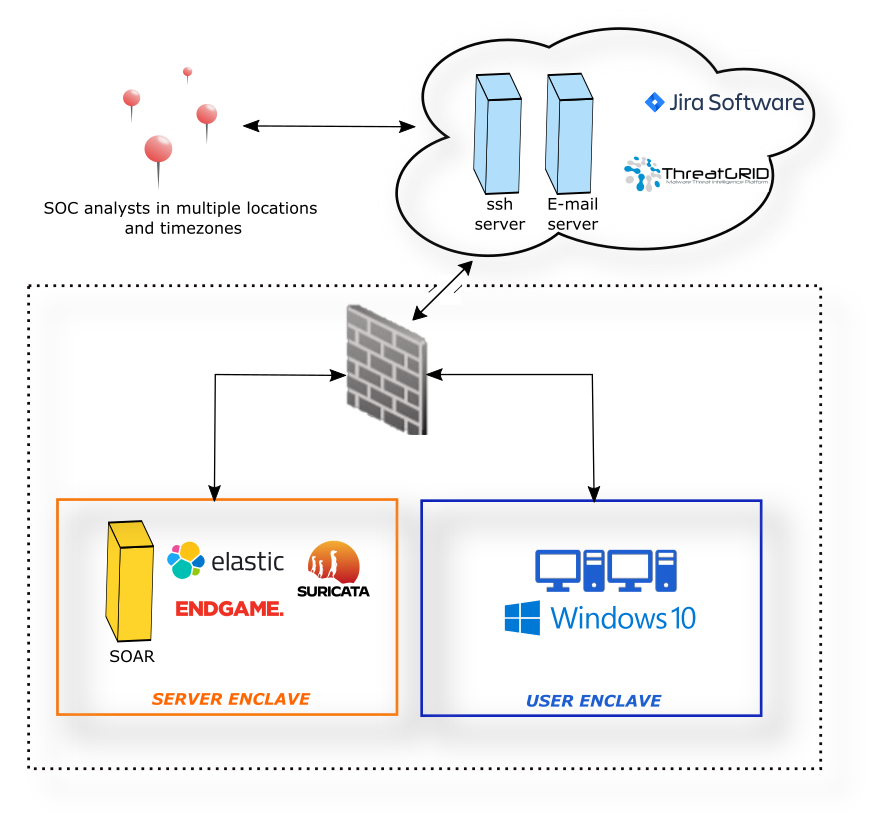}
    \caption{Diagram of the SOAR experiment environment depicted.}
    \label{fig:network-diagram}
    \vspace{-.35cm}
\end{figure}

\section{Building the Test Environment}
\label{sec:test-network}
To perform the tests, a test environment was designed to meet many constraints and challenges.  
Overall, the test environment must provide many SOC workstations and allow many representative security investigations to be completed with and without a SOAR tool while measurements are taken. 
To this end, we met repeatedly with three different SOCs to understand their sensor-/toolsuites, common investigations, methods for collaboration, etc. in order to design the network environment with sufficient generality and realism. 
Here, we itemize these challenges and our approach. 
Since the majority of the participants are US Navy analysts residing in diverse geographic locations, remote access from a secure SOC into the environment is a constraint for our application. 

% To successfully carry out the experiment with this design and scope, we had to create solutions for many small challenges.  Here, we describe the challenges and how each were addressed.

\subsection{Establish a  Uniform Baseline Environment for Operator Testing and SOAR Installations that Is Reproducible, Easily Learnt, and Representative}
To accommodate a  user study of SOC analysts, we established a baseline environment with a representative SOC tool suite. 
Our goal was  to establish a SOC work environment in which the analysts could, with minimal training, become equally familiar (in comparison to each other) prior to engaging with the SOAR tools so that their focus could be on the investigation.
% , and a guide for providing training to the operators. 
% This provides familiarity with the baseline environment prior to engaging with the SOAR tools so that focus could be mostly on the investigation. 
% located around the globe completing investigations ``hands-on'' with and without the SOAR tools, 
% Because of limitations on analyst availability, we needed to ensure the focus could be solely on tool-testing rather than navigating an unfamiliar network environment and procedures.  

To solve this challenge, we built a test network that simulated a small enterprise composed of approximately 50 virtual machines complete with servers, workstations, and networking with a full security software stack. See Figure \ref{fig:network-diagram} and \ref{fig:network-diagram2}. 
The IP addresses for each of the core components were fixed, which enabled consistency across tools and assisted the analysts in developing ``muscle memory'' in the environment.
Participants learned the environment by following a detailed tutorial of the baseline environment prepared by the research team. Multiple instances of the baseline are needed to facilitate concurrent tests and ensure the same starting conditions for each SOAR tool. 
Once established, this baseline environment was duplicated onto separate nodes in our environment such that each vendor's tool installation began from a consistent state.  

The baseline environment consisted of a Windows Server 2019--based Active Directory domain with a De-Militarized Zone (DMZ), internal enclaves, and cloud-provided services.  We used Elasticsearch with a Kibana UI as the SIEM, Suricata as the NIDS tool, and Endgame as the Endpoint Detection and Response (EDR) tool. 
In the cloud, we used Jira instance for ticketing, 
ThreatGrid for malware investigations, and a dedicated email system with an associated, registered domain. 
Several Linux systems provided common server functions including database, web, and file transfer, as well as entry points for compromising the environment in detectable ways, which was needed to produce the scenarios. 
We joined to the domain Windows 10 desktops as user workstations and SOC operator workstations. Agents for endpoint management and protection were installed on the workstations.  The user workstations were variously used as targets for simulated attacks.  Full network vulnerability scans and their associated logs were made by Nessus on 30-minute intervals.

Rather than network traffic, SOAR tools act on events forwarded by the monitored cybersecurity tools. Originally we had designed to simulate user and network activity on the workstations, because we wanted the generated security events of interest to be potentially hidden by other, benign events.  With so many network security tools set producing alerts and monitoring logs, even with this relatively small network there were more than enough benign network security events for monitoring by the SOC operators.  
All logs and events were consolidated by the SIEM at approximately 100K recorded events per hour.

\subsection{Train the Analysts to Perform Scenarios Using Unfamiliar Tools and Procedures}
The same training materials developed for the analysts were provided to a vendor along with a set of slides describing the expectations for the vendor.  In addition, the vendors were provided with the original training documents, so they could be modified into analyst training for their SOAR tool.  The vendors were responsible for the training materials and SOPs provided to the analysts for their own tools.  This method served as a proxy for training provided by a vendor.

\subsection{Enable Remote, Secure Connections to the Test Environment for Remote Analysts}
For our particular sponsor, we needed to enable simple, fast, secure, and flexible remote access to our experimental environment for both vendors and SOC analysts. 
Analysts participated, for example, across 18 hours of time zones  with various bandwidth, times, and locations affecting their Internet access.  
We also needed to prevent unfettered access to the environments to ensure the integrity of our experiment.  
Additionally, each time the analysts participated, they needed to connect to a new test environment. 

% The solution we ultimately settled was elegant and satisfied these requirements.  
As analysts were onboarded, a single account, long passphrase, and unique port number were assigned to each of them.  
They used this single set of credentials to establish an SSH tunnel to a single, pass-through SSH server located in the cloud, which was the only system with access into our firewalls.  
We controlled when the accounts were enabled and to which  SOC workstation destination the tunnels were connected. 
The analyst was unaware of the complexity and would simply establish the tunnel and connect their RDP client to a consistent port on their localhost.

\subsection{Testing Multiple Analysts at the Same Time on the Same Tool}
Logistics around SOC operator availability meant that we need to host multiple operators testing different tools simultaneously. The assigned user accounts were enabled on the Windows Active Directory in each environment with the same passwords, so the individual analysts used their user account in each vendor's environment, which greatly simplified collecting and recording results.  
With multiple, identical SOC workstations in each environment, we could test multiple analysts at the same time. 
%Because of  the scale of this project, we needed a method in place to be able to rapidly collect data from user tests, leading to our team 

\subsection{Sequentially Evaluate Vendors While Preventing the Later Vendors from Gaining Advantage}

There was a risk that tools scheduled to be tested later in the experiment would experience an advantage due to increased familiarity of the research team and of the participants with the functionality of the network environment and the testing process. 
To help mitigate these effects, we randomly assigned a pair of research team members to one or two tools such that our internal test leads for each tool were all roughly at the same level of experience. 
In addition, we also limited the exact, repetitive investigations that operators performed across the tools to prevent memorization from test to test. 

\subsection{Design and Build Investigation Scenarios}
 \label{sec:scenarios}
To test a SOAR tool, representative investigations needed to be possible for the analysts to work. Our initial efforts including meeting with many SOC operators to learn about scenarios that were often repeated, as this informed us of the SOC operators' target actions that SOAR tools could assist. 
Ultimately, the goal was to create multiple types of scenarios, each with corresponding Standard Operating Procedures (SOPs)--step-by-step guides to complete and document each investigation---to be performed by participants.
Further, we required a method of reproducing the scenarios (e.g., by curating all the necessary artifacts in the alert and monitoring logs to allow a participant to perform the investigation), but with slight variations. 
It was important to produce similar but not identical versions of each scenario so, as the same participant tested different tools, they were not working identical events and simply remembering details. 
While the details of the specific scenarios may differ, analysts all still followed the same SOP to guide their investigation. 
Care was taken during the analysis of the collected data to normalize by any discrepancies in difficulty between the scenarios. 
As an example, for quantitative results such as time to completion, we normalize for each scenario type (subtracting the minimum, then dividing by maximum minus minimum, which maps the data into the interval [0,1]). This allows comparability of the quantitative data from two different types of scenarios that would have otherwise incomparable measurements.  Details are in Section \ref{measurement-categories}.

Five types of scenarios were completed by participants. They are listed below, with more detail provided in the subsections. 
\begin{itemize}
    \item \textbf{Network Intrusion Detection System  (NIDs) Investigation:} Operators were tasked with determining the nature of a detected intrusion event and identifying hosts on which it was present. Two NIDS scenarios were completed in each test---one with a CVE reference and one without. 
    \item \textbf{Malware Triage:} Operators performed this scenario twice with two malware samples. For each sample, they were instructed to obtain the VirusTotal score and record a decision of benign, suspicious, or malicious for the samples. Of the two samples, one was benign, and the other was determined to be either suspicious or malicious.
    \item \textbf{Malware Investigation:} The nature of the malware sample that was determined to be suspicious or malicious was further investigated. 
    \item \textbf{IP Reporting:} Given a time range and a description of suspicious activity for an internal IP address, analysts generated a report after analyzing packet capture (PCAP) files. 
    \item \textbf{Across-SOC Sharing Add-on:} After completion of a scenario, the report was shared using the SOAR tool's sharing mechanisms. 
\end{itemize}

\subsubsection{NIDS Scenario}
The NIDS scenarios were the first type of investigations designed to be representative of common tasks for a Tier 1 (i.e., junior) operator. 
For these scenarios, the analyst investigated an alert generated by Suricata, the NIDS component of the test environment.  The analyst had to understand the given NIDS alert and determine if it was a true or false indicator of compromise on the network.  If a true positive, the list of affected machines was determined using the SIEM to gather evidence.  Some of the NIDS tests contained information such as common vulnerability and exposure (CVE) numbers that could be easily enriched, while others were more vague.  With these types of scenarios, the SOAR tools could assist the analysts by enriching the content of the alert and performing common data-gathering tasks.  In particular, these scenarios exercised the SOAR tools' ability to interact with the SIEM and to follow complex playbooks. Unassisted, these scenarios took the analysts about an hour to complete.

All of our scenarios had to be real so the SOAR tools could apply their resources to enrich the events for the analyst.  To develop the NIDS scenarios, we surveyed recent CVEs, selecting 20 that we could implement in our baseline environment.  Often this involved adding misconfigured servers or software.  Once created, inside the baseline test environment, we triggered the compromise and worked out the complete scenario through all cybersecurity systems.  These were then turned into events that were triggered for the analysts to investigate.

To generate an alert initiating an NIDS scenarios, we hosted vulnerable services on the local environment. Suricata was then configured to monitor network traffic interacting with those vulnerable services. Alerts were generated either by exploiting the services, which we deemed as a true positive, or by interacting with the services as intended, resulting in a false positive.

\subsubsection{Malware Triage Scenario}

Malware Triage scenarios were the second of the common tasks.  For these scenarios, a potential malware sample was copied to a Windows system, causing the EDR tool, Endgame, to alert and quarantine the sample.  Unassisted, the analyst was expected to retrieve the sample from Endgame, submit it to the VirusTotal website to obtain a consensus score, and mark the sample as benign, suspicious, or malicious based on the score.  Depending on this determination, the analyst directed Endgame to take certain actions with the samples.  If the sample was determined to be suspicious, for example, a ticket was generated in Jira to elevate it for further investigation.  These scenarios exercised the abilities of the SOAR tools to interact with Endgame's API to carry out automated responses.  Unassisted, these scenarios typically took analysts about 15 minutes.  Assisted by SOAR tools, these became almost instantaneous because the complete chain could be automated.

To build these scenarios, benign, suspicious, and malicious software samples were selected from a repository we built for previous challenges.  The first qualification for a sample was that Endgame alerted on it; finding benign samples that would elicit a reaction from Endgame required extra effort.  This pool of samples was each submitted to VirusTotal for its ratings.  As expected, known malicious samples received high scores, and benign samples received low scores; finding samples that lead to a ``suspicious'' determination required identifying files that received scores inbetween, which again, required extra effort.  In the end, 20 samples were selected and developed into scenarios for testing by the analysts.  The same samples were used on all test environments.   

A malware event was initiated by triggering an alert. To accomplish this, we hosted our previously selected samples on a web server out of band from the local network. A member of our team then logged into one of the test environments and navigated to the web address where the malware sample was hosted to download the sample. Once downloaded, this caused Endgame to alert, allowing the analyst to proceed through a predefined SOP to determine if the sample was benign, suspicious, or malicious.
The scenario ends with actions taken in the EDR interface to handle benign and suspicious samples and documentation in a ticket, and for suspicious samples, a creating a ticket documenting findings and elevating the investigation to a higher tier analyst's workflow

\subsubsection{Malware Investigation Scenario}
In this scenario, samples triaged as suspicious, from previously worked Malware Triage scenarios, were submitted for further analysis using ThreatGrid, a cloud-based malware sandbox testing environment.  These scenarios evaluated the SOAR tools' ability to interact with a foreign tool (i.e., a tool for which no native integration existed for these SOAR tools), enrich the analysis of suspicious samples beyond VirusTotal, and to interface this information to the analyst.  After evaluation, the analyst would report a final determination on the associated ticket.

\subsubsection{IP Reporting Scenario}
This scenario was derived from reporting tasks, which currently require ample manual effort by senior analysts to dig through PCAPs to investigate a questionable IP and then report findings in a precisely formatted report.  
Our analogous testing scenarios involved detected, suspicious activity by a machine at a given IP address, warranting the investigation and report.  
To build these scenarios, a set of PCAPs were captured containing various suspicious interactions by machines at predetermined IP addresses.  
We created suspicious activity by including interactions with foreign websites and various data infiltration and exfiltration.  
A formatted inquiry request was created for each scenario, and a required report template was provided.  
Unassisted, these scenarios were tedious and took at least an hour to complete.

\subsubsection{Across-SOC Sharing (Add-On)}
The final scenario type was created to allow SOAR tool vendors to demonstrate sharing and interaction between their software at different SOC locations.  
By definition, the baseline environment did not have any sharing capabilities whereas the SOAR tools all purported to provide an extensive variety of sharing and interaction abilities.  The vendors determined which sharing abilities were demonstrated. 
To evaluate the sharing abilities, the participant, upon completion of one of the above scenarios, would proceed through a set of steps to share the report and/or extend an invitation to a colleague located at another SOC to collaborate on an incident.  The analyst would next switch their role in the investigation to the colleague receiving the report by logging in as the remote analyst and retrieve the shared report.  Similar to the previous scenarios, quantitative data in the form of time to completion and counts of window swaps were collected. 
\begin{wraptable}[7]{r}{0.65\textwidth}
\vspace{-.2cm}
\caption{Counts of configuration ratings across all six tools.} % with counts of tools in each configuration category with each configuration status.}
\label{tab:config}
\begin{tabular}{lccc}\\\toprule  
 \multirow{2}{1.1in}{Configuration rating}& \multirow{2}{*}{\shortstack{Data\\ingestion}} & \multirow{2}{*}{\shortstack{Ticketing/Jira\\integration}} & \multirow{2}{*}{\shortstack{Playbook\\setup}}\\
\\
\midrule
Well (1) & 2 & 1 & 1\\  
Neutral (0) & 3 & 2 & 3 \\
Poorly (-1) & 1 & 3 & 2 \\  
\bottomrule
\end{tabular}
\end{wraptable}

%  \hl{Put the videos explaining envrionment SOPs docs online (github) and mention + reference them??} 

% The Network Intrusion Detection System (NIDS) rules were probed to gather information on CVE's that would trigger an alert. 
% Once a list of CVE's were gathered, they were vetted for reproducibility. 
% VMs were designed as targets for triggering the alerts with the appropriate services and versions installed that associated with their corresponding CVE. 
% Once established, this capability allowed for quickly triggering an alert and generating appropriate logs for an analyst to investigate.

% We developed these different scenarios, described in detail below, on the baseline environment using our colleagues as test analysts as we developed the SOPs.  Once these were internally tested, the documentation and training videos were created for the analysts and vendors that demonstrated one of each scenario type in the baseline environment.  A number of each type of scenario was developed so that the analysts could be tested on each tool with both familiar and unfamiliar scenarios.  

\section{Results}
\label{sec:data-results}
%Phase 2 of this challenge consisted of hands-on evaluation of the SOAR tools by Navy operators. 

This section presents the results  of our evaluation.

    \subsection{Vendor Installation and Ingestion}
\label{sec:install-results}
 Section \ref{sec:install-test} discusses data collection and analysis used to evaluate installation and ingestion. 
To respect tool anonymization requirements, we are not permitted to release per-tool  scores. 
Most vendors required two to four hours of customization per tool, although some took many hours of extra effort and iteration for basic functionality.

\textit{Observation: Configuration of SOAR tools involves creative design by the installation team, especially with regard to displaying information, automating tasks, and facilitating effective user interaction with the SOAR tool. Making this an iterative process with users is critical to avoid pitfalls such as inconsistent data placement across playbooks that leads to confusion. 
It is important to have clearly defined procedures for playbook/workflow creation and SOAR automation, especially to determine what is ``OK'' to automate and what requires the analysts' attention.} %vendor installs and config data ratings  
    \subsection{DDIL Results} 
\label{sec:ddil-results}
This section presents the results of the tools' abilities to navigate a DDIL environment. See Section \ref{sec:ddil-test} for corresponding  experimental design and data collection procedures. To protect tool anonymity we are limited to reporting only aggregate statistics with respect to the bandwidth values.

\subsubsection{Denied Internet}
Of the six SOAR tools tested, one tool ran its playbook and completed a ticket, along with providing warnings of queries that were unsuccessful. Three tools executed their playbooks in some capacity but eventually stalled and were unable to complete. Of these, two displayed an error message, and one did not. For the remaining two tools, one ran its playbook through to completion but produced an incorrect result. The other, a cloud-based tool, was unable to initiate its playbook at all. 

\textit{Observation: The results of Denied Internet testing varied widely. Most tools (excluding the cloud-based tool) were able to at least initiate their playbook, though investigation completeness and accuracy were inconsistent.}

\subsubsection{Disrupted/Intermittent Internet}
Only one tool did not recover network events once its internet connection was restored. The other five tools were able to do so, with one tool given special attention for alerting with an on-screen notification that there had been a lapse in internet service. 

\textit{Observation: All but one tool were able to recover and report network events that were generated during a lapse in internet connection.} 

\subsubsection{Steady-State Limited Bandwidth}
Values are reported in the format of \texttt{mean} $\pm$ \texttt{standard deviation}. Per 1,000 events, tools required on average $0.373 \pm 0.621$ MB.

\textit{Observation: Total network traffic was measured in MB for each SOAR tool during a period of rest while no actions were being taken on the network by users; the SOAR tool was simply monitoring the network. The amount of required bandwidth to monitor the network varied widely across tools.}

\subsubsection{Working-Test Limited Bandwidth}
On average, the tools required $22.80 \pm 15.73$ MB to perform the investigation.
\textit{Observation: Even more so than steady-state bandwidth, bandwidth required to work scenario with the tools vary widely.}

% The results of this test varied from tool to tool. Some SOAR tools were able to run through their respective playbook and skip any information that it would normally gather from the internet and fill out their tickets with the limited information it could obtain. We considered these successful.  The other tools completely failed to execute a playbook, stalled, or came to incorrect conclusions.  Cloud based SOAR tools, could not see the events caused by the executed scenario, so they were not able to immediately run a playbook.  It is interesting to note that the cloud-based tools did recognize the events after the Internet was reconnected and completed their playbooks as normal.

% DDIL thorough description moved to appendix. It has both the test description/setup with the results. We need to extract the results and present them here. (table? or text?) 

% \textit{Observation:\hl{Need concise takeaway}} 
    \subsection{Qualitative Evaluation}
\label{sec:qualitiative-results} 
\begin{figure}
    \centering
    \hbox{\hspace{-1em}\includegraphics[scale = .8]{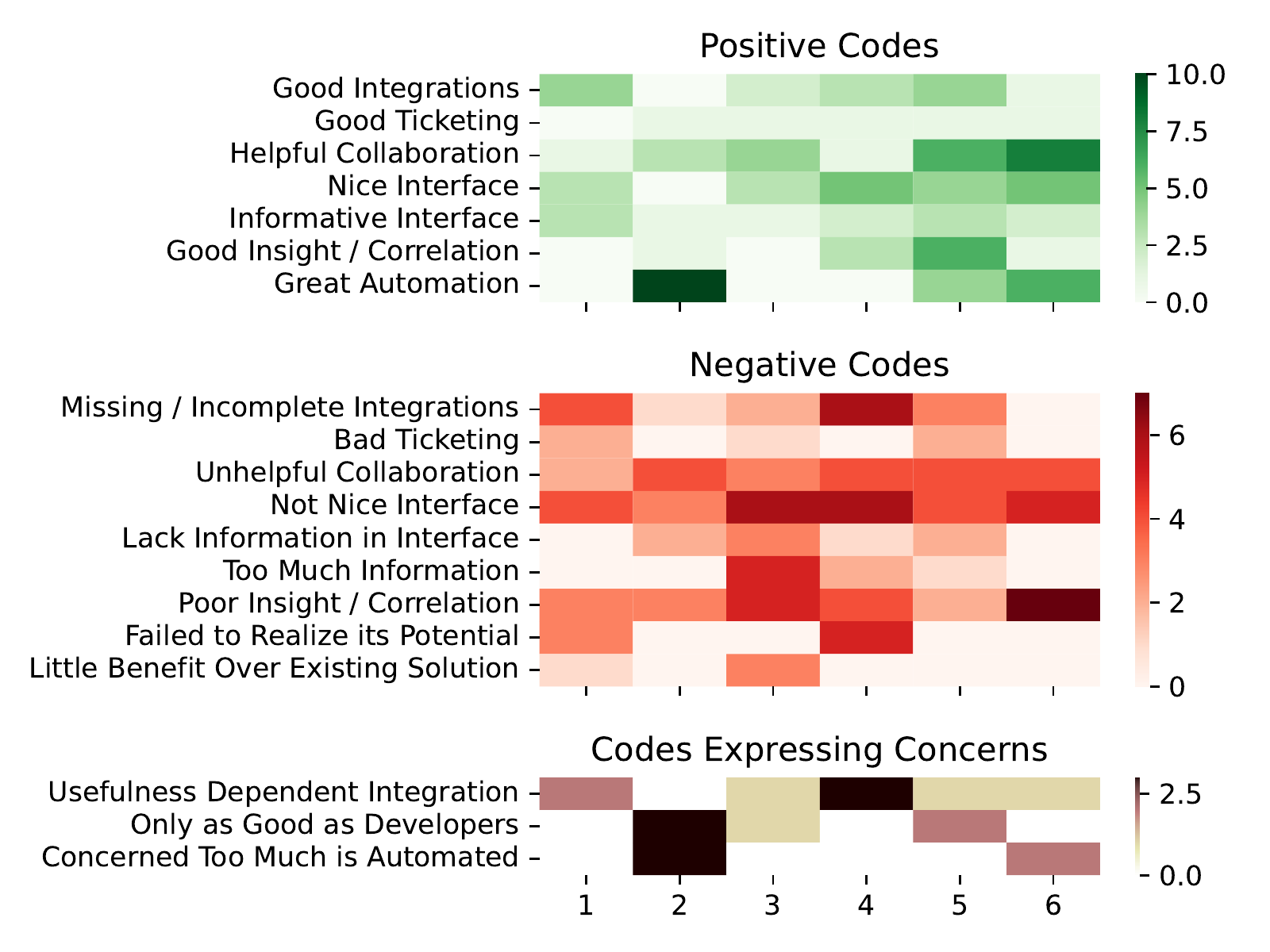}}
    \caption{Participant responses coded to the semistructured interviews for Tools 1--6.
    This figure provides a heatmap with the count of each code for each tool, where the count represents the number of participants who had at least one statement with a given code.}
    \label{fig:codes}
\end{figure}

See Section \ref{sec:usability-test} for description of the data collection and coding methodology. 
Figure~\ref{fig:codes} summarizes the prevalence of codes expressing positive sentiment, negative sentiment, and reasons for concern. 
This qualitative evaluation provides insight into how analysts perceived the tools, the benefit the tools might provide, and the potential downsides to SOAR tool adoption. Our evaluation showed that, even if an analyst clearly preferred a specific tool, they might not perform best with that tool. 
Quantitatively, in terms of measures such as time to complete tasks and ticket completeness, they may have performed better with a tool different from the one they preferred. 
This disparity can be expected given that analysts' preferences may be based on many factors besides performance, such as the usability or attractiveness of the tool's UI or a specific feature of the tool.

Notably, several analysts recognized that, given more time to optimize each tool for the environment, several of the tools have more potential than was demonstrated in this evaluation\textemdash see a quote from participant 5 (P5) below. 
Given that we did not have time or resources to fully configure each tool, it was an inherent limitation of our approach that the full capabilities of each tool would not be exercised. 

\begin{quote}
    Tool has more potential -- P5 -- \textit{`While there are some concerns with how to manage cases and collaborate amongst analysts, the capabilities of the tool would certainly increase our SOC’s efficiency.  Many of my concerns regarding triage and data visualization e.g. dashboard elements, are likely a result of how the system is configured to support the challenge and not a limitation of the technology itself.  We could likely tailor our own dashboards to better get after APT and effective triage.'}
\end{quote}

\subsubsection{Appropriate Role of Automation}
While SOAR tools are intended to take analysts out of the loop to increase efficiency, there are voices of caution related to automation.
Schneier \cite{schneier} reminds us that we cannot automate what we do not know and, therefore, can never completely remove humans from the loop because uncertainty is a given in cybersecurity; similarly, Baxter~\cite{baxter_ironies_2012} echoes Schneier's assertion, that the more we depend on technology the more we also depend on highly skilled individuals to ensure that our technology is both resilient and properly configured. 
Several participants in our study, as demonstrated by the quote below from P12, also expressed concerns that too much automation could have unintended negative consequences. 
It is interesting to note that while our initial survey of analysts, conducted prior to testing, showed that automation was the most-sought-after feature in a SOAR tool, once analysts actually utilized the tools, they expressed concerns regarding too much automation.
Determining the appropriate place for automation in SOCs is an open area of research ripe for further study. 

\begin{quote}
	Concerned Too Much is Automated -- P12 -- \textit{`There should be a double check option, in case there was an automation error. We need to review tickets before closing it out.'}
\end{quote}

\subsubsection{Usefulness Dependent on Integration and Training}
Across solutions, multiple participants expressed concern that these tools would be difficult to integrate into their environment and require highly trained developers/analysts, as demonstrated below by quotes from P14 and P5. 
It should be noted that these concerns match the experiences of companies that currently utilize SOAR solutions. 
The complexities of integrating SOAR tools are such that Gartner recommends beginning with several clear use cases to ensure that integration has a clear objective to help prioritize data sources for effective ingestion and quick impact \cite{gartner2019}. 
Not all organizations have the necessary personnel or existing infrastructure to utilize SOAR solutions, which is why defining clear use cases is so important. 

\begin{quote}
    Usefulness Dependent -- on Integration -- P14 -- \textit{`Highly dependent on Automation functionality and integration into our current architecture. Current toolset is already baked in and analysts are well trained.'}
\end{quote} 

\begin{quote}
	Only as Good as Developers -- P5 -- \textit{`The tool would certainly enhance our SOC by reducing the amount of time taken to conduct analytic functions.  I do however feel that we would have to spend a considerable amount of time configuring, updating, maintaining the playbooks to ensure that it was working as expected.'}
\end{quote}

\begin{figure*}
    \includegraphics[width = \textwidth]{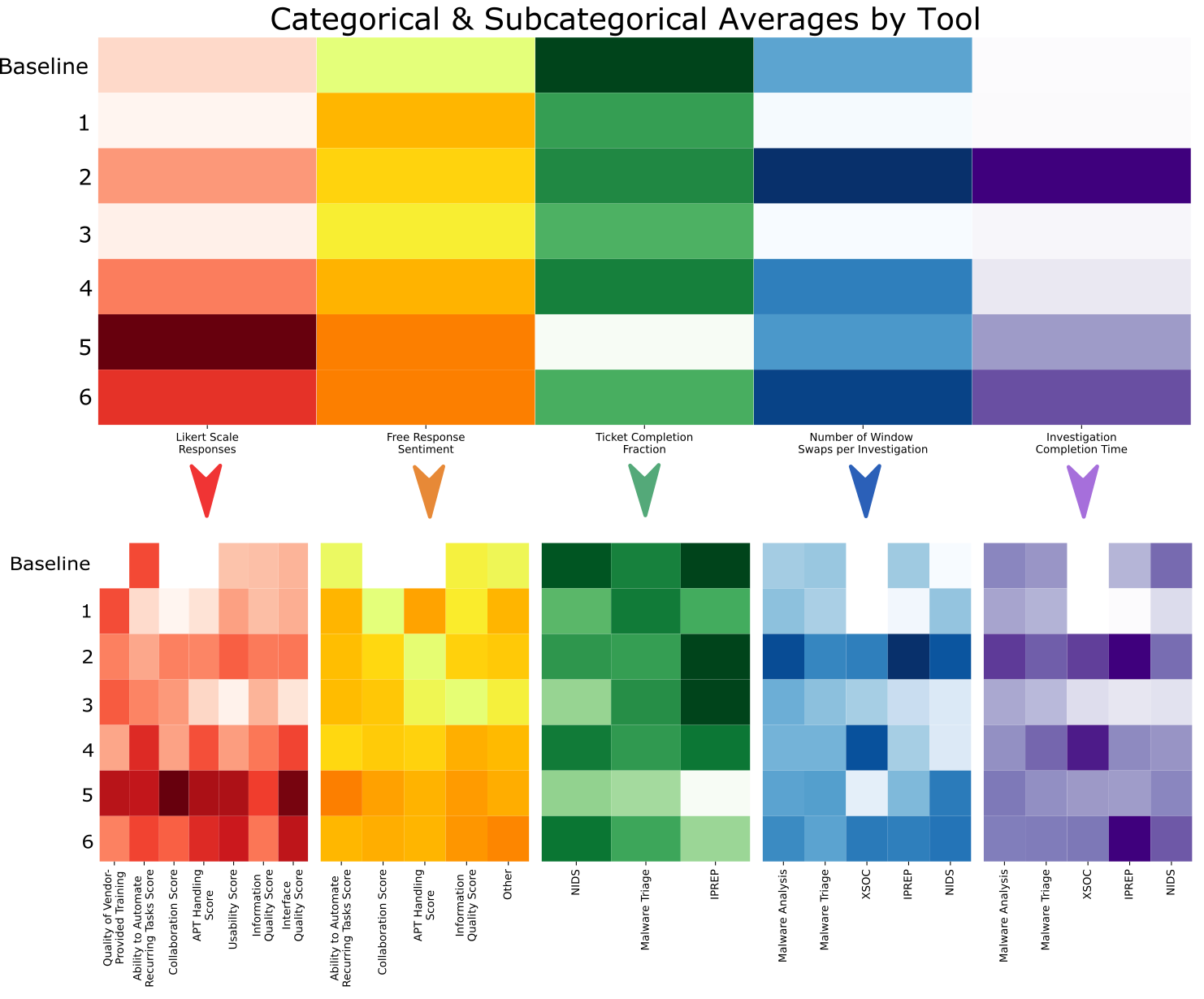}
    \caption{
    Tool performance for each type of data collected. Dark colors imply strong performance; light colors imply weak performance. Top: Broad categories of qualitative and quantitative data collected (see Section \ref{measurement-categories}) for each tool. Bottom: Subcategory details indicate the contribution of each to the overall score for the tool in the category. For instance, Tool 5 performed the best in the Likert Response category, and it is largely driven by its high scores in the collaboration, usability, and interface quality subcategories. Similarly, Tool 2 performed well in the investigation time category, with that mostly attributed to the IP Reporting (IPREP), Malware Analysis, and Across-SOC (XSOC) scenarios. Due to the nature of the XSOC investigation, qualitative data was not collected for this scenario.} 
    \label{fig:bigfig} 
\end{figure*}

\subsection{Quantitative Results Summary per Tool}
\label{sec:averages}
Figure \ref{fig:bigfig} summarizes all collected data including  per-category and per-subcategory averages for each tool. 
Notably, Figure \ref{fig:bigfig} shows a discrepancy between some of the qualitative and quantitative measurements. 
For example, Tool 5 scored very well in the Likert response and sentiment data but lacked efficiency, as it required more context switching and on average longer times to complete tasks. In contrast, Tool 2 was the most efficient in terms of context switching and investigation time, but was not a preferred tool based on Likert response and sentiment data. Ticket completion fractions are inconsistent across tools for a number of reasons. Baseline (i.e., test investigations performed with no SOAR tool) had the most complete tickets overall, but it is important to recognize that this was accomplished in the absence of automation and the operators had to manually complete every ticket. This is reflected in the time data for baseline. Furthermore, auto-population of tickets occasionally exhibited great inconsistency for the same tool: some fields were populated for some investigations but not for others. 
%Such information is critical for understanding what is driving a tool's performance, both in terms of quantitative analysis as well as operator preferences. 

\textit{Observation: Two SOAR tools on average received poorer Likert results than the baseline, and all SOAR tools ticket accuracy scores were worse than the baseline. Four SOAR tools decreased the number of window swaps required for investigations, and all SOAR tools improved or did not exhibit degraded investigation times and received better sentiment responses compared to the baseline.
}

\textit{Observation: Areas of strengths and weaknesses varied greatly across these six SOAR tools. 
Many of the problems and limitations discovered in the test were seemingly fixable with iterative tuning (e.g., insufficient content in automatic ticket population,  bugs in integrations, confusing layout of configurable displays); hence, we suggest iterative operator use, feedback, and reconfiguration to be expected upon deployment. }
% Other issues are inherent to the tool design.}

 % averages of data by tool, category, subcategory
    \begin{figure}
    \centering
    \includegraphics[width = \textwidth]{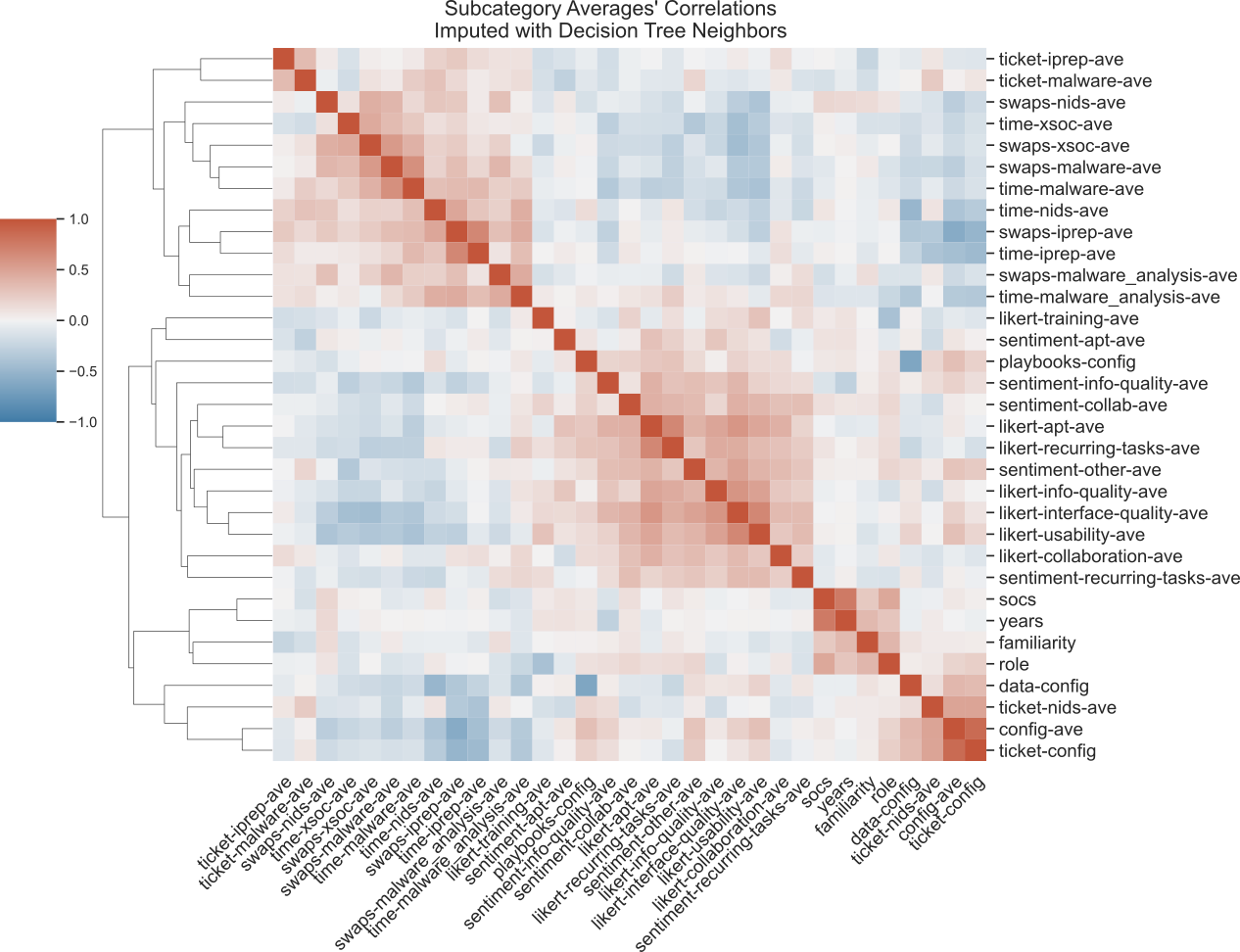}
    \caption{The 75 normalized measurements are coarsened into 25 subcategories by averaging measurements that fall in the same subcategory (e.g., multiple Likert responses pertaining to usability are averaged to obtain the subcategory `likert-usability-ave') for each  test (user, tool). Correlation of these subcategory scores across all tests (user, tool) is depicted. Multiple imputation is used to prevent artificial inflation of correlation caused by missing data. Missing data is sampled from kernel densities estimated from the missing values' peers, and the average pairwise correlation across $m=1,000$ samples is depicted. On the left of the heatmap is a dendrogram showing hierarchical clustering of all subcategories based on their the mutual correlations. Notably, the data breaks into two main clusters. The first cluster (from bottom) included every Likert and free response subcategory, showing user sentiment is on average consistent throughout the questionnaires. Meanwhile, the second cluster included every subcategory of the Ticket Quality Ratios, Window Swapping Counts, and Time per Investigation data, indicating that performance with the tools trend together. Finally, we note that, broadly, the correlations of these two clusters (top right/bottom left of heatmap) are light blue, indicating a slightly negative correlation. While users may express favorable/unfavorable sentiment in questions about a SOAR tool, their performance with the tool may trend in the opposite direction. This accentuates the importance of using objective measurements of performance in addition to the qualitative data collected.}
    \label{fig:subcat-clustermap}
\end{figure}

\begin{wrapfigure}[35]{r}{.5\textwidth}
    \vspace{-.2cm}
    \centering
    \includegraphics[width = \linewidth]{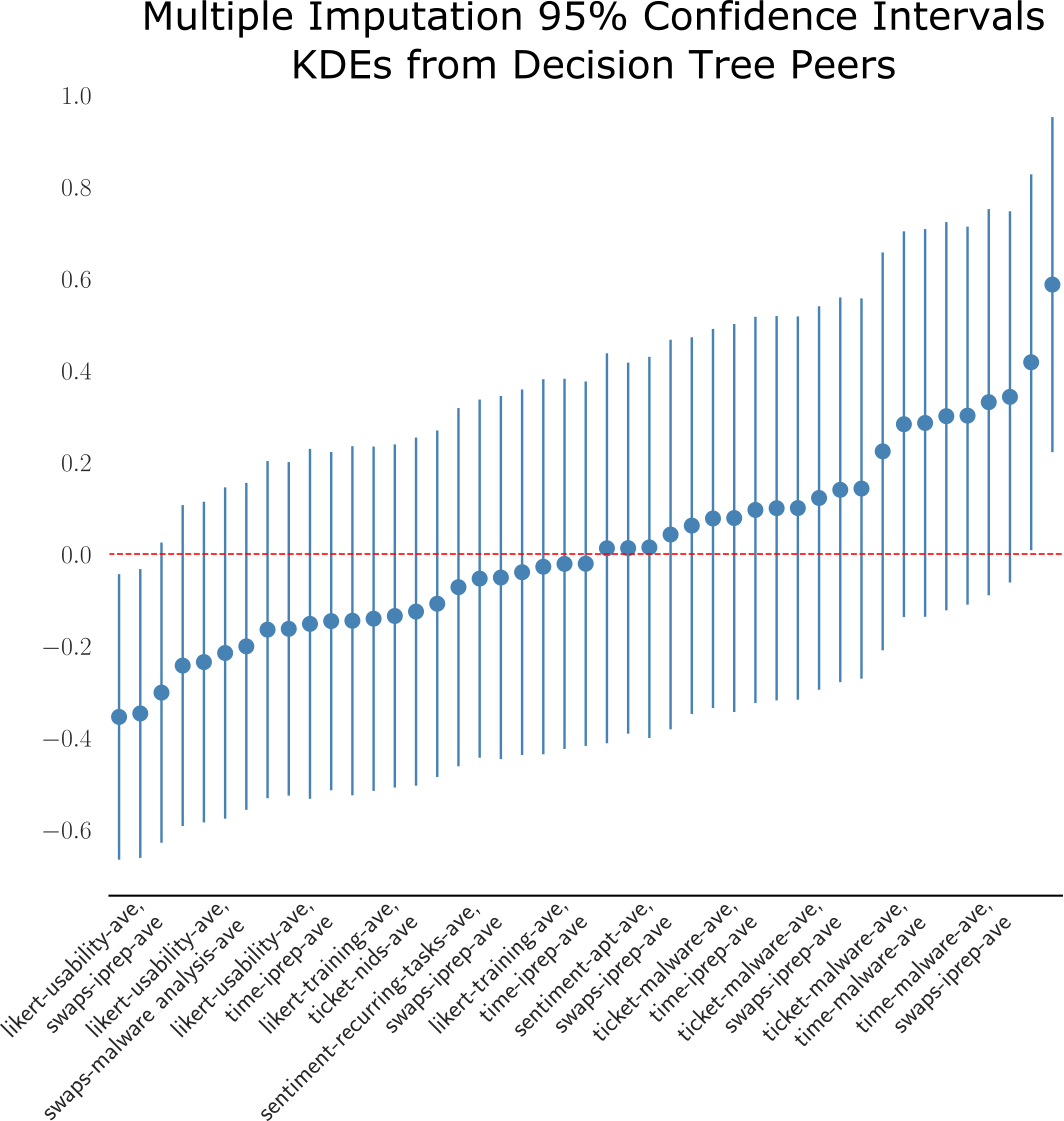}
    \caption{Images depicting 95\% confidence interval for correlation of subcategory average values based on $m=1,000$ imputations of missing data cells sampled from KDEs fit to missing values peers. Due to space constraints, out of the 300 pairs of Subcategory Average columns, we randomly selected two subcategories from each category and displayed the confidence intervals for 45 pairs', a subset of which are labeled on the x-axis. Notably, all the pairs of subcategories from Likert and free response and, separately, pairs of the objective measurement subcategories have 95\% confidence intervals above or nearly above 0, indicating they are almost certainly mutually correlated. Confidence intervals for subcategory pairs crossing these two groups lie mostly and sometimes fully below 0. These results provide evidence that the two main clusters of Figure \ref{fig:subcat-clustermap} indeed have positive within-group and small or even negative between-group correlations.}
    \label{fig:mi-confidence-intervals}
\end{wrapfigure}
\subsection{Subcategory Correlation Results}
\label{correlation-results} 
We detail our correlation approach in Subsection 
\ref{sec:correlation-approach}. 
Figure \ref{fig:subcat-clustermap} shows a heatmap correlating subcategory scores across all tests. 
The subcategory averages form two main clusters. The first cluster includes every Likert and free response subcategory, showing user sentiment is on average consistent throughout the questionnaires. 
The second cluster included every subcategory of the Ticket Quality Ratios, Window Swapping Counts, and Time per Investigation data, indicating that performance with the tools trend together. 
Additionally, note that, broadly, the correlations of these two clusters (top right/bottom left of heatmap in Figure \ref{fig:subcat-clustermap}) is light blue, indicating a slightly negative correlation. While users may express favorable/unfavorable sentiment in questions about a SOAR tool, their performance with the tool may trend in the opposite direction. This accentuates the importance of using objective measurements of performance in addition to the qualitative data collected.
% We provide more details and data for this correlation analysis in Section \ref{sec:appendix-mi} of the appendix. 

To further investigate this finding, we  computed 95\% confidence intervals for our imputed correlation samples, following Schafer \cite{schafer1999multiple} (in our case, with $m=1,000$, this is a $z-$test for a normal distribution). See a sample of confidence intervals in Figure \ref{fig:mi-confidence-intervals}.

% We display confidence intervals for a subset of the 300 total Subcategory pairs in Figure \ref{fig:mi-confidence-intervals}.
% This confirms that correlations of pairs of questionnaire subcategories and pairs of performance measurement subcategories provide at least 95\% of being positively correlated, while the between-group pairs are, if not confidently negative, below 0.05 in correlation. 
% As displayed in Figure \ref{fig:mi-confidence-intervals}, we can compute confidence intervals and perform hypothesis tests from our imputation \cite{schafer1999multiple, austin2021missing}.

\textit{Observation: The users' sentiment and Likert responses for the tools were positively correlated with each other. Similarly, the performance measurements with the tools were positively correlated. Yet, these two groups of measurements were slightly negatively correlated, indicating that user preferences about the tools do not match how well they performed with the tools. Our data implies that SOAR tool selection will require a balance between performance (e.g., investigation quality, efficiency, and context switching) and user preferences.}

% \subsubsection{Correlations in Subcategory Averages}
% \label{sec:subcat-corr}
% Figure \ref{fig:subcat-clustermap} displays the correlations between all pairs of the 25 subcategory measurements alongside hierarchical clustering of each measurement based on their mutual correlations. 
% The takeaway from this 
% analysis is immediate from the plot: all questionnaire subcategories cluster together with high correlation, as do the performance measurement subcategories (Ticket Completion Fraction, Window Swapping Counts, and Investigation Time), but the between-group correlations are slightly negative. 
% This indicates that, for a given test, the participant expressed broadly consistent Likert responses and sentiments when responding to the  questionnaires, but their feedback was on average slightly in opposition to their performance with the tool, as indicated by our measurements. 

Considering the coded responses in Figure \ref{fig:codes} and the per-tool averages in Figure \ref{fig:bigfig}, we can find specific evidences of the subcategory correlation results. 
For example, Tool 2, which according to the coded responses (Figure \ref{fig:codes}) was praised by many operators for ``great automation" also raised concerns regarding overautomation. The average responses (Figure \ref{fig:bigfig}) for Tool 2 show that this tool was, on average, best at reducing context switching (as evidenced by the lowest number of  window swaps) and the quickest (as evidenced by the lowest investigation times), but received middle-of-the-road scores in the Likert and free response categories. 
Furthermore, the average Likert and free response results for Tool 5 show that it was overwhelmingly preferred by users, but it scored far worse in ticket completion and delivered mediocre performance in window swap count and investigation time. Hence, even though operators provided strong feedback, ticket quality suffered, and reduction in context switching and investigation time were mediocre. 

\textit{Observation: These results underscore the importance of using quantitative measurements of the efficiency, quality, and context switching enabled by the SOAR tools alongside and as a complement to user feedback when testing a SOAR tool.}

 % correlation of subcategory results 
    \begin{figure}[!ht]
    \centering
    \includegraphics[width = \textwidth]{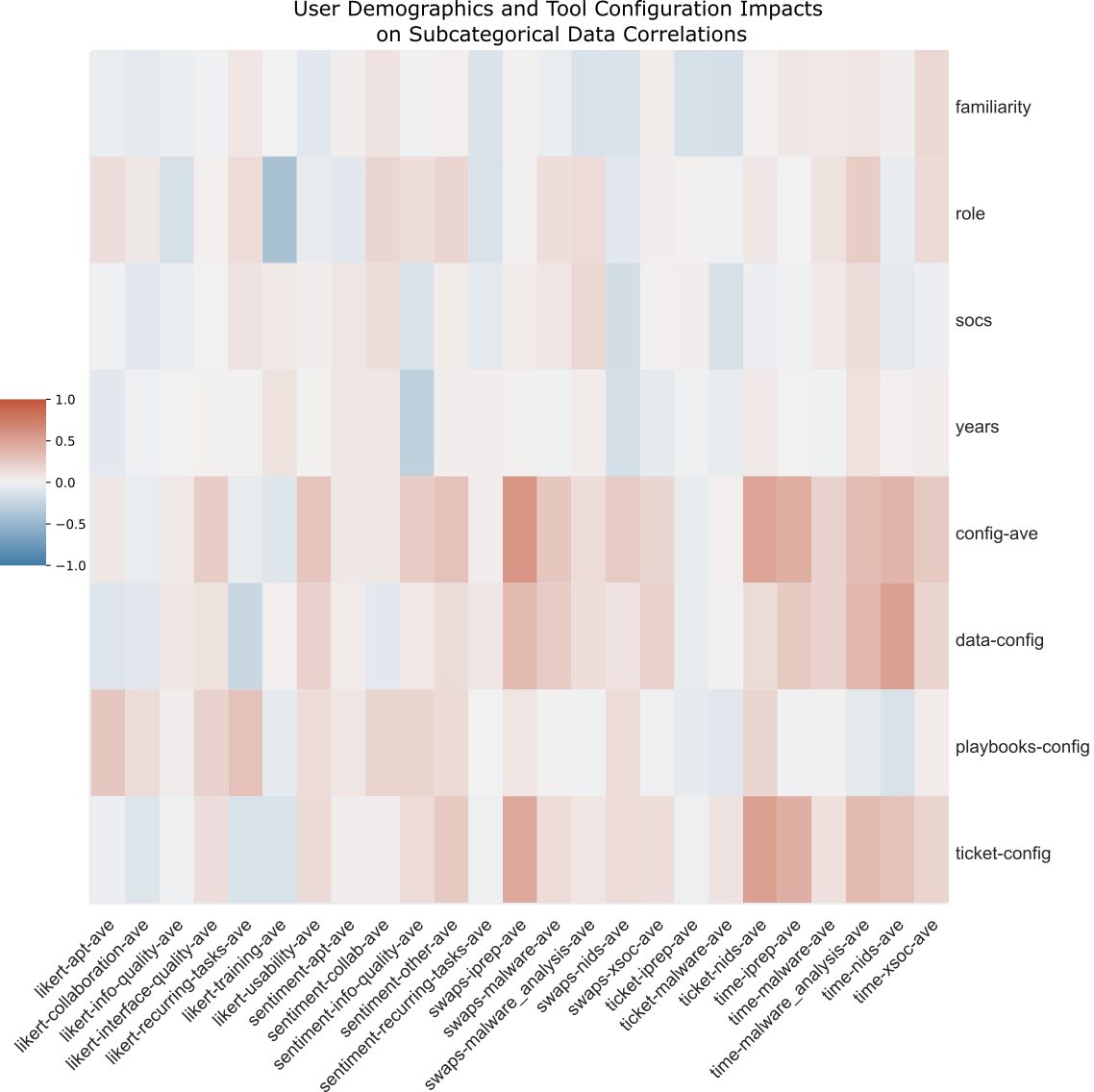}
    \caption{Heatmap detailing the impacts of user demographics and tool configuration on SOAR tool performance in each of the data subcategories. The strongest correlation of a user demographic comes from "Job Role," where users functioning as an SOC operator felt less trained following vendor-provided training than did those functioning as an NOC operator or in some other role. In terms of the quantitative metrics, (i.e., ticket completion, window swapping, and investigation time), the overall tool configuration (config-ave) plays a bigger role than for the qualitative Likert and sentiment scores. At a finer level, playbook configuration has a slight positive correlation with some of the Likert questions, though the effects of this are neutralized in the overall configuration due to the weak and opposite correlations of data configuration and ticket configuration.}
    \label{fig:user-tool-results-corr}
\end{figure}

\subsection{Correlation of Test Data with User Demographics and Tool Configuration Results}
\label{sec:correlation-test-user-data-results}
See Section \ref{sec:data-and-preprocessing} for an overview of the data tables and data analysis methodologies, specifically \ref{sec:corr-user-data-results-methodology}. 
Results are visualized in Figure \ref{fig:user-tool-results-corr}. 
Recall ``Job Role'' question was encoded as a 1 for the response ``SOC operator'' and as a 0 for ``NOC (network operation center) Operator'' or ``Other.'' 
The ``config-ave'' column is the average of the three different configuration ratings given upon installation of each tool (i.e., the overall configuration). `Data-config' references `Data Integration,' `playbooks-config' references `Playbook Setup,' and `ticket-config' references `Jira/Ticket Integration,' all of which are described in Section \ref{sec:install-results}. 

\textit{Observation: Correlations of test results with user demographics are negligible, except for  ``Job Role," which negatively correlated with questions intended to assess how well trained to use the SOAR tool the participant believed they were. (SOAR training was provided by the vendors in a video). This suggests that non-SOC users (e.g, those trained for network operations or other related IT roles)  will need more training to feel equally comfortable (as SOC operators)  with a SOAR tool. 
}

\textit{Observation: Configuration ratings (recorded at time of installation) for a tool's ability to integrate with diverse data sources and assist in ticketing were highly correlated with efficiency metrics; whereas, playbook configuration ratings were negatively correlated. 
This suggests that the appearance at time of installation of well-configured playbooks may not be representative of their functionality in practice. }

% Our experimental framework allowed for data collection on SOAR tools that has never been done before. 
% From this data, we can draw conclusions about the benefits of SOAR tools in the SOC, as well as quantify their enhancement of investigation. However, it is critical that we present this information within the context that it was collected, and as such we must consider user demographics and how well the tool was configured (Fig. \ref{fig:subcat-clustermap}). 

\subsubsection{Do SOAR Tools Improve Efficiency in an SOC?}
\label{sec:results-improved-efficiency} 
Figure \ref{fig:bigfig} (top, right column) shows that, compared to the baseline, four of the six tested SOAR tools clearly decreased investigation time (top, second to right column) and the number of window swaps on average, and all tools were as good as the baseline.   
Given that our results indicated most SOAR tools help, and none hurt efficiency metrics, we overlaid configuration data to check for correlations. 
Referring still to results in Figure \ref{fig:bigfig}, of the two tools that, on average, showed no improvement to the baseline in these categories (i.e., Tools 1 and 3), one was rated poorly in configuration scores (-1 for ticketing and data integration, 0 for playbooks), while the other was given configuration scores (0 for ticketing, -1 for data integration, and 1 for playbook configuration).

To investigate efficiency metrics correlation with configuration, we set up the following dichotomy: tools that received an average configuration score less than 0 were termed ``not well-configured whereas tools with an average configuration score at least 0 were considered ``well-configured.''

\begin{landscape}
\begin{figure}
    % \centering
    \includegraphics[scale = 0.6]{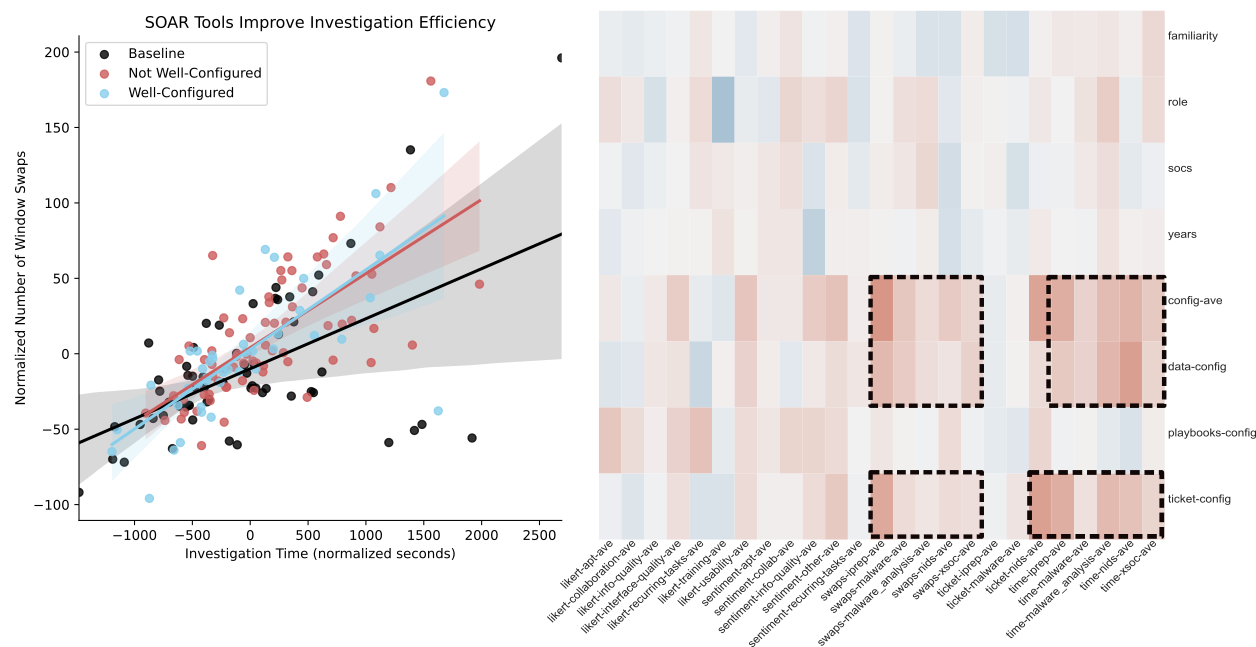}
    \caption{\textit{Left:} Plot comparing efficiencies of SOAR tools and baseline environment. Both window swaps and investigation time have been inverted to reflect favorable investigation conditions with fewer swaps and shorter investigation times. We note that, regardless of configuration, SOAR tools offer an edge over an environment with no SOAR tool in terms of efficiency. This is observed in the greater slope for the SOAR tools as opposed to the baseline environment, in which more context switching (denoted by window swaps) occurs in a shorter amount of time. \textit{Right:} Subcategory averages from Figure \ref{fig:subcat-clustermap}, where the black dotted lines highlight notable correlations of windows swaps and investigation time data with user demographics and tool configurations. We observed distinct correlations in the window swaps data for the IPREP and Malware scenario with the overall configuration at a coarse level, but then we note that is heavily influenced by the playbook configuration and the ticket configuration, or Jira integration. For the investigation time data, we note overall stronger correlations with the overall configuration of the SOAR tool, with IP Reporting, Malware Analysis, and NIDS scenario times being influenced by playbook configuration and Jira integration. In terms of user demographics, the Malware Analysis scenario may have a slight correlation with job role, but otherwise there appears to be no real influence of user demographics on investigation efficiency.} 
    \label{fig:swapscorr}
\end{figure}
\end{landscape}

Figure \ref{fig:swapscorr} regresses the counts of window swaps onto investigation time for well-configured SOAR tools, poorly configured SOAR tools, and our no-SOAR-tool baseline environment. 
The primary takeaway of the left plot is that, when comparing investigations with similar amounts of required window swapping,  SOAR tools took less time than the baseline, on average, regardless of configuration rating. 

\textit{Observation: For an investigation with a fixed number of window swaps, all SOAR tools provided an efficiency gain in our data, with greater gains for well-configured tools. Our data implies that average configuration is correlated with a SOAR tool's impact on efficiency.}
%key to maximizing efficiency with SOAR tools.}

\subsubsection{Do SOAR Tools Improve Investigation Quality?}
In addition to efficiency gains, we evaluated the claim that SOAR tools improve investigation quality. The tickets created and populated by SOAR tools provide documentation allowing us to examine investigation quality and to determine whether increased efficiencies derived from using a SOAR tool might come at the expense of investigation quality (defined as a thorough investigation).   

% \hl{Ashley, see comment} 
% \textit{Observation: Neither average SOAR tool configuration nor operator years of experience do not seem to affect ticket completeness, a measure of investigation quality, indicating that SOAR tools may help bridge the gap in operator expertise.} 

\begin{landscape}
\begin{figure}[ht!]
    \centering
    \includegraphics[scale = 0.6]{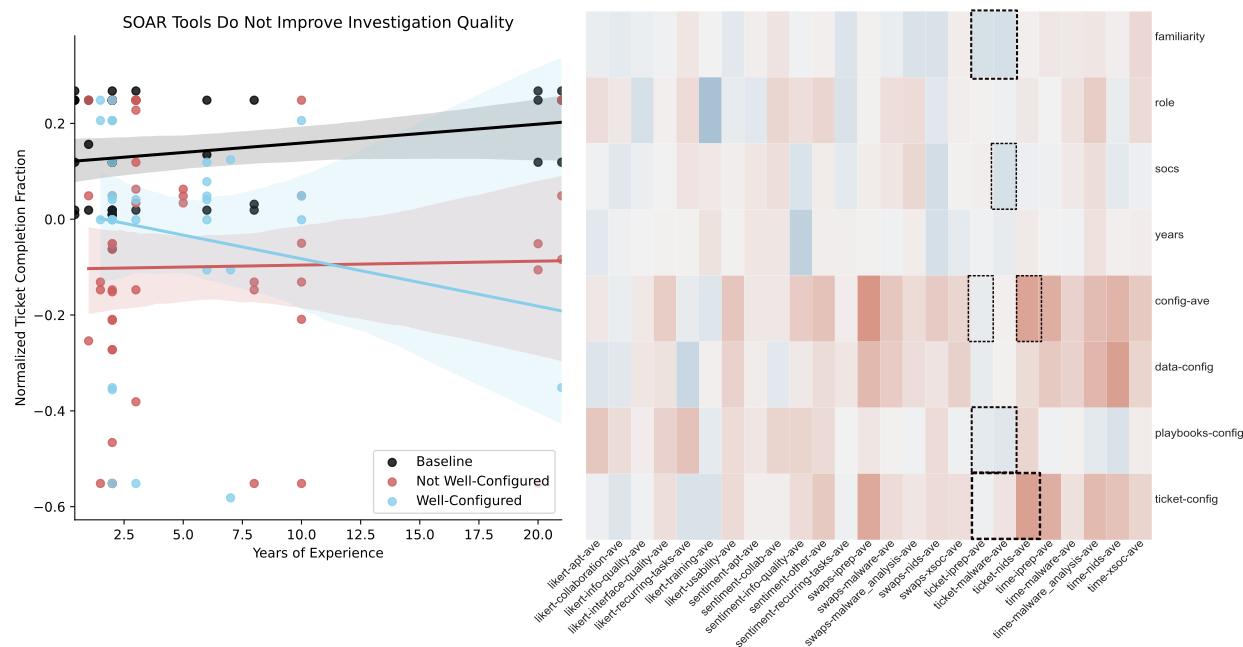}
    \caption{Left: Ticket completion fraction vs. operator years of experience. There appears to be no strong correlation between ticket completion and experience for any of the configuration cases. On average, baseline tickets were completed at a higher fraction than tickets completed using SOAR tools. This is largely due to some inconsistencies that plagued SOAR tool ticket population, as is reflected in the NIDS scenario's correlation with ticket and playbook configuration status. Right: There are possible slight negative correlations with some aspects of user demographics, such as number of SOCs they worked in and familiarity, indicated by the blue boxes highlighted with black dotted lines.} 
    \label{fig:tickets-experience}
\end{figure}
\end{landscape}

As noted in the top center column of Figure \ref{fig:bigfig}, our baseline environment produced the most complete tickets for the investigation. However, this result was expected given that tickets were completed manually for the baseline. In the case of SOAR tools, there were some instances in which the tools were inconsistent with the fields they populated in the ticket, leading to less-than-thorough documentation of an investigation when compared to the manual recording with the baseline environment.

Similar to our examination of configuration effects on efficiency gains, we probed the effect of SOAR tool use on ticket completion, along with accounting for analysts' years of experience to determine whether more experienced operators completed more thorough investigations. Figure \ref{fig:tickets-experience} shows that configuration status does not seem to impact ticket completeness, with both configurations of SOAR tools performing similarly. Furthermore, no strong relationship appears to exist between operator experience and ticket completeness.

\textit{Observation: When following an SOP step by step with no SOAR tool (i.e., in the baseline test), operators produced far better investigation tickets (graded on correctness and completeness) than with SOAR tools. Inconsistencies in SOAR tools' ticketing automation were problematic for ticketing quality.}

 % overlaying correlation of quantitative data with user, tool data 

\section{Discussion} 
\label{sec:discussion}
This section presents limitations, takeaways learned about SOAR tools, and practical experience lessons for testing them. 
\subsection{Limitations}
\label{sec:limitations} 
There are several limitations of note involved with this study. Here, we discuss both limitations in the design of the experiment, as well as intrinsic limitations due to the nature of this study. 
\subsubsection{Experimental Limitations}
During the initial training period for the operators in which they were guided through use of our baseline test environment and the SOAR tools they would be using, it is possible that there was inconsistency in the instruction as the operators were trained by different members of our research team. While the general protocols were the same, the thoroughness of the training could have differed between training sessions. 

We acknowledge that in an ideal case, we would have had every user test every tool. However, due to the limited availability of the SOC operators and the time requirements for participation, this was not possible. Previous user studies (e.g., involving SOC operators) also report that SOC operator time constraints limited their ability to assist in such studies (e.g., \cite{goodall2004, botta2007towards, werlinger2009security}). We did ensure that each tool had at least eight users, and most users tested two or three tools. 
To offset this disparity, we had three Tier 3 analysts explore all the tools in a hands-on manner in our environment. 
To account for different analysts testing each tools where some analysts may have been more likely to give lower or higher tool ratings, we normalized the scores according to each user. 

To prevent analysts from working identical scenarios over time, and thereby potentially performing better with tools that they tested later in the experiment, many variants of each scenario type (NIDS, Malware Triage, ...)  were employed (See Section \ref{sec:scenarios}).
There are two limitations here: 
(1) repeating similar, even if not identical scenarios, may still result in better performance in tools tested later by each user; 
(2) each user-tool test is not identical resulting in a loss of consistency---some adjustments may have complicated the scenarios more so than others. 
We designed the experiment to combat these two pitfalls. 
First, the training involved working scenarios and providing SOP guides with the hope that sufficient familiarity would result before testing. 
Second, we designed the scenario variants to combat variation in difficulty.  
Specifically, each user-tool test involved an ``easy'' and a ``hard'' NIDS scenario whereby the analyst did not require / did require reading and processing information in the related CVE. The CVE scenarios were chosen by our technical team from the national database to have similar details available for the operator.  The non-CVE scenarios were similarly chosen to be equal in complexity.  The example scenario provided to all of the vendors did have a CVE.  As a test of the installers' attention to detail, we did not disclose---unless asked during installation---that the tested scenarios may not have CVEs associated with them.
Similarly, while the two particular Malware Triage scenarios varied among many examples in each user-tool test, the two chosen for each user always involved an investigation of a file to be deemed ``malicious'' or ``benign'' (both require identical steps in the SOP) and the second scenario a file to be deemed ``suspicious'' (this required a few extra steps compared to ``malicious'' and ``benign'' determinations).

%Similar to normalizing by ``tough'' and ``easy'' users, here we normalized according the difficulty of the scenarios to offset this---

Regarding the Malware Triage scenario, presenting the users with one file at a time and advising them to initiate a malware investigation introduced some bias, and our design ignores the fact that file conviction decisions are often too numerous to all be manually investigated in practice. 
Perhaps a better design is to present the analyst with set of several files allowing them to prioritize and investigate the files in a set amount of time. 

While the six SOAR tools tested were popular commercial solutions, they may not be a representative sample of all SOAR solutions. 

Lastly, we note that 21 of 24 analysts were recruited from US Navy security and/or network professionals, with the remaining 3 coming from the research organization; this makes it difficult to generalize results beyond a specific population. %For best results, our experimental framework should be used to conduct additional tests in the target environment. 

\subsubsection{Intrinsic Limitations}
In addition to limitations with the experiment, there were also several limitations inherent to this study. 

Firstly, our sample population was  small, consisting only of 24 analysts. While this study was specifically about the use of SOAR tools within Navy SOCs, generalizability can certainly be improved with testing by a larger, more diverse population. 

Secondly, the vendors were given a limited amount of time to configure their tool in our environment, so our study is not representative of long-term operation of SOAR tools in a real SOC. 
%As such, the tools here may not be representative of their performance in environments wherein vendors have extra time and can ensure optimal configuration---
In practice multiple iterations with the vendor-provided installation team and the SOC would take place to refine tool configurations. 
By the same token, operators would use the tools on a daily basis and SOCs would have the ability to tweak configurations and optimize it to their specific environments and use cases. This is something that could not be tested in this experiment. 

Thirdly, as the test was performed in our experimental network environment, it cannot provide the fidelity of a real network SOAR tool deployment. 
% Regarding the data itself that was generated through a simulated environment, it lacks the fidelity of data associated with a real network. However, realistic data was not necessarily as important as realistic logs and scenarios. 

Fourthly, we acknowledge that individual data collected on investigation efficiency and quality could vary depending on the effort given by the participant during the tests. We observed a seemingly wide variance in the level of interest, effort, and thoroughness of investigation across participants. 
It is unclear if this is indicative of the variance of SOC operators in practice, or is perhaps correlated with something spurious, e.g., their interest in the experiment or in learning about SOAR tools.

Finally, the skill and effort put forth by each particular vendor's installation team affected the results for their tool. Whether or not their efforts match real deployments or how much it may vary across different members of the vendor's organization at unknown. 

%in the test seemed to  thorough the user was during the test and would not necessarily be correlated to proficiency with the tool. 
%and compared their rankings against our overall results. 
%(Their ratings confirmed our findings from the majority of analysts who only used several tools.) 

% \hl{Of a similar notion, constraints on operator time for testing were what facilitated the need for Phase 1's down-selection methods. In this phase, we learned what the preferences of the operators were in SOAR tool features. We noted that automation was (one of?) the top preferred feature. However, Phase 2's findings indicated that in practice, operators preferred tools that were not too opaque in their automation procedures such that they could still easily verify that the tool was performing as expected. As such, it is possible that Phase 1 eliminated preferred tools that were not evaluated in Phase 2.} 

    \subsection{SOAR Takeaways}
Configuration of SOAR tools is critical, and it requires creative design by the installation team as well as iteration with the SOC to ensure the tool is usable and the automation satisfies SOC needs. 
The SOC should have well-defined procedures for automation.  
One of the main lessons learned shared by the senior analysts involved in the study is that SOAR tools can indeed overautomate pertinent processes, so, SOCs should help curate a semiautomated process that expedites analyst activities while prompting the user to complete investigation steps that require human attention (and ideally the SOAR tool will present the needed correlated  information so the analyst can focus on the investigation).

Our average results indicated that analysts preferred SOAR tools to the baseline. 
Importantly, our results also confirm that SOAR tools increase efficiency while also decreasing the number of window swaps. 
Interestingly, even though the number of window swaps per investigation is lower with SOAR tools than without, for the same number of window swaps, SOAR tools enable analysts to work faster (Figure \ref{fig:swapscorr}). 
The only ``sore spot'' was ticket quality, which likely can be addressed with SOAR configuration and analyst training, given we are now aware that these influential preliminaries can suffer.

Interestingly, our results revealed negative correlations between analyst-preferred tools and performance. This was in part driven by overautomation concerns and poor ticketing of a highly preferred tool, but shows that analysts preferred SOAR tools with which they performed worse {\textit more} than those with which they performed better.

Additionally, if a SOAR tool will be used in a DDIL bandwidth environment, it should be tested for how well it performs when an internet connection is unavailable or limited, as some tools failed under this condition while others exhibited graceful declines in abilities. 

One question of interest that remains is: Do SOAR tools allow junior operators to perform at a higher level? 
Our testing method was unable to provide any answers to this question for two reasons.
First, ticketing quality was the artifact allowing quantification of investigation quality, yet all SOAR tools tested were worse in ticket quality than the baseline. 
This leads us to believe that analyst quality may be obscured by trusting ticket auto-population that was not configured to perform to the same level of thoroughness as the analysts alone. 
Second, as seen in Figure \ref{fig:tickets-experience}, there was no correlation of ticket completeness (our measure of investigation quality) with experience level.

% \hl{\subsection{What are the implications of a Cloud-based architecture?}}

    \subsection{Testing Takeaways} 
In terms of testing SOAR tools, we recommend designing investigation scenarios at the beginning of the process, in conjunction with SOC operators and then building the environment to establish all of the necessary tools an SOC analyst will need to work investigations, populating with necessary logs and alerts, and so on. 
As many baseline environments are needed, and (separately) many similar but different versions of each investigation scenario are needed, establishing these components and then automating these recurring processes is a necessary and worthwhile milestone. 

Investigations for testing SOAR tools arguably must require querying external intelligence feeds (e.g., we leveraged VirusTotal, CVE, NVD), yet some of the external information required for a particular scenario may change throughout the testing period. 
We recommend taking note of this need as tests are run and  accuracy measurements are being gathered. 

We allowed vendors 13 days to review materials and prepare for installation, and the research team provided a week of hands-on support. 
We suggest planning a longer setup/training period per tool, as many of our takeaways and results reflect need for more configuration effort. 

Much of the data collection was manual (e.g., recording Likert scores and free responses by hand). 
Establishing very consistent practices across team members is necessary for both uniformity in testing and manipulating the raw data into machine-readable data.  

Our results also showed that, once established, the test environment can be used for training analysts and red team members, testing other SOC tools in conjunction with a full tool suite, and creating simulated SIEM datasets.

\section{Conclusion}
In this work we present a novel user study involving experimental testing and evaluation of SOAR tools. We collaborated with SOAR tool vendors to configure their tool in our cyber range test environment. 
Once the tools were configured, operators from four US Navy SOCs as well as experienced participants from our internal team used the SOAR tools to perform five types of structured investigations that were designed to be similar in nature to investigations conducted regularly within SOCs. 
Data from several evaluation points, both qualitative and quantitative, was collected for comparison amongst the tools. 
Important takeaways from interpretation of the results include: 
(1) user success and SOAR tool performance is heavily dependent on configuration of the tool; 
(2) SOAR tools increase investigation efficiency and reduced context switching during investigations, although ticket accuracy and completeness (potentially indicating investigation quality) decreased with SOAR use; 
(3) user performance with a tool is slightly negatively correlated with their preference for the tool; and 
(4) internet dependence among the tools varies widely. 
In addition to these findings, we have produced a public version of our collected data for additional analysis.
We hope this study provides worthwhile lessons for future researchers.

% We found that SOAR configuration is critical for success, as it involves creative design for data display and automation and should involve iteration with users and vendors.
% We also found that SOAR tools increased efficiency and reduced context switching during investigations, although ticket accuracy and completeness (potentially indicating investigation quality) decreased with SOAR use. 
% Our findings indicated that user preferences from usability studies are slightly negatively correlated with their performance with the tool; 
% overautomation was a concern of senior analysts, and SOAR tools that balanced automation with assisting a user to make decisions were preferred. 
% We also found that SOAR dependence on constant internet varies widely with the tool. 

\section*{Acknowledgment}
The authors thank:  Mike Karlbom for ongoing support and leadership through the AI ATAC Challenge series; 
Jonathan Hodapp for the bureaucracy; 
Jessica Briesacker for the counseling (legal and otherwise); 
Jaimee Janiga and Laurie Varma for assistance with the infographic; 
Laurie Varma for editoral review; 
Mingyan Li for technical review, and translation of non-English written previous works. 

The research is based upon work supported by the US Department of Defense (DOD), Naval Information Warfare Systems Command (NAVWAR), via the US Department of Energy (DOE) under contract  DE-AC05-00OR22725.
The views and conclusions contained herein are those of the authors and should not be interpreted as representing the official policies or endorsements, either expressed or implied, of the DOD, DOE, NAVWAR, or the US Government.
The US Government is authorized to reproduce and distribute reprints for Governmental purposes notwithstanding any copyright annotation thereon.

\small
\bibliographystyle{IEEEtran}
\bibliography{references}

\begin{thebibliography}{10}
\providecommand{\url}[1]{#1}
\csname url@samestyle\endcsname
\providecommand{\newblock}{\relax}
\providecommand{\bibinfo}[2]{#2}
\providecommand{\BIBentrySTDinterwordspacing}{\spaceskip=0pt\relax}
\providecommand{\BIBentryALTinterwordstretchfactor}{4}
\providecommand{\BIBentryALTinterwordspacing}{\spaceskip=\fontdimen2\font plus
\BIBentryALTinterwordstretchfactor\fontdimen3\font minus
  \fontdimen4\font\relax}
\providecommand{\BIBforeignlanguage}[2]{{%
\expandafter\ifx\csname l@#1\endcsname\relax
\typeout{** WARNING: IEEEtran.bst: No hyphenation pattern has been}%
\typeout{** loaded for the language `#1'. Using the pattern for}%
\typeout{** the default language instead.}%
\else
\language=\csname l@#1\endcsname
\fi
#2}}
\providecommand{\BIBdecl}{\relax}
\BIBdecl

\bibitem{islam_multi-vocal_nodate}
C.~Islam and M.~A. Babar, ``\BIBforeignlanguage{en}{A {Multi}-{Vocal} {Review}
  of {Security} {Orchestration}},'' \emph{\BIBforeignlanguage{en}{ACM Computing
  Surveys}}, vol.~52, no.~2, p.~45, 2019.

\bibitem{fink2006bridging}
G.~A. Fink, V.~Duggirala, R.~Correa, and C.~North, ``Bridging the host-network
  divide: Survey, taxonomy, and solution.'' in \emph{LISA}, 2006, pp. 247--262.

\bibitem{goodall2004}
J.~Goodall, W.~Lutters, and A.~Komlodi, ``The work of intrusion detection:
  rethinking the role of security analysts,'' \emph{AMCIS 2004 Proceedings}, p.
  179, 2004.

\bibitem{botta2007towards}
D.~Botta, R.~Werlinger, A.~Gagn{\'e}, K.~Beznosov, L.~Iverson, S.~Fels, and
  B.~Fisher, ``Towards understanding it security professionals and their
  tools,'' in \emph{Proceedings of the 3rd symposium on Usable privacy and
  security}.\hskip 1em plus 0.5em minus 0.4em\relax ACM, 2007, pp. 100--111.

\bibitem{kokulu2019matched}
F.~B. Kokulu, A.~Soneji, T.~Bao, Y.~Shoshitaishvili, Z.~Zhao, A.~Doup{\'e}, and
  G.-J. Ahn, ``Matched and mismatched {SOCs}: A qualitative study on security
  operations center issues,'' in \emph{Proceedings of the 2019 ACM SIGSAC
  Conference on Computer and Communications Security}, 2019, pp. 1955--1970.

\bibitem{bridges2018information}
R.~A. Bridges, M.~D. Iannacone, J.~R. Goodall, and J.~M. Beaver, ``How do
  information security workers use host data? a summary of interviews with
  security analysts,'' \emph{arXiv preprint arXiv:1812.02867}, 2018.

\bibitem{gartner2019}
\BIBentryALTinterwordspacing
Gartner, ``Market guide for security orchestration, automation and response
  solutions.'' [Online]. Available:
  \url{https://www.gartner.com/en/documents/3942064/market-guide-for-security-orchestration-automation-and-r}
\BIBentrySTDinterwordspacing

\bibitem{aiatac3}
\BIBentryALTinterwordspacing
ORNL, ``{AI ATAC} 3 challenge description.'' [Online]. Available:
  \url{https://www.challenge.gov/challenge/AI-ATAC-3-challenge/}
\BIBentrySTDinterwordspacing

\bibitem{norem2021mathematical}
S.~Norem, A.~E. Rice, S.~Erwin, R.~A. Bridges, S.~Oesch, and B.~Weber, ``A
  mathematical framework for evaluation of {SOAR} tools with limited survey
  data,'' in \emph{Computer Security. ESORICS 2021 International
  Workshops}.\hskip 1em plus 0.5em minus 0.4em\relax Springer, 2022.

\bibitem{werlinger2009integrated}
R.~Werlinger, K.~Hawkey, and K.~Beznosov, ``An integrated view of human,
  organizational, and technological challenges of it security management,''
  \emph{Information Management \& Computer Security}, vol.~17, no.~1, pp.
  4--19, 2009.

\bibitem{werlinger2009security}
R.~Werlinger, K.~Hawkey, D.~Botta, and K.~Beznosov, ``Security practitioners in
  context: Their activities and interactions with other stakeholders within
  organizations,'' \emph{International Journal of Human-Computer Studies},
  vol.~67, no.~7, pp. 584--606, 2009.

\bibitem{werlinger2010preparation}
R.~Werlinger, K.~Muldner, K.~Hawkey, and K.~Beznosov, ``Preparation, detection,
  and analysis: the diagnostic work of it security incident response,''
  \emph{Information Management \& Computer Security}, vol.~18, no.~1, pp.
  26--42, 2010.

\bibitem{sundaramurthy2016turning}
S.~C. Sundaramurthy, J.~McHugh, X.~Ou, M.~Wesch, A.~G. Bardas, and S.~R.
  Rajagopalan, ``Turning contradictions into innovations or: How we learned to
  stop whining and improve security operations,'' in \emph{Proc. 12th Symp.
  Usable Privacy and Security}, 2016.

\bibitem{adetoye2023building}
B.~Adetoye and R.~C.-w. Fong, ``Building a resilient cybersecurity workforce: A
  multidisciplinary solution to the problem of high turnover of cybersecurity
  analysts,'' in \emph{Cybersecurity in the Age of Smart Societies: Proceedings
  of the 14th International Conference on Global Security, Safety and
  Sustainability, London, September 2022}.\hskip 1em plus 0.5em minus
  0.4em\relax Springer, 2023, pp. 61--87.

\bibitem{nyre-yu_identifying_nodate}
M.~Nyre-Yu, ``\BIBforeignlanguage{en}{Identifying {Expertise} {Gaps} in {Cyber}
  {Incident} {Response}: {Cyber} {Defender} {Needs} vs. {Technological}
  {Development}}.''

\bibitem{nyre2021identifying}
------, ``Identifying expertise gaps in cyber incident response: Cyber defender
  needs vs. technological development,'' in \emph{Proceedings of the 54th
  Hawaii International Conference on System Sciences}, 2021, p. 1978.

\bibitem{young_automated_2019}
\BIBentryALTinterwordspacing
S.~Young, ``\BIBforeignlanguage{en}{Automated systems only: why {CISOs} should
  switch off their dumb machines},'' \emph{\BIBforeignlanguage{en}{Network
  Security}}, vol. 2019, no.~9, pp. 6--8, Sep. 2019. [Online]. Available:
  \url{https://linkinghub.elsevier.com/retrieve/pii/S1353485819301060}
\BIBentrySTDinterwordspacing

\bibitem{brewer_could_2019}
\BIBentryALTinterwordspacing
R.~Brewer, ``\BIBforeignlanguage{en}{Could {SOAR} save skills-short {SOCs}?}''
  \emph{\BIBforeignlanguage{en}{Computer Fraud \& Security}}, vol. 2019,
  no.~10, pp. 8--11, Oct. 2019. [Online]. Available:
  \url{https://linkinghub.elsevier.com/retrieve/pii/S136137231930106X}
\BIBentrySTDinterwordspacing

\bibitem{tankard_pandemic_2020}
\BIBentryALTinterwordspacing
C.~Tankard, ``\BIBforeignlanguage{en}{Pandemic underpins need for {SOAR}},''
  \emph{\BIBforeignlanguage{en}{Network Security}}, vol. 2020, no.~5, p.~20,
  May 2020. [Online]. Available:
  \url{https://linkinghub.elsevier.com/retrieve/pii/S135348582030057X}
\BIBentrySTDinterwordspacing

\bibitem{fransen2019security}
F.~Fransen, R.~Wolthuis, and N.~el~Ouajdi, \emph{Security at machine
  speed}.\hskip 1em plus 0.5em minus 0.4em\relax Groningen: TNO, 2019.

\bibitem{Microsoft}
\BIBentryALTinterwordspacing
J.~Trull, ``Top 5 best practices to automate security operations,'' 2017.
  [Online]. Available:
  \url{https://www.microsoft.com/security/blog/2017/08/03/top-5-best-practices-to-automate-security-operations/}
\BIBentrySTDinterwordspacing

\bibitem{Markets}
\BIBentryALTinterwordspacing
Rohan, ``Security orchestration market with 1682.4 million usd by 2021,'' 2018.
  [Online]. Available:
  \url{https://www.marketsandmarkets.com/PressReleases/security-orchestration.asp}
\BIBentrySTDinterwordspacing

\bibitem{schneier}
\BIBentryALTinterwordspacing
``Security {Orchestration} and {Incident} {Response} - {Schneier} on
  {Security}.'' [Online]. Available:
  \url{https://www.schneier.com/blog/archives/2017/03/security_orches.html}
\BIBentrySTDinterwordspacing

\bibitem{nyre-yu_informing_2019}
M.~Nyre-Yu, K.~A. Sprehn, and B.~S. Caldwell,
  ``\BIBforeignlanguage{en}{Informing {Hybrid} {System} {Design} in {Cyber}
  {Security} {Incident} {Response}},'' in \emph{\BIBforeignlanguage{en}{{HCI}
  for {Cybersecurity}, {Privacy} and {Trust}}}, ser. Lecture {Notes} in
  {Computer} {Science}, A.~Moallem, Ed.\hskip 1em plus 0.5em minus 0.4em\relax
  Cham: Springer International Publishing, 2019, pp. 325--338.

\bibitem{njogu2013comprehensive}
H.~W. Njogu, L.~Jiawei, J.~N. Kiere, and D.~Hanyurwimfura, ``A comprehensive
  vulnerability based alert management approach for large networks,''
  \emph{Future Generation Computer Systems}, vol.~29, no.~1, pp. 27--45, 2013.

\bibitem{sadamatsu2016practice}
T.~Sadamatsu, Y.~Yoneyama, and K.~Yajima, ``Practice within fujitsu of security
  operations center: Operation and security dashboard,'' \emph{Fujitsu Sci.
  Tech. J.}, vol.~52, no.~3, pp. 52--58, 2016.

\bibitem{luo2016orchestration}
S.~Luo and M.~B. Salem, ``Orchestration of software-defined security
  services,'' in \emph{2016 IEEE International Conference on Communications
  Workshops (ICC)}.\hskip 1em plus 0.5em minus 0.4em\relax IEEE, 2016, pp.
  436--441.

\bibitem{koyama2015security}
T.~Koyama, B.~Hu, Y.~Nagafuchi, E.~Shioji, and K.~Takahashi, ``Security
  orchestration with a global threat intelligence platform,'' \emph{NTT
  Technical Review}, vol.~13, no.~12, 2015.

\bibitem{islam_architecture-centric_2020}
C.~Islam, M.~Ali~Babar, and S.~Nepal, \emph{Architecture-centric {Support} for
  {Integrating} {Security} {Tools} in a {Security} {Orchestration} {Platform}},
  Sep. 2020.

\bibitem{erdur_thesis_nodate}
\BIBentryALTinterwordspacing
E.~S. Erdur, ``\BIBforeignlanguage{en}{Continuous improvement on maturity and
  capability of security operation centers},''
  \emph{\BIBforeignlanguage{en}{{Masters Thesis, The Middle East Technical
  University}}}, p.~61. [Online]. Available:
  \url{https://open.metu.edu.tr/handle/11511/45040}
\BIBentrySTDinterwordspacing

\bibitem{kaur_introduction_2021}
\BIBentryALTinterwordspacing
G.~Kaur and A.~H. Lashkari, ``An {Introduction} to {Security} {Operations},''
  in \emph{Advances in {Cybersecurity} {Management}}, K.~Daimi and C.~Peoples,
  Eds.\hskip 1em plus 0.5em minus 0.4em\relax Cham: Springer International
  Publishing, 2021, pp. 463--481. [Online]. Available:
  \url{https://doi.org/10.1007/978-3-030-71381-2_21}
\BIBentrySTDinterwordspacing

\bibitem{singh_application_nodate}
K.~Singh, ``\BIBforeignlanguage{en}{Application of {SIEM}/{UEBA}/{SOAR}/{SOC}
  ({Cyber} {SUSS}) {Concepts} on {MSCS} 6560 {Computer} {Lab}},'' p.~82.

\bibitem{perera_next_2021}
A.~Perera, S.~Rathnayaka, N.~D. Perera, W.~Madushanka, and A.~N. Senarathne,
  ``The {Next} {Gen} {Security} {Operation} {Center},'' in \emph{2021 6th
  {International} {Conference} for {Convergence} in {Technology} ({I2CT})},
  Apr. 2021, pp. 1--9.

\bibitem{mavroeidis2020nonproprietary}
V.~Mavroeidis and J.~Brule, ``A nonproprietary language for the command and
  control of cyber defenses---{OpenC2},'' \emph{Computers \& Security},
  vol.~97, p. 101999, 2020.

\bibitem{leite2023actionable}
C.~Leite, J.~den Hartog, D.~Ricardo~dos Santos, and E.~Costante, ``Actionable
  cyber threat intelligence for automated incident response,'' in \emph{Secure
  IT Systems: 27th Nordic Conference, NordSec 2022, Reykjavic, Iceland,
  November 30--December 2, 2022, Proceedings}.\hskip 1em plus 0.5em minus
  0.4em\relax Springer, 2023, pp. 368--385.

\bibitem{sworna}
T.~Z. Sworna, I.~Chandi, and A.~B. Muhammad, ``{APIRO}: A framework for
  automated security tools {API} recommendation,'' \emph{submitted to
  Association for Computing Machinery}, 2021.

\bibitem{johnson2023soar4der}
J.~Johnson, C.~B. Jones, A.~Chavez, and S.~Hossain-McKenzie, ``Soar4der:
  Security orchestration, automation, and response for distributed energy
  resources,'' in \emph{Power Systems Cybersecurity: Methods, Concepts, and
  Best Practices}.\hskip 1em plus 0.5em minus 0.4em\relax Springer, 2023, pp.
  387--411.

\bibitem{cheng2022adopting}
\BIBentryALTinterwordspacing
Y.~Cheng-Hsiang, ``Adopting security orchestration automation and response to
  improve the efficiency of security operation center---a case study of {C}
  company,'' \emph{{Masters thesis, National Sun Yat-sen University in
  Taiwan}}, 2022. [Online]. Available:
  \url{https://ethesys.lis.nsysu.edu.tw/ETD-db/ETD-search/view_etd?URN=etd-0618122-114723}
\BIBentrySTDinterwordspacing

\bibitem{bilali2023iris}
V.-G. Bilali, D.~Kosyvas, T.~Theodoropoulos, E.~Ouzounoglou, L.~Karagiannidis,
  and A.~Amditis, ``Iris advanced threat intelligence orchestrator-a way to
  manage cybersecurity challenges of iot ecosystems in smart cities,'' in
  \emph{Internet of Things: 5th The Global IoT Summit, GIoTS 2022, Dublin,
  Ireland, June 20--23, 2022, Revised Selected Papers}.\hskip 1em plus 0.5em
  minus 0.4em\relax Springer, 2023, pp. 315--325.

\bibitem{miles-board}
T.~Al-Jody, ``Bearicade: A novel high-performance computing user and security
  management system augmented with machine learning technology,'' p. 183, 2021.

\bibitem{gibadullin}
R.~F. Gibadullin and V.~V. Nikonorov, ``Development of the system for automated
  incidient management based on open-source software,'' \emph{International
  Russian Automation Conference}, 2021.

\bibitem{vast_artificial_2021}
R.~Vast, S.~Sawant, A.~Thorbole, and V.~Badgujar, ``Artificial {Intelligence}
  based {Security} {Orchestration}, {Automation} and {Response} {System},'' in
  \emph{2021 6th {International} {Conference} for {Convergence} in {Technology}
  ({I2CT})}, Apr. 2021, pp. 1--5.

\bibitem{donevski_survey_2018}
M.~Donevski and T.~Zia, ``A {Survey} of {Anomaly} and {Automation} from a
  {Cybersecurity} {Perspective},'' in \emph{2018 {IEEE} {Globecom} {Workshops}
  ({GC} {Wkshps})}, Dec. 2018, pp. 1--6.

\bibitem{dutta_cybersecurity_2020}
S.~D. Dutta and R.~Prasad, ``Cybersecurity for {Microgrid},'' in \emph{2020
  23rd {International} {Symposium} on {Wireless} {Personal} {Multimedia}
  {Communications} ({WPMC})}, Oct. 2020, pp. 1--5, iSSN: 1882-5621.

\bibitem{zonouz2015}
S.~A. Zonouz, R.~Berthier, H.~Khurana, W.~H. Sanders, and T.~Yardley,
  ``Seclius: An information flow-based, consequence-centric security metric,''
  \emph{IEEE Transactions on Parallel and Distributed Systems}, vol.~26, no.~2,
  2015.

\bibitem{forte2017}
\BIBentryALTinterwordspacing
D.~Forte, ``Security orchestration \& automation: Parsing the options.''
  [Online]. Available:
  \url{https://www.darkreading.com/threat-intelligence/security-orchestration-and-automation-parsing-the-options/a/d-id/1329886?piddl_msgid=329392}
\BIBentrySTDinterwordspacing

\bibitem{rapid7}
\BIBentryALTinterwordspacing
J.~C. Montero, ``Accelerate incident response with security orchestration and
  automation.'' [Online]. Available:
  \url{https://www.rapid7.com/blog/post/2018/09/06/accelerate-incident-response-with-security-orchestration-and-automation/}
\BIBentrySTDinterwordspacing

\bibitem{forescout}
\BIBentryALTinterwordspacing
ForeScout, ``Forescout agentless visibility and control.'' [Online]. Available:
  \url{https://www.forescout.com/wp-content/uploads/2018/08/Agentless-Visibility-and-Control-ForeScout-White-Paper.pdf}
\BIBentrySTDinterwordspacing

\bibitem{greenfield}
\BIBentryALTinterwordspacing
Greenfield, ``Should {OT} follow {IT}'s centralized security orchestration?''
  [Online]. Available:
  \url{https://www.automationworld.com/products/networks/blog/13317415/should-ot-follow-its-centralized-security-orchestration}
\BIBentrySTDinterwordspacing

\bibitem{barbieri2020tweeteval}
F.~Barbieri, J.~Camacho-Collados, L.~E. Anke, and L.~Neves, ``Tweeteval:
  Unified benchmark and comparative evaluation for tweet classification,'' in
  \emph{Proceedings of the 2020 Conference on Empirical Methods in Natural
  Language Processing: Findings}, 2020, pp. 1644--1650.

\bibitem{schafer1999multiple}
J.~L. Schafer, ``Multiple imputation: a primer,'' \emph{Statistical methods in
  medical research}, vol.~8, no.~1, pp. 3--15, 1999.

\bibitem{austin2021missing}
P.~C. Austin, I.~R. White, D.~S. Lee, and S.~van Buuren, ``Missing data in
  clinical research: a tutorial on multiple imputation,'' \emph{Canadian
  Journal of Cardiology}, vol.~37, no.~9, pp. 1322--1331, 2021.

\bibitem{baxter_ironies_2012}
\BIBentryALTinterwordspacing
G.~Baxter, J.~Rooksby, Y.~Wang, and A.~Khajeh-Hosseini,
  ``\BIBforeignlanguage{en}{The ironies of automation: still going strong at
  30?}'' in \emph{\BIBforeignlanguage{en}{Proceedings of the 30th {European}
  {Conference} on {Cognitive} {Ergonomics} - {ECCE} '12}}.\hskip 1em plus 0.5em
  minus 0.4em\relax Edinburgh, United Kingdom: ACM Press, 2012, p.~65.
  [Online]. Available:
  \url{http://dl.acm.org/citation.cfm?doid=2448136.2448149}
\BIBentrySTDinterwordspacing

\bibitem{scikit}
F.~Pedregosa, G.~Varoquaux, A.~Gramfort, V.~Michel, B.~Thirion, O.~Grisel,
  M.~Blondel, P.~Prettenhofer, R.~Weiss, V.~Dubourg \emph{et~al.},
  ``Scikit-learn: Machine learning in python,'' \emph{Journal of machine
  learning research}, vol.~12, no. Oct, pp. 2825--2830, 2011.

\bibitem{asuero2006correlation}
A.~G. Asuero, A.~Sayago, and A.~Gonzalez, ``The correlation coefficient: An
  overview,'' \emph{Critical reviews in analytical chemistry}, vol.~36, no.~1,
  pp. 41--59, 2006.

\end{thebibliography}
% \addbibresource{references.bib}

\break
\normalsize
\appendix
    \begin{figure}
    \vspace{-.2cm}
    \centering
    \includegraphics[scale = 0.35]{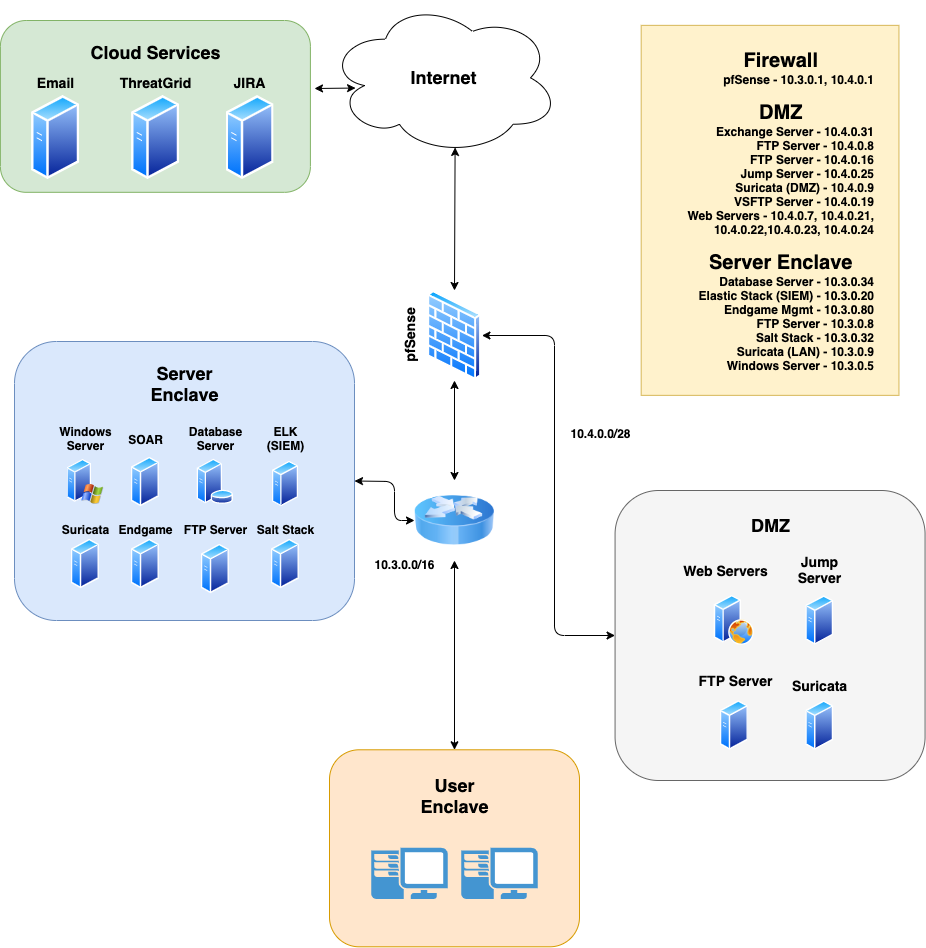}
    \caption{Network diagram of the experimental environment depicted.}
    \label{fig:network-diagram2}
    \vspace{-.35cm}
\end{figure}

\section{Appendix: Network Diagram} 
\label{sec:appendix-network} 
This diagram is a representative topology and set of network services and SOC tools that were used for our experimental environment. It complements Figure \ref{fig:network-diagram} by providing extra details, e.g., server IPs. The De-Militarized Zone (DMZ) is a firewall protected zone that contains a network of servers and/or services accessible from the internet. This zone is used to allow external network traffic access to said servers/services without the same traffic entering the user and server enclaves. By isolating these servers/services, the user and server enclaves are protected should one of the servers/services become compromised.

    \section{Appendix: SOAR Tool  Down-Selection Overview (AI ATAC 3 Phase 1)}
% \hl{this is a lot about phase 1 that i think we can condense and just site the phase 1 paper, if space becomes an issue}
\label{sec:phase1}
\label{sec:appendix-phase1}
This section of the appendix provides an overview of the main components of the Phase 1 experimental design. It describes the pipeline we have developed to carefully select the top contenders qualifying for the hands-on evaluation (i.e., Phase 2) of the challenge. 
Norem et al. \cite{norem2021mathematical} %\hl{[redacted]},
focuses exclusively on the Phase 1 study, providing full details of our Phase 1 methodology and findings. % and provides accompanying \hl{code } \cite{githubpage}, but a general overview of Phase 1 will be provided here. 
%Norem et al. \cite{norem2021mathematical}

%This phase kicked off the AI ATAC 3 Challenge 
Phase 1 began with the analysis of two videos, one ``overview'' and one ``demonstration,'' submitted by each of 11 SOAR tool vendors. 
These videos were viewed by US Navy SOC personnel and were meant to provide the necessary context of each tool and demonstrate its functionality.
After watching the videos, operators reviewed each tool according to a questionnaire. 
Prior to evaluating the tools based on the videos, operators were asked to rank six features of a SOAR tool in order of importance: ticketing, collaboration, automation, data ingestion, playbooks, and alert ranking. Tools were evaluated with a Likert-scale survey based on their capabilities in these areas. 
Operators were also asked to provide an overall score, also using a Likert scale, and a ranking of the tools that they reviewed. 
Free response answers on the questionnaire were converted to Likert values with sentiment analysis.
Because of the constraints on operator time, it was not feasible for every operator to review all 11 tools. 
Therefore, we requested that operators review a minimum of eight tools. 

Due to the sensitivity of statistical analyses on population size, we first identified the number of operators needed to ensure statistical validity of the results. 
To do this, we designed and implemented a simple set of simulations consisting of just two tools and assuming only one rating per tool per operator. We performed a sensitivity analysis to quantify how accurately we could use a hypothesis test with p-value 5\% defining statistical significance and to determine when one tool is preferred over another and when both tools are preferred equally, as the number of ratings varies. 

\begin{wrapfigure}[18]{l}{.65\linewidth}
    \vspace{-.2cm}
    \centering
    \includegraphics[width=\linewidth]{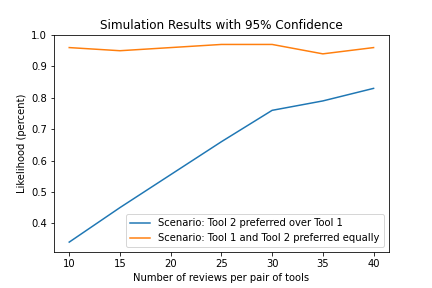}
    \caption{Results of simulation to quantify our ability to accurately determine tool preference using a two-sided $t$-test as the number of reviews changes.}
    \label{fig:simulation}
\end{wrapfigure}

Multiple hypothesis tests (t-, z-, $\chi^2-$, and Binomial) were used in a Monte Carlo simulation, so we could first quantify which tests provide the lowest Type I and Type II errors, simultaneously. The two-sided t-test was determined to be most accurate. 
We plotted the number of reviews per pair of tools against the likelihood of correctly concluding user preferences in Figure \ref{fig:simulation}. 
Secondly, we used these results to answer questions about the number of reviewers needed, as follows. 
Our study examined  [11 choose 2] = 55 pairs of tools, and each reviewer rated 8 tools, providing 28 = [8 choose 2] pairwise tool reviews. Hence, we received $n=k\times 28/55$ or approximately $n = k/2$ reviews per pair of tools. 

Using Figure \ref{fig:simulation}, and following discussions with the sponsor, we targeted at least $n=25$ reviews per pair of tool, requiring $k=50$ or more reviewers. Despite a widespread effort by our sponsor to recruit participants, only 18 total participants including one non--US Navy SOC operator participated in Phase 1. 

As mentioned above, operators reviewed only a subset of the submitted tools, which presents a challenge regarding assigning tools to ensure evenly distributed reviews. 
When a single operator rated two tools, we obtained a ``direct comparison''; 
hence, we sought to implement an assignment method that maximized the number of direct comparisons of all [11 choose 2] = 55 tool pairs. 
To do this, a greedy optimization procedure was implemented to approximate the optimal solution, or the case in which the direct comparisons are uniformly evaluated across the set of operators. 
The algorithm we used for this research was equipped with an objective function $f$ that analytically measures how far the current set of assignments is from the optimal set, and at each step it adds an assignment that minimizes $f$. 
This approach was sufficiently fast as well as flexible in terms of assigning tools to any number of operators. 
The assignment set generated was very close to optimal; specifically, each pair of tools received within one review of an optimal assignments number of reviews per pair. 
In practice, we ran the algorithm 100 times with more assignments than needed and selected the most optimal of all 100 sets.

Because operators did not review all 11 tools, the collected data contained missing values in some areas where no survey responses were obtained from any operator for a specific tool. 
A collaborative filtering algorithm was implemented and customized to predict unknown aspect ratings. 
From the complete set of aspect ratings, the unknown overall ratings were predicted with the best-performing supervised machine learning algorithm from the several tested.  
To account for all the collected data and ensure that our methods for filling in missing data were not interfering with the results, four methods for producing an overall ranking were used: (1) per-user raw overall ratings; (2) per-user raw rankings represented as a directed graph, with a diffusion algorithm used to determine the overall tool rankings; (3) per-user raw and predicted overall ratings; and (4) per-user predicted rankings (which are complete, as in ranking all tools) were used with the graph diffusion process.

% A collaborative filtering mechanism was designed to predict unknown aspect ratings, and many supervised learning algorithms (including the collaborative filtering) were tested and the best used to predict unknown overall ratings from the aspect ratings. 
% In order to take into account all the collected data, four methods for producing an overall ranking were used: (1) the per-user raw overall ratings were used; (2) the per-user (incomplete, raw) rankings were represented as a directed graph, and a diffusion algorithm was used to determine the overall ranking of tools; (3) the per-user raw and predicted overall ratings were used; (4) the per-user predicted (complete) rankings were used with the graph diffusion process. 
Notably, there was a clear dichotomy in the ranking from the four methods, providing confidence in the tools promoted to Phase 2 and those eliminated. Tools that were consistently low-scoring were eliminated from Phase 2. 
Statistical analysis showed that results were not correlated with operators' reported years of experience nor occupation but were correlated with familiarity with SOAR tools in general and with the quality of SOAR tools' videos. 
% In the end, through a combination of methods, we generated a complete dataset with a rating for each feature, an overall score, and an overall ranking of every tool for all operators. 

    \section{Appendix: Demographic Survey}
\label{demographicsurvey}
\label{sec:appendix-demographic-survey}

\quest{1}{How familiar are you with SOAR tools?}{
	\item Never used
	\item Somewhat familiar
	\item Expert user
}

\quest{2}{Which of these best fits your role?}{
	\item Security operator
	\item Network operator
	\item Systems administrator
	\item Other
}

\questionS{3}{How many years have you been in that role?}

\questionS{4}{How many different SOCs have you worked in?} 

\section{Appendix: Ease of Installation Survey}
\label{installsurvey}
\label{sec:appendix-install}

Each two-person research team assisting a SOAR vendor during their installation in our environment completed the following survey. 
They also noted any observations they felt would be relevant during analysis. 

\questionS{1}{Was the tool able to ingest all new data sources? Y / N. If no, provide details.}

\questionS{2}{Approximately how long did it take to configure the tool to ingest the new data source?}

\quest{3}{Please rate the following statements on a scale of 1 to 5, where 1 is ``very dissatisfied'' and 5 is ``very satisfied'':}{
	\item Overall, I am satisfied with the ease of ingesting new data sources with this tool. 
	\item Overall, I am satisfied with the amount of time it took to ingest new data sources with this tool. 
	\item Overall, I am satisfied with the support information (on-line help, messages, documentation) when ingesting new data sources with this tool. 
	\item Overall, I am satisfied with the support provided by the vendor when ingesting new data sources with this tool. 
}

\section{Appendix: Usability Rating and Ranking Sheet}
\label{usabilityandranking}
\label{sec:appendix-likert}
Analysts completed the following for each tool they tested and then ranked them by order of preference.

Please rate the following statements on a scale of 1 to 5, where 1 is ``not at all'' and 5 is ``very much so'':

\questionC{1}{Training}{
	\item I felt prepared for testing after the training period. 
	\item The training videos covered what I needed to know to use the tool.
}

\questionC{2}{1,000x per day}{
	\item It was easy to triage common events with this tool.  
	\item This tool helped me find the information I needed to triage an event. 
	\item This tool helped me understand the importance and meaning of alerts. 
	\item It was easy to create, update, and close tickets with this tool.  
}

\questionC{3}{Collaboration}{
	\item It was easy to collaborate with another analyst with this tool. 
	\item This tool helped me share my findings with my collaborator. 
	\item This tool made it easy to communicate nonverbally (via chat, for example) with my collaborator.  
	\item It was easy collaborate on tickets with this tool.  
}

\questionC{4}{Advanced persistent threats}{
	\item It was easy to triage an APT with this tool. 
	\item I was easy to initially create and populate a ticket with this tool.   
	\item It was easy to update/pass off a ticket with this tool. 
	\item This tool would significantly decrease the amount of time I spend triaging an APT. 
	\item This tool helped me find the information I needed to trace an attacker’s entry, movement within, and objectives in my network. 
	\item This tool helped me quickly understand the importance and meaning of alerts. 
	\item This tool helped me prioritize alerts and events so that I could focus on the most important threats.  
}

\questionC{5}{System usefulness}{
	\item Overall, I am satisfied with how easy it is to use this tool. 
	\item It was simple to use this tool. 
	\item I was able to complete the tasks and scenarios quickly using this tool. 
	\item I felt comfortable using this tool. 
	\item It was easy to learn to use this tool. 
}

\questionC{6}{Information quality}{
	\item The tool gave error messages that clearly told me how to fix problems. (“n/a” if did not experience error messages) 
	\item Whenever I made a mistake using the tool, I could recover easily and quickly. 
	\item The information (such as online help, onscreen messages, and other documentation) provided with this tool was clear. 
	\item It was easy to find the information I needed using this tool. 
	\item The information available in the tool was effective in helping me complete the tasks and scenarios. 
	\item The organization of information in the tool windows was clear. 
}

\questionC{7}{Interface quality}{
	\item The interface of this tool was pleasant. 
	\item I liked using the interface of this tool. 
	\item This tool has all the functions and capabilities I expect it to have. 
	\item Overall, I am satisfied with this tool.  
}

\questionC{8}{Ranking}{
	\item Rank tools in order (1 = best) based on which tool you want to see in your SOC.
}

\section{Appendix: Semistructured Interview Guide}
\label{semistructured}
\label{sec:appendix-sentiment}
We used the following questions for a semistructured interview with each analyst for each tool they tested and then coded their answers using open coding, as well as conducting a sentiment analysis on the results. 

\questionC{1}{1,000$\times$ per day}{
	\item What could be improved about this tool to help you triage common events? 
    \item What do you feel were the greatest strengths of this tool? 
    \item What do you feel were the greatest weaknesses of this tool? 
    \item Do you think that this tool would improve your SOC? How/in what ways? 
}

\questionC{2}{Collaboration}{
    \item Which parts of this tool helped you collaborate with other analysts the most? 
    \item What could be improved about this tool to help you collaborate with other analysts?  
    \item How did collaborating with this tool differ from the way that you normally collaborate in your SOC? 
}

\questionC{3}{Attack scenarios}{
    \item What do you feel was the greatest benefit this tool provided over your existing tool suite when triaging an APT? 
    \item What could be improved about this tool to help you triage APTs? 
    \item What could be improved about this tool to help with ticket management (create, update, pass off)? 
    \item Did this tool help you connect the dots to understand the actions and goals of an attacker? 
}

\questionC{4}{Usability and wrap-up}{
    \item Did the SOAR tool prioritize information in the way it was displayed so that it was useful? 
    \item Did the SOAR tool display information in a way that separates noise (too much data) from signal (needed information)? If so, does it throw away or occlude/exclude too much? 
    \item Did the SOAR tool group data so that multiple related events could easily be understood? 
    \item Does the tool promote standardization of processes (e.g., via playbooks)? 
    \item Would you want this tool in your SOC? Why/why not?  
    \item Is there anything else about this tool you think would be helpful for us to know? 
}

For data analysis, the free response questions listed below were reorganized thematic  subcategories with large overlap, but slight differences compared with the sections above. 
 
\questionB{1}{Recurring tasks}{
    \item What could be improved about this tool to help you triage common events?
    \item What could be improved about this tool to help with ticket management (create, update, pass off)?
    \item Does the tool promote standardization of processes (e.g., via playbooks)?
}
                                                                             
\questionB{2}{Collaboration}{
    \item Which parts of this tool helped you collaborate with other analysts the most?
    \item What could be improved about this tool to help you collaborate with other analysts?
    \item How did collaborating with this tool differ from the way that you normally collaborate in your SOC?
}

\questionB{3}{Attack scenarios}{
    \item What do you feel was the greatest benefit this tool provided over your existing tool suite when triaging an APT?
    \item what could be improved about this tool to help you triage APTs?
    \item Did this tool help you connect the dots to understand the actions and goals of an attacker?
}

\questionB{4}{Information quality}{
    \item Did the SOAR tool prioritize information in the way it was displayed so that it was useful?
    \item Did the SOAR tool display information in a way that separates noise (too much data) from signal (needed information)? If so, does it throw away or occlude/exclude too much?
    \item Did the SOAR tool group data so that multiple related events could easily be understood?
}

\questionB{5}{General}{
    \item (Prompt to participant to discuss anything they wish about the tool between Likert response questions and free response questions.)
    \item What do you feel were the greatest strengths of this tool?
    \item What do you feel were the greatest weaknesses of this tool?
    \item Do you think that this tool would improve your SOC? How/in what ways?
    \item Would you want this tool in your SOC? Why/why not?
    \item Is there anything else about this tool you think would be helpful for us to know?
    \item What tools are missing from this tool suite that you would normally use?
    \item How does the process used in this exercise differ from the way that you normally triage common events?
}

\section{Appendix: Multiple Imputation  for Correlation with Missing Data}
\label{sec:appendix-mi} 
We employed multiple imputation (MI) of missing data when computing correlations of two vectors. 
We let $\vec{x},\vec{y}$ be two vectors of length $N$ with missing data. 
Note that direct computation of correlation of $\vec{x},\vec{y}$, denoted $\text{corr}(\vec{x},\vec{y})$ by ignoring missing values, would have overweighted the contributions of the data. 
MI seeks to identify the expected correlation given the observed data through a Monte Carlo simulation. 
Our treatment followed methods described in Schafer \cite{schafer1999multiple} and Austin et al. \cite{austin2021missing}, involving sampling each missing data value $m$ times and computing imputed correlation values $\{\text{corr}(\vec{x}^i,\vec{y}^i)\}_{i=1}^m$, where $\vec{x}^i, \vec{y}^i$ denoted the now-filled vectors $\vec{x},\vec{y}$ with the $i$-th simulated missing values. 
The resulting estimate was simply the sample mean, $(1/m)\sum_i \text{corr}(\vec{x}^i,\vec{y}^i)$, and this Monte Carlo simulation allowed us to investigate statistical power of the estimates through hypothesis tests and confidence intervals.  
We chose $m=1,000$ by observing the average $\ell_1$ change of the resulting correlation matrix from the first $i$ samples to the $i+1$. 
\begin{wrapfigure}[19]{r}{.35\textwidth}
    \includegraphics[width = \linewidth]{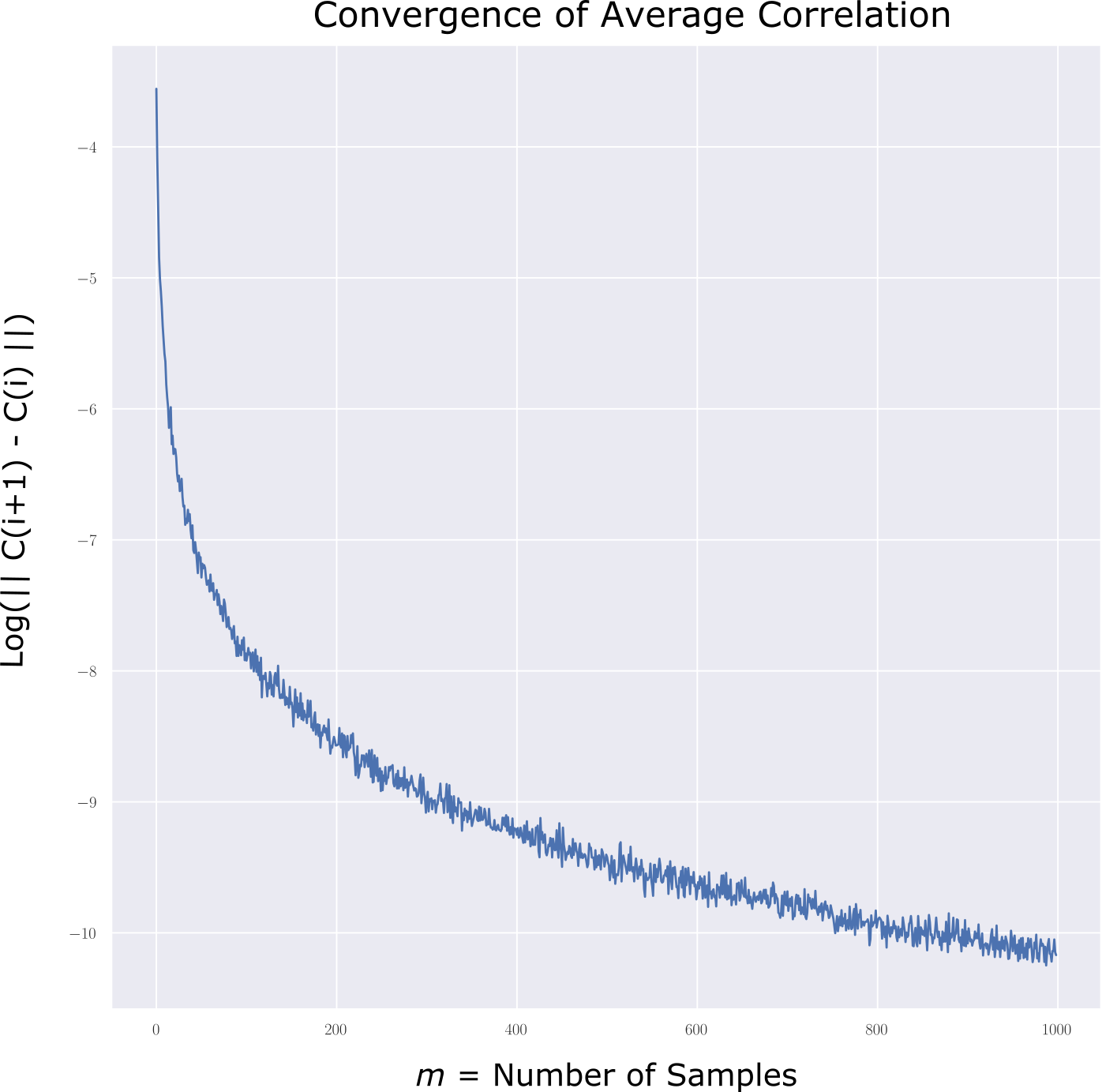}
    \caption{Log plot of average $\ell_1$ change in the imputed estimate (average correlation matrix) from the first $i$ imputed samples vs. $i+1$-st samples using the KDE for sampling. Similar convergence is observed for uniform random sampling.}
    \label{fig:convergence}
\end{wrapfigure}

\subsection{Which Density Function? Uniform Random  or Kernel Density from Peers}
In our case, all data was normalized to $[0,1]$. Generating samples for the missing components of $\vec{x},\vec{y}$ required a distribution on $[0,1]$. 
We considered two approaches. 
In the first approach, we sampled all missing data uniformly at random (probability density function $p(x) = 1$ for $x\in[0,1]$). 
As it may seem unbelievable that all possible values for a missing data point are equally likely, we considered a second approach in which a  distribution for each missing cell in our data table is learned from its peers' values. 
We first joined our user demographics and tool configuration data to the subcategory averages table so each row  (comprising the subcategory average measurements from a user testing a tool) includes information for a specific user and a specific tool, which may be pertinent to predicting missing values. 
For each subcategory (i.e., column in the original table), we trained a decision tree regressor to predict that target subcategory from the values in the other columns. 
Each leaf node of the fit decision tree was then considered a ``peer'' group as the rows of the data lying in that leaf are the subset of the table that has been learned to best predict the value of the target column for those with similar other columns. 
Armed with a decision tree's fit to predict a target column, we created a kernel density estimate (KDE) from the target column values in each leaf; that is,  probability density function $p(x) = \sum_{x_i\in \text{leaf}(x)} \phi(x)/|\text{leaf}(x)|$ with $\phi(x)$ the ``tophat'' (square wave) kernel with bandwidth $b = \min(.05, \text{dist(leaf}(x), \{0,1\}))$ and shifted to keep positive values inside $[0,1]$.  
We leveraged Scikit's \cite{scikit} Decision Tree and  passed the constraint that each leaf node (i.e., peer group) must have at least five training data points.  
Figure \ref{fig:kdes} depicts three such KDEs.

\begin{figure}
    \centering
    \includegraphics[width = \textwidth]{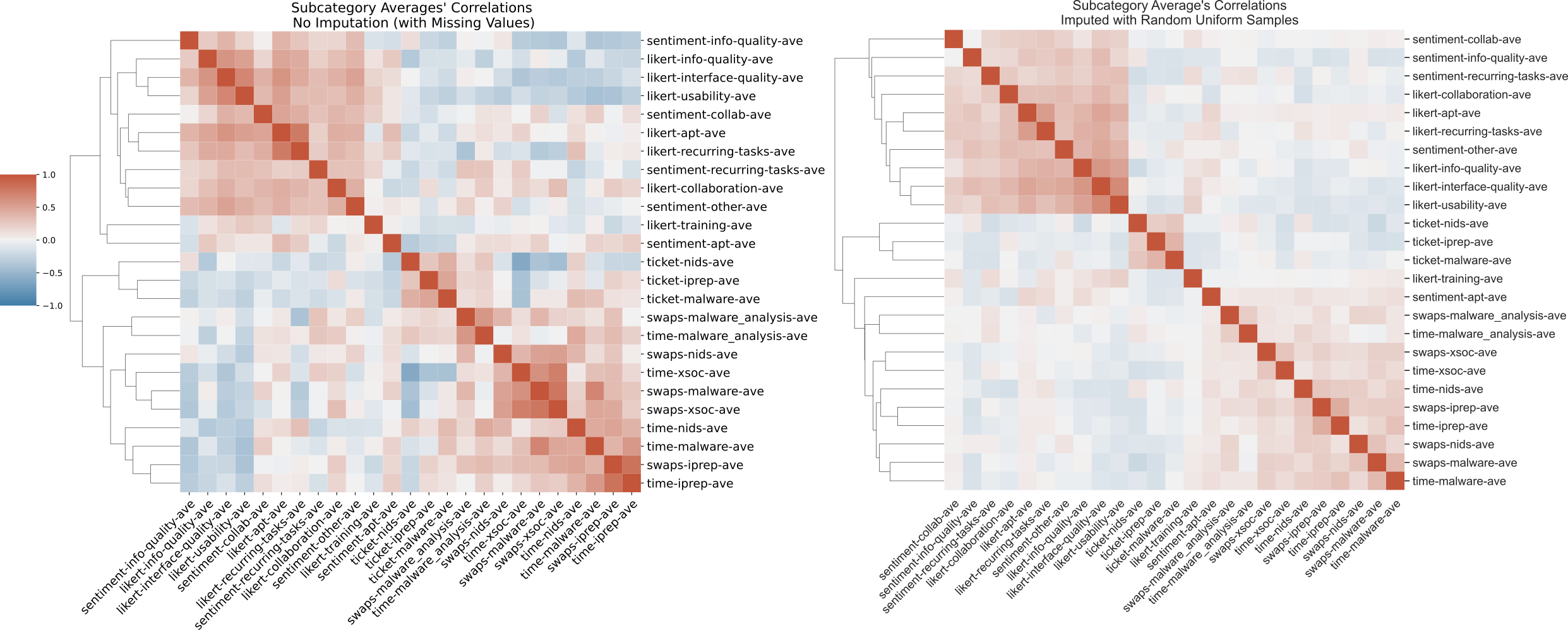}
    \caption{Heatmap and hierarchical clustering of subcategory average's correlations computed  (left) with missing data values and (right) missing values imputed uniformly at random. Compare with Figure \ref{fig:subcat-clustermap}, which provides the same plot using kernel densities for more precise sampling. 
    Viewing side-by-side shows the dampening effect imputation has to counteract overweighting caused by missing data. Importantly, while the groupings are slightly different, the main trend holds: the two primary clusters (Likert and Sentiment vs. all other subcategories) are consistent, with positive within-cluster correlation but slightly negative between-cluster correlations on average.}
    \label{fig:compare-imputations}
\end{figure}

% \begin{figure} 
% % \begin{wrapfigure}{r}{.65\textwidth}
%     \includegraphics[width = .99\linewidth]{images/random-and-kde-imputation-dampening.png}
%     % \includegraphics[width = .49\linewidth]{images/random-imputation-delta-scale-hue.png}
%     % \includegraphics[width = .49\linewidth]{images/kde-imputation-delta-scale-hue.png}
%     \caption{Plots show difference of imputed correlation from correlation with missing values (delta) vs. correlation with missing values. Linear models are displayed for different numbers of missing components. Plot supports the hypothesis uniform random sampling scales correlations by $k/N$ with $k$ the missing number of components and $N$ the length of the vector, while KDE imputation scales by less.}
%     \label{fig:mi-dampening}
% \end{figure}

\begin{figure} 
% \begin{wrapfigure}{r}{.65\textwidth}
    \includegraphics[width = .49\linewidth]{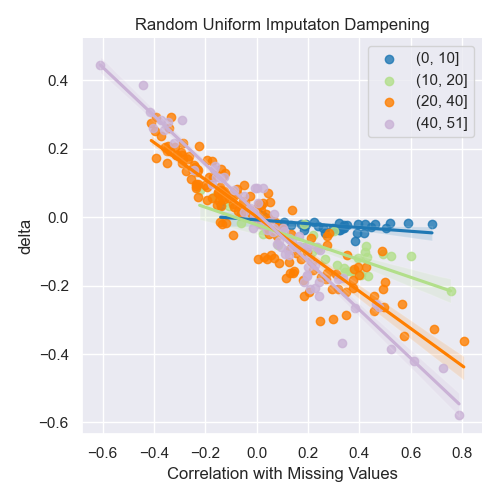}
    \includegraphics[width = .49\linewidth]{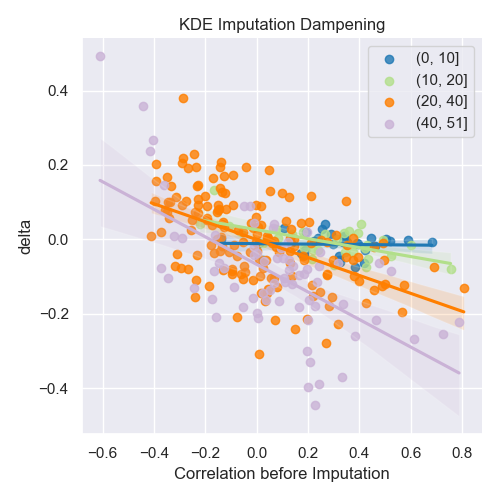}
    \caption{Plots show difference of imputed correlation from correlation with missing values (delta) vs. correlation with missing values. Linear models are displayed for different numbers of missing components. Plot supports the hypothesis uniform random sampling scales correlations by $k/N$ with $k$ the missing number of components and $N$ the length of the vector, while KDE imputation scales by less.}
    \label{fig:mi-dampening}
\end{figure}

\subsubsection{(More) Subcategory Averages' Correlation Results}
In our application to subcategory average correlations, we included the naive (i.e., no imputation, simply ignore missing values) correlation heatmap and clustering dendrogram alongside the uniform random sampling imputation in  Figure \ref{fig:compare-imputations}, and both can be compared with the KDE-imputation version discussed in the paper (Figure \ref{fig:subcat-clustermap}). 
\subsubsection{How MI Dampens Correlations}
We found that the uniform random imputations roughly mapped $\text{corr}(\vec{x},\vec{y}) \mapsto (-k/N)\text{corr}(\vec{x},\vec{y})$, where $k$ is the number of missing components in $x$ or $y$, and $N=68$ is the length of our vectors while, using the KDE sampling, the scale factor was roughly half as much. 
See Figure \ref{fig:mi-dampening}.

\subsubsection{Confidence Intervals} 
To compute confidence intervals with the imputed statistics, we converted the correlations via Fisher's transform ($f(r) := (1/2)\log{(1+r)/(1-r)}$ to an approximately normal random variable with known variance $U:= 1/(N-3)$, where $N=68$ is the length of our vectors \cite{asuero2006correlation}. 
We let $Q^i:= f( \text{corr}(\vec{x}^i,\vec{y}^i) $, $\bar{Q} = (1/m)\sum_i Q^i$ (our sample mean), $U^i = U = 1/(N-3) = 1/65$ the within-sample variance (i.e., the variance of $Q^i$), and  $B = \sum(Q^i - \bar{Q})^2/(m-1)$ our between-sample variance (i.e., the sample variance of observations caused by imputation). 
Our total variance, following Schafer \cite{schafer1999multiple}, was $U+B+B/m$, although the $B/m$ term was negligible for our data with $m = 1,000.$
Further, the literature sets up a t-test for our statistic, although  the degrees of freedom were so large in our case (>1,000) that the t-distribution became a normal distribution (again because $m=1,000$). 
Hence, we simply computed our confidence interval as $\bar{Q}\pm 1.96\sqrt{U+B}$ (1.96 corresponding to two-sided 95\% confidence from a standard normal distribution) and pulled these intervals back into $[0,1]$ via the $f^{-1}.$
See Figure \ref{fig:mi-confidence-intervals} for confidence intervals from the KDE imputations. 
Note that when using the uniform distribution to sample, the confidence intervals were much larger, and nearly all crossed 0.

% \begin{threeparttable}
% \centering
% \caption{Table provides statistics to accompany Figures \ref{fig:subcat-clustermap} and \ref{fig:mi-confidence-intervals}, which  present correlation results of subcategory averages. This table depicts how multiple imputation (MI) of missing values when computing correlation dampens the propensity of missing data to amplify correlation values from the observed data.}
% \label{tab:mi}
%   \begin{tabular}{ccc}
% \toprule
% \multirow{2}{*}{\shortstack{\textbf{Correlation with}\\\textbf{Missing Data}}} & \multirow{2}{1.2in}{\shortstack{\textbf{Average Change}\\ \textbf{by MI}}} &  \multirow{2}{*}{\textbf{Count}} \\
% \\
% \midrule
% (-1.0,-0.5]                      &                 0.00 &     0 \\
% (-0.5,-0.2]                    &                 0.18 &    30 \\
% (-0.2,0.0]                       &                 0.06 &    82 \\
% (0.0,0.2]                        &                -0.05 &    72 \\
% (0.2,0.5]                      &                -0.15 &    99 \\
% (0.5,1.0)                        &                -0.25 &    17 \\
% \cdashlinelr{1-3}
% (-1.0,1.0)                         &                -0.04 &   300 \\
% \bottomrule
% \end{tabular}
% \end{threeparttable}
% \begin{figure}
%     \centering
%     \includegraphics[width=\textwidth]{images/imputation-before-clustermap-Subcategory-Average-Data-inkscaped.png}
%     \caption{Heatmap and hierarchical clustering of subcategory averages with missing values, i.e., with no imputation.}
%     \label{fig:my_label}
% \end{figure}

\end{document}